\documentclass[twocolumn]{aastex631}

\usepackage[printonlyused,withpage]{acronym}
\graphicspath{{./}{figures/}}

\begin{document}

\title{Compact Binary Merger Rate with Modified Gravity in Dark-Matter Spikes}

\correspondingauthor{Saeed Fakhry}
\email{s\_fakhry@sbu.ac.ir}

\author[0000-0002-6349-8489]{Saeed Fakhry}
\email{s\_fakhry@sbu.ac.ir} 
\affiliation{Department of Physics, Shahid Beheshti University, 1983969411, Tehran, Iran}

\author{Sara Gholamhoseinian}
\email{s.gholamhoseinian@gmail.com} 
\affiliation{Department of Physics, Shahid Beheshti University, 1983969411, Tehran, Iran}

\author{Marzieh Farhang}
\email{m\_farhang@sbu.ac.ir} 
\affiliation{Department of Physics, Shahid Beheshti University, 1983969411, Tehran, Iran}
 
\begin{abstract}
In this study, we investigate the impact of modified gravity (MG) on the merger rate of compact binaries within dark-matter spikes surrounding super-massive black holes (SMBHs). Specifically, we calculate binary merger rates involving primordial black holes (PBHs) and/or neutron stars (NSs) in Hu-Sawicki $f(R)$ gravity and the normal branch of Dvali-Gabadadze-Porrati (nDGP) gravity, with three SMBH mass functions, Benson, Vika, and Shankar. The results show consistently higher merger rates predicted for PBH-PBH and PBH-NS binaries in these gravity models compared to general relativity (GR), in particular at lower SMBH masses and for steeper dark-matter spike density profiles. The predicted merger rates are compared to the LIGO-Virgo-KAGRA observations to constrain the parameters of the theory. In particular we find steeper dark-matter spike density profiles in the MG scenarios compared to GR. When compared to current observational constraints on PBH abundance, the mass ranges allowed by Hu-Sawicki $f(R)$ models are found to be wider than by nDGP models, for given merger rates. The results are highly dependent on the choice of SMBH mass function, with Vika and Shankar mass functions predicting lower abundances. The considerable sensitivity of the results on the assumed gravity scenario and SMBH mass function demonstrates the necessity of incorporating the corresponding theoretical uncertainties in making relatively robust predictions on compact binary merger rates and, as a result on PBH properties.
\end{abstract}

\section{Introduction} \label{sec:i}
Gravitational waves (GWs) have revolutionized our understanding of the Universe, offering unprecedented insights into cosmological and astrophysical phenomena. Since the direct detection by the LIGO-Virgo-KAGRA (LVK) observatories, GWs have unveiled the dynamical Universe through events such as compact binary mergers \citep{2016PhRvL.116f1102A, 2016PhRvL.116x1103A, 2016PhRvL.116v1101A, 2020ApJ...896L..44A, 2020PhRvL.125j1102A}. These detections primarily fall into three categories: binary black holes (BBHs), black hole-neutron star (BH-NS) binaries, and binary neutron stars (BNSs). Notably, the majority of the GW events detected are BBH mergers, with masses ranging in $(10\mbox{-}100)\,M_{\odot}$ \citep{2019PhRvX...9c1040A, 2021PhRvX..11b1053A, 2021arXiv211103606T}. Although several studies have discussed the origins of these BHs, their origin is a subject of ongoing debate \citep[see, e.g.,][]{2019JCAP...02..018R, 2021MNRAS.507.5224B, 2021PhRvL.126b1103N, 2021PhRvL.127o1101N, 2022PhR...955....1M}. They might have originated from the collapse of massive stars via various channels or be primordial black holes (PBHs) formed in the early Universe, e.g., through the collapse of primordial density fluctuations. Distinguishing between these formation pathways is crucial for comprehending the processes of BH formation and the conditions of the early Universe. The hypothesis that many of the BHs detected by the LVK collaboration are primordial has gained attraction, supported by cosmological theories that suggest PBHs form from significant density peaks exceeding a critical threshold \citep[see, e.g.,][]{2016PhRvL.116t1301B, 2017PDU....15..142C, 2016PhRvL.117f1101S, 2021PhRvD.103l3014F, 2022PhRvD.105d3525F, 2022ApJ...941...36F, 2023PDU....4101244F, 2023PhRvD.107f3507F, 2023arXiv230811049F, 2024ApJ...966..235F}.

Nonetheless, due to potential uncertainties regarding the contribution of PBHs to these merger events, careful consideration is warranted. For example, an analysis of the GW catalog suggests that astrophysical formation models may greatly influence the proportion of PBHs within this subpopulation \citep{2022PhRvD.105h3526F}. Additionally, it is crucial to acknowledge that other formation theories might also be compatible with the mergers detected by LVK detectors \citep{2022LRR....25....1M}. Primordial black holes are notable for their diverse mass range and are considered potential candidates for dark matter \citep[e.g.,][]{2016ApJ...823L..25K, 2016JCAP...11..036B, 2018JCAP...01..004B, 2021FrASS...8...87V, 2021arXiv211002821C}. Nevertheless, other potential candidates for dark matter are under serious discussion, see, e.g., \citep{2000PhRvL..84.3760S, 2000ApJ...542..281A, 2012LRR....15....6L, 2018RPPh...81f6201R, 2020PhRvL.124j1303B, 2020PhRvD.102h4063D, 2021CQGra..38s4001D}. This aspect integrates them into the broader cosmological models where dark sectors are pivotal \citep[see, e.g.,][]{2011PhRvL.107b1302S, 2014JCAP...08..034Y, 2018RPPh...81a6901H, 2020PDU....3000666D, 2021arXiv210700562G, 2023PDU....3901144G, 2023PDU....4101259D, 2024ARep...68...19D, 2024PDU....4601544F}. Recent advancements in observational methods have been instrumental in constraining the abundance of PBHs across different mass ranges, providing valuable data on the early Universe at smaller scales \citep{2017PhRvD..96b3514C, 2021RPPh...84k6902C}. This adds another layer to our understanding of the cosmological significance of PBHs. 

Black holes can also form binary systems with NSs in dense environments like star clusters, active galactic nuclei, and central regions of dark-matter halos. These BH-NS mergers have the potential to emit GW and electromagnetic signals, providing valuable insights for multimessenger astronomy \citep{2020EPJA...56....8B, 2021FrASS...8...39R}. However, not all BH-NS mergers can generate an electromagnetic counterpart. For instance, if the mass ratio between the components exceeds a certain threshold, the NS will be entirely swallowed by the BH without any mass ejection \citep[see, e.g., ][]{2015PhRvD..92h1504P, 2018PhRvD..98h1501F, 2020PhRvD.101j3002K}. Gravitational wave detectors can provide insights into the NS nuclear equation of state, and BH accretion processes. Gravitational wave observations of two such mergers revealed component masses of $(8.9^{+1.2}_{-1.5}, 1.9^{+0.3}_{-0.2}) M_{\odot}$ and $(5.7^{+1.8}_{-2.1}, 1.5^{+0.7}_{-0.3}) M_{\odot}$ \citep{2021ApJ...915L...5A}. Despite uncertainties in the formation and merging processes of BH-NS binaries, further study of their evolution is promising. 

Substantial evidence supports the existence of supermassive black holes (SMBHs) at the centers of galactic halos \citep[e.g.,][]{2021NatRP...3..732V, 2021AstL...47..515P, 2022PhRvD.106d3018S, 2022MNRAS.511.5436D}. This is drawn from observing the Keplerian velocity dispersion of stars in the innermost regions of galactic halos \citep[For more details, see][]{2005SSRv..116..523F, 2013ARA&A..51..511K}. Central SMBHs are believed to enhance the density of nearby dark-matter particles. It is suggested that a dense region, termed the dark-matter spike, forms around a central SMBH if it evolves adiabatically from an initial power-law cusp \cite{1999PhRvL..83.1719G}. The high dark-matter density within these spikes might imply a significant number density of PBHs. The merger rate of compact binaries within dark-matter spikes has been examined through the lens of GR in several studies \citep{2019PhRvD..99d3533N, 2022ApJ...931....2S, 2023ApJ...947...46F}. 

The interaction between SMBHs and dark matter halos plays a critical role in the formation and evolution of dark matter spikes. As SMBHs grow through accretion or mergers, their gravitational influence causes adiabatic contraction, concentrating dark matter toward the galactic center and forming a spike region \citep{1999PhRvL..83.1719G}. A slow SMBH growth rate allows dark matter to redistribute smoothly, enhancing this spike, whereas SMBH mergers may disrupt it through GW recoil, ejecting dark matter from the center \citep{2024MNRAS.532.4681P}. Moreover, the specific structure of the dark matter halo, such as the cuspy Navarro-Frenk-White (NFW) profile or cored halos, significantly influences the spike's strength, with denser halos producing more pronounced spikes \citep{2014PhRvD..89f3534L}. Also, Some key factors like halo mass, concentration, and external effects such as stellar interactions, supernovae feedback, galaxy mergers, and tidal forces can further modulate the structure of dark matter spikes \citep[see, e.g.,][]{2001PhRvD..64d3504U, 2018ARA&A..56..435W, 2024MNRAS.527.3196S}.

In recent years, due to various theoretical challenges and observational tensions faced by GR \citep[e.g.,][]{2014JCAP...08..034Y, 2018RPPh...81a6901H}, modified gravity (MG) theories have been proposed \citep[e.g.,][]{2007IJGMM..04..115N, 2012PhR...513....1C, 2021arXiv210512582S}. These theories incorporate additional mechanisms to satisfy observational criteria in extensively tested high-density environments and matching the expansion history of the Universe \citep[see, e.g.,][]{2015Symm....7..220B, 2018arXiv181002652A, 2019LRR....22....1I, 2022PDU....3701106S, 2023PhLB..84337988O, 2024ApJ...971..138S}. The Chameleon and Vainshtein effects are two of the most frequently studied mechanisms in this context \citep{1972PhLB...39..393V, 2013ApJ...779...39J}.

Among MG theories, the Hu-Sawicki $f(R)$ model \citep{2007PhRvD..76f4004H} and the normal branch of Dvali-Gabadadze-Porrati (nDGP) model \citep{2000PhLB..485..208D} have garnered significant attention due to their ability to provide viable modifications to GR. These theories introduce additional degrees of freedom that can potentially account for observed cosmic acceleration without the need for a cosmological constant \citep{2008PhRvD..77b4024C, 2016MNRAS.459.3880H, 2021MNRAS.508.4140M}. The Hu-Sawicki $f(R)$ model modifies the Ricci scalar $R$ in the Einstein-Hilbert action, introducing a scalar degree of freedom known as the scalaron, which can drive the late-time acceleration of the Universe. The primary advantage of the Hu-Sawicki model lies in its ability to reproduce the expansion history of the Universe while remaining consistent with solar system tests through the Chameleon mechanism. This mechanism ensures that deviations from GR are minimal in high-density regions, thereby passing stringent local gravity tests \citep{2013JCAP...04..029B, 2016ARNPS..66...95J, 2018LRR....21....1B}. On the other hand, the nDGP model, a braneworld gravity scenario, modifies GR by embedding the four-dimensional Universe within a higher-dimensional space, naturally incorporating self-acceleration and eliminating the need for a dark energy component. The additional graviton modes in the nDGP model contribute to the accelerated expansion observed in the Universe without conflicting with local gravity constraints. Moreover, both MG models can address dark matter phenomena without introducing a cosmological constant. These theories extend the landscape of possible modifications to gravity, providing alternative explanations that can be tested against observational data. Here, an important question emerges: what do MG models predict about the merger rate of compact binaries within dark-matter spikes? Note that the choice of theoretical models is crucial, as they offer distinct predictions regarding binary merger rates, influenced by their unique formulations and parameter sensitivities. This raises the question of degeneracy with other MG theories. Naturally, different models can yield varying predictions due to their specific characteristics.

In this work, we propose to study the merger rates of compact binaries in dark-matter spikes, utilizing Hu-Sawicki $f(R)$ and nDGP MG theories. By comparing these results with those obtained from GR, we aim to enhance our understanding of how MG theories impact the dynamics of compact binaries within dark-matter spikes. Compact binaries in galactic halos can form through various channels. In this work, we focus particularly on determining the merger rate of compact binaries formed through dissipative two-body dynamical encounters. In these cases, the binaries are expected to emit gravitational radiation immediately and merge. However, three-body interactions can also lead to the formation of compact binaries. These processes typically result in the creation of wide binaries within dark matter halos, whose binding energy is insufficient to cause immediate decay through GW emission. Consequently, binaries formed through these interactions generally have merger times longer than the Hubble time \citep{1989ApJ...343..725Q} and remain undetectable by LVK observatories. Nevertheless, some binaries formed through three-body interactions can survive and represent the dominant mechanism of binary formation in collisional star clusters \citep{1993ApJ...403..271G, 2015ApJ...800....9M, 2024ApJ...970..112A}. In such cases, binary-single interactions within stellar environments cause these binaries to harden, shrinking their semimajor axis, until they enter the GW-dominated regime \citep{1996NewA....1...35Q}.

The structure of this work is organized as follows: In Sec.\,\ref{sec:ii}, we describe the theoretical foundations of the MG models. Next, in Sec.\,\ref{sec:iii}, we present a suitable model for dark-matter spikes, discussing essential parameters such as the spike density profile, the SMBH mass function, and the concentration parameter. Then, in Sec.\,\ref{sec:iv}, we focus on calculating the merger rate of compact binaries in dark-matter spikes using MG models and comparing these results with those obtained from GR. Additionally, we compare our findings with data from the LVK detectors to constrain the power-law index. We also compare the predictions of our analysis with observational data on the abundance of PBHs. Finally, in Sec.\,\ref{sec:v}, we review the results and summarize the conclusions.
\section{Modified Gravity Models} \label{sec:ii} 
In this section, we introduce two pivotal models prevalent in the MG literature: the Hu-Sawicki $f(R)$ model and the nDGP model. These models exemplify the screening mechanisms known as the Chameleon and Vainshtein classes.

\subsection{Hu-Sawicki $f(R)$ Model}
The Hu-Sawicki $f(R)$ model incorporates a non-linear modification function, denoted as $f(R)$, into the traditional Einstein-Hilbert action \citep{2007PhRvD..76f4004H}:
\begin{equation}\label{frgravity}
S=\int d^{4}x \sqrt{-g}\left[\frac{R+f(R)}{2\kappa}+\mathcal{L}_{\rm m}\right],
\end{equation}
where $R$ is the Ricci scalar, $\kappa$ is the Einstein gravitational constant, $g$ is the metric determinant, and $\mathcal{L}_{\rm m}$ is the matter Lagrangian. By setting $f = -2\Lambda$, where $\Lambda$ represents a cosmological constant, GR can be restored.

By applying a conformal transformation, Eq.\,(\ref{frgravity}) can be reformulated into a scalar-tensor theory involving the scalaron, denoted as $f_{\rm R} \equiv {\rm d}f(R)/{\rm d}R$. The scalaron represents the degree of freedom introduced by MG. Choosing an appropriate functional form for $f(R)$ is essential to facilitate cosmic acceleration in the late-time Universe while adhering to the constraints imposed by solar system tests \citep{2008PhRvD..78j4021B}. A class of broken power-law models, known as the Hu-Sawicki model, effectively addresses these requirements \citep{2007PhRvD..76f4004H}. The Hu-Sawicki model is described as follows:
\begin{equation}
f_{R}=-m^2\frac{c_1(R/m^2)^n}{c_2(R/m^2)^n+1},
\end{equation}
where $m^2=\kappa\bar{\rho}_{\rm{m}0}/3$ represents the characteristic mass scale, and $\bar{\rho}_{\rm{m}0}$ denotes the present-day background matter density. The parameters $c_1$, $c_2$, and $n>0$ are dimensionless free parameters that must be chosen carefully to reproduce the expansion history and pass solar-system tests using the chameleon mechanism.

It is important to maintain the stability of the solution in high-density regions, where $R \gg m^2$. Additionally, cosmological experiments based on the $f(R)$ model should be consistent with those based on GR. To meet this criterion, $f_{RR}={\rm d}^2 f/{\rm d}R^{2} > 0$ must hold. Thus, the Hu-Sawicki model can be expanded as follows:
\begin{equation}\label{limfr}
\lim_{m^2/R \rightarrow 0} f(R) \approx -\frac{c_1}{c_2}m^2+\frac{c_1}{c_2^2}m^2\left(\frac{m^2}{R}\right)^n.
\end{equation}

Although the Hu-Sawicki model does not include an actual cosmological constant, it exhibits features similar to a cosmological constant in both large-scale and local experiments. Furthermore, the finite value of $c_1/c_2$ results in a constant curvature that remains unaffected by changes in matter density. As a result, a class of models can be formulated that accelerates the expansion of the Universe, mirroring the behavior observed in the standard model of cosmology. Therefore, relation\,(\ref{limfr}) can be rewritten as:
\begin{equation}
f(R) \approx -\frac{c_1}{c_2}m^2-\frac{f_{R_{0}}}{n}\frac{\bar{R}_{0}^{n+1}}{R^n},
\end{equation}
where $\bar{R}_{0}$ denotes the present-day background curvature, and $f_{R0} \equiv f_R(\bar{R}_0)$ is the field strength. Note that if $|f_{R0}| \rightarrow 0$, one can obtain $(c_1/c_2) m^2 = 2\kappa \bar{\rho}_{\Lambda}$, where $\bar{\rho}_{\Lambda}$ represents the inferred background energy density attributed to dark energy. Cosmological and solar-system tests have constrained the field strength $f_{R0}$ \citep{2020PhRvD.102j4060D}. Various values of $|f_{R0}|$, ranging from $10^{-4}$ to $10^{-8}$, have been investigated in the literature \citep{2019JCAP...09..066M}. In our study, we focus on the Hu-Sawicki model with $n = 1$ and $|f_{R0}| = 10^{-4}, 10^{-5}$, and $10^{-6}$, denoted as $f4$, $f5$, and $f6$, respectively. 
\subsection{nDGP Model}
The nDGP model of gravity is a MG theory proposed in \cite{2000PhLB..485..208D}. The nDGP model envisions the Universe as a four-dimensional brane embedded in a five-dimensional Minkowski space. The action comprises two terms:
\begin{equation}\label{ndgpgravity}
S=\int d^{4}x \sqrt{-g}\left[\frac{R}{2\kappa}+\mathcal{L}_{\rm m}\right] + \int d^{5}x \sqrt{-g_5}\frac{R_{5}}{2\kappa_{5} r_{\rm c}},
\end{equation}
where $R_5$, $g_5$, and $\kappa_5$ are the Ricci scalar, metric determinant, and Einstein gravitational constant of the fifth dimension, respectively. Moreover, $r_{\rm c}=(\kappa_5/2 \kappa)$ is the crossover distance, representing the scale below which GR is effective in a four-dimensional framework. For scales larger than $r_{\rm c}$, the second term in the action is no longer insignificant, leading to deviations from GR.

The general DGP model consists of two branches: the ``normal" branch (nDGP), and the ``self-accelerating" branch (sDGP). We focus on the former due to its freedom from ghost instabilities \citep{2009PhRvD..80f3536L}. At larger scales, gravity strengthens, while at smaller scales, gravity behaves like GR due to Vainshtein screening. Consequently, the nDGP model can be explored to replicate the $\Lambda$CDM expansion history. This approach is promising due to extensive prior simulations. In this case, the sole adjustable parameter to be constrained is $n = H_0 r_{\rm c}$ (where $H_0$ denotes the Hubble constant), with values between $1$ and $5$ being thoroughly investigated. Note that GR is recovered if $n \rightarrow \infty$, corresponding to a steep gradient of gravitational force in Vainshtein screening. Numerous studies have explored the nDGP model's implications for structure formation and cosmology through numerical simulations and observational data comparisons. This work examines three values of $n=1, 2$, and $5$, referred to as nDGP(1), nDGP(2), and nDGP(5), respectively.
\section{Dark-Matter Spike Model} \label{sec:iii} 
In this section, we present a theoretical framework to describe the structural and statistical characteristics of dark-matter spikes, enabling us to calculate the merger rate of compact binaries in these regions.
\subsection{The density profile}\label{sec.iia}
In cosmological perturbation theory, dark matter halos are recognized as nonlinear structures that form hierarchically and are distributed throughout the Universe due to the collapse of linear cosmological fluctuations. Indirect observations, such as the rotation curves of galaxies, suggest that dark matter particles are not uniformly distributed within galactic halos \citep[see, e.g.,][]{2010AdAst2010E...8K}. This observation is particularly relevant for SMBHs located at galactic centers. While ordinary BHs may form from stellar collapse, the formation of SMBHs at high redshifts challenges standard astrophysical scenarios. It is hypothesized that the mass of an SMBH is correlated with the mass of its host dark matter halo, indicating a coevolution of SMBHs and their host halos \citep{2000ApJ...539L...9F}. Self-interacting dark matter halo models predict that early SMBH seeds may form through the gravothermal catastrophe \citep{2019JCAP...07..036C, 2021ApJ...914L..26F, 2022JCAP...08..032F}. As a result, an SMBH is expected to be surrounded by a dense spike of dark matter at the center of the galactic halo.

Assuming $M_{\rm SMBH}$ is the mass of the SMBH at the galactic center, the halo density profile can be expressed as $\rho(r) \simeq \rho_0(r_0/r)^\gamma$, where $\rho_0$ and $r_0$ are characteristic parameters of the halo, and $\gamma$ is the power-law index. Numerical simulations for dark matter suggest that for the power-law index we have $0.9 < \gamma < 1.2$ \citep{2008Natur.454..735D, 2010MNRAS.402...21N}. However, baryonic matter collapsing into a baryonic disk can result in steeper power-law indices \citep{2004ApJ...616...16G, 2006PhRvD..74l3522G}. For instance, in the central regions of the Milky Way, the power-law index is estimated to be around $\gamma \sim 1.6$ \citep{2015JCAP...12..001P}. Based on this, the radius of the dark-matter spike can be described by the following relation \citep{1999PhRvL..83.1719G}:
\begin{equation}
r_{\rm sp} = a_{\gamma}r_{0}\left(\frac{M_{\rm SMBH}}{\rho_{0} r_{0}^{3}}\right)^{1/(3-\gamma)},
\end{equation}
where $a_{\gamma}$ is determined numerically for each power-law index $\gamma$.

By incorporating a Newtonian approach to the allowed energies and angular momenta of the dark matter, the density profile of the dark-matter spike for $r$ in the range of $4r_{\rm s}<r< r_{\rm sp}$ is given by \footnote{Subsequently, in \cite{2022PhRvD.106d4027S}, a comprehensive relativistic analysis was conducted, with selected scale parameter values revealing notable qualitative differences in spike characteristics compared to \cite{1999PhRvL..83.1719G}. In the relativistic framework, spikes can be significantly closer to the central BH, indicating a heightened overdensity near it. This may result in an increased density of compact objects within the spike regions, potentially enhancing the merger rate of compact binaries, presenting an intriguing avenue for further research.} \citep{1999PhRvL..83.1719G, 2019PhRvD..99d3533N}:
\begin{equation}
\rho_{\rm sp}(r) = \rho_{0}\left(\frac{r_{0}}{r_{\rm sp}}\right)^{\gamma}\left(1-\frac{4r_{\rm s}}{r}\right)^{3}\left(\frac{r_{\rm sp}}{r}\right)^{\gamma_{\rm sp}},
\end{equation}
where $\gamma_{\rm sp} = (9-2\gamma)/(4-\gamma)$ and $r_{\rm s}$ is the Schwarzschild radius of the SMBH, 
\begin{equation}
r_{\rm s} = \frac{2GM_{\rm SMBH}}{c^2} \simeq 2.95\,{\rm km}\, \left(\frac{M_{\rm SMBH}}{M_{\odot}}\right).
\end{equation}
Here $G$ is the gravitational constant and $c$ is the speed of light in vacuum. Numerical simulations and analytical approaches show that the density profile at small radii follows a power-law \citep{2006AJ....132.2685M, 2009MNRAS.398L..21S, 2010MNRAS.402...21N}. In this study, we consider $\gamma$ in the range $0<\gamma\leq 2$. To describe dark matter distribution in galactic halos, the NFW profile is often used \citep{1996ApJ...462..563N}, described by:
\begin{equation}
\rho_{\rm NFW}(r)=\frac{\rho_0}{\left(r/r_{0}\right) \left(1+r/r_{0}\right)^2}.
\end{equation}

\begin{figure}[t!] 
\includegraphics[width=0.5\textwidth]{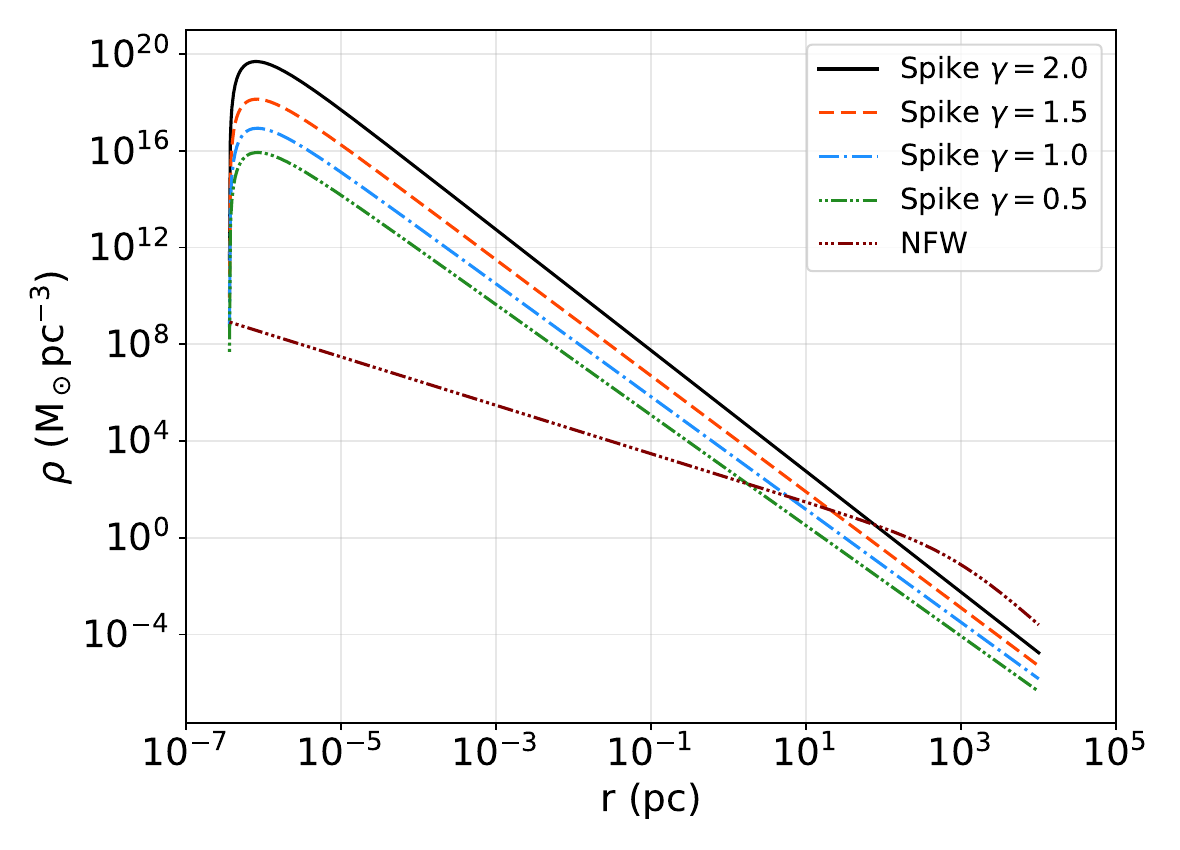}
\caption{ Comparison between the NFW density profile and the density profile of a dark-matter spike characterized by indices $\gamma=0.5, 1.0, 1.5,$ and $2$, surrounding a SMBH with a mass of $M_{\rm SMBH}=10^6 M_{\odot}$.}
\label{fig1}
\end{figure}

In Fig.\,\ref{fig1}, we have demonstrated the variation in the behavior of the dark-matter spike density profile compared to the NFW profile, given a SMBH mass of $M_{\rm SMBH}=10^6 M_{\odot}$, and various values for $\gamma$. The figure reveals that the density profile within the dark-matter spike is considerably higher than that of the NFW profile. This observation highlights the interest in calculating the merger rate of compact binaries in dark-matter spikes. Moreover, it is crucial to recognize that the radius $r=r_{\rm sp}(\gamma, M_{\rm SMBH})$ delineates the area within which the merger rate of compact objects should be computed, as this is where the dark-matter spike profiles intersect the NFW profile.

\subsection{The $M_{\rm SMBH}\mbox{-}\sigma$ Relation}
Multiple lines of evidence suggest a close link between the growth of SMBHs and the evolution of their host halos \citep[e.g., ][]{2002MNRAS.331..795M, 2002ApJ...574..740T, 2004ApJ...604L..89H, 2007MNRAS.378..198G, 2007ApJ...669...67H}. It has been proposed that the mass of a SMBH is strongly correlated with the velocity dispersion of dark matter particles in a galactic halo, $\sigma$, known as the $M_{\rm SMBH}\mbox{-}\sigma$ relation. This implies that the halo's characteristic parameters, $\rho_0$ and $r_0$, can be related to $M_{\rm SMBH}$ through this relation. A convenient form of the $M_{\rm SMBH}\mbox{-}\sigma$ relation is given by \citep{2000ApJ...539L...9F, 2000ApJ...539L..13G}:
\begin{equation} \label{masssmbh}
\log\left(\frac{M_{\rm SMBH}}{M_{\odot}}\right) = a + b \log\left(\frac{\sigma}{200\,{\rm km s^{-1}}}\right),
\end{equation}
where $a$ and $b$ are empirically determined. According to \cite{2009ApJ...698..198G}, the values $a=8.12 \pm 0.08$ and $b=4.24 \pm 0.41$ provide a reasonable fit for various types of galactic halos \citep{2013ARA&A..51..511K}. The NFW profile is assumed to describe the density profile outside the spike regions, i.e., $r\gg r_{\rm sp}$, up to the virial radius $r_{\rm vir}>r_{0}$\footnote{The virial radius is defined as the radius enclosing a volume where the average halo density reaches $200$ to $500$ times the critical density of the Universe.}. Under these assumptions, the total mass enclosed within a radius $r$ is given by:
\begin{equation}
M(r)=4 \pi \rho_{0} r_{0} \int_{0}^{r} \frac{r' dr'}{\left(1+r'/r_{0}\right)^2}=4 \pi \rho_{0} r_{0}^{3} g(r/r_{0}),
\end{equation}
where $g(x) = \ln(1+x) - x/(1+x)$. The contributions of the dark-matter spike and the central SMBH are negligible compared to the total mass of the dark matter halo. The concentration parameter, defined as $C \equiv r_{\rm vir}/r_{0}$, determines the central density of dark matter halos. Thus, the virial mass is:
\begin{equation}
M_{\rm vir}=4 \pi \rho_0 r_0^3 g(C).
\end{equation}

Additionally, the maximum circular velocity of dark matter particles occurs at a distance $r_{\rm m} = C_{\rm m}r_{0} = 2.16\,r_{0}$, which is obtained from the maximum of $M(r)/r$. The one-dimensional velocity dispersion of dark matter particles:
\begin{equation} \label{sigmaaa}
\sigma^2 = \frac{GM(C_{\rm m}r_{0})}{C_{\rm m}r_{0}}=4\pi G\rho_0 r_0^2 \frac{g(C_{\rm m})}{C_{\rm m}}.
\end{equation}
Thus, a relation between $\rho_0$, $r_0$, and $M_{\rm SMBH}$ can be established through Eqs. (\ref{masssmbh}) and (\ref{sigmaaa}). According to $N$-body simulations, the concentration parameter decreases with halo mass and varies with redshift at a constant mass \citep[e.g.,][]{2012MNRAS.423.3018P, 2014MNRAS.441.3359D, 2016MNRAS.460.1214L, 2016MNRAS.456.3068O}, consistent with the expected dynamics from dark matter halo evolution. In this study, we calculate the merger rate of compact binaries in the present-time Universe by using the concentration parameter from \cite{2016MNRAS.456.3068O} for dark matter halo models in GR, the parameter derived from \cite{2019MNRAS.487.1410M} for those in Hu-Sawicki $f(R)$ models, and the one obtained in \cite{2021MNRAS.508.4140M} for those in nDGP models.
\subsection{The mass function of SMBHs}
Understanding the growth and evolution of SMBHs is a key challenge in extragalactic astronomy. The mass function of SMBHs offers crucial insights into their masses and evolutionary patterns within galactic halos. Consequently, the SMBH mass function serves as an invaluable tool for studying the growth of SMBHs and testing theoretical models. Additionally, this mass function is pivotal for structuring future surveys by providing predictions for the mass distribution of SMBHs \citep{2012AdAst2012E...7K}. However, obtaining a precise mass function for SMBHs is challenging, and current estimates involve significant theoretical uncertainties, which can influence the accuracy of merger rate calculations for compact binaries in dark-matter spikes. A practical approach to address these uncertainties is to compare results from various empirical SMBH mass functions.

In a study by \cite{2007MNRAS.379..841B}, the {\it Galactica} code is used to analyze a sample of $8839$ SDSS galaxies. The main goal is to extrapolate the luminosity functions for spheroid and disc galaxies, leading to the derivation of the following SMBH mass function:
\begin{equation}\label{massfunc1} 
\phi(M_{\rm SMBH}) = 10^{9}\left(\frac{\phi_{0} M_{\rm SMBH}^{\alpha}}{M_{*}^{\alpha+1}}\right)\exp\left[-\left(\frac{M_{\rm SMBH}}{M_{*}}\right)^{\beta}\right], 
\end{equation} 
which is now referred to as the Benson mass function. In the above relation, $\alpha=-0.65$, $\beta = 0.6$, $\phi_{0}=2.9\times 10^{-3}\,h^3{\rm Mpc^{-3}}$, and $M_* = 4.07 \times 10^{7}\,h^{-2}M_{\odot}$.

Similarly, in \cite{2009MNRAS.400.1451V}, a different mass function, referred to as the Vika mass function, is derived for SMBHs using data from the {\it Millennium Galaxy Catalogue} \citep{2003MNRAS.344..307L} for $1743$ galaxies. This mass function is based on the empirical relation between SMBH mass and host spheroid luminosity, expressed as:
\begin{equation}\label{massfunc2}
\phi(M_{\rm SMBH}) = \phi_* \left(\frac{M_{\rm SMBH}}{M_*}\right)^{\alpha +1}\exp\left[ 1-\left(\frac{M_{\rm SMBH}}{M_*}\right)\right],
\end{equation}
where $\log\phi_*=-3.15\,h^{3}{\rm Mpc^{-3} dex^{-1}}$, $\log (M_*/M)=8.71$, and $\alpha=-1.20$. This mass function is applicable for masses in the range $10^6M_{\odot}<M_{\rm SMBH}< 10^{10}M_{\odot}$.

Additionally, in \cite{2004MNRAS.354.1020S}, another mass function for SMBHs is introduced, which will be called the Shankar mass function. This function is derived from the observed correlation between SMBH mass and halo velocity dispersion, utilizing both kinematic and photometric data. The Shankar mass function is formulated as:
\begin{equation}\label{massfunc3}
\phi(M_{\rm SMBH}) = \phi_* \left(\frac{M_{\rm SMBH}}{M_*}\right)^{\alpha +1}\exp\left[ 1-\left(\frac{M_{\rm SMBH}}{M_*}\right)^\beta\right],
\end{equation}
where $\phi_*=7.7 \times 10^{-3}\,{\rm Mpc^{-3}}$, $M_*=6.4\times 10^{7}\,M_{\odot}$, $\beta=0.49$, and $\alpha=-1.11$. This mass function is valid for the mass range $10^{6}M_{\odot}\leq M_{\rm SMBH}\leq 5\times10^{9}M_{\odot}$.
\section{Compact Binary Merger Rate} \label{sec:iv}
\subsection{Binary Formation Scenario}
As mentiond in Sec.\,\ref{sec:i}, compact objects in galactic halos can form binary systems via different channels. In this work, relying on minimal assumptions, we limit ourselves to calculating the merger rate of compact binaries through two-body dynamical encounters. However, we recall that in the central regions of galaxies dominated by a nuclear star cluster, considering other channels such as three-body encounters could potentially be more complete hypotheses \citep[see, e.g.,][]{2020MNRAS.492.2936A}.

Consider a scenario within a dark-matter spike where a compact object with mass $m_1$ encounters another with mass $m_2$ on a hyperbolic trajectory, with a relative velocity at large separation of $v_{\rm rel} = |v_1 - v_2|$. During a two-body scattering event, significant gravitational radiation is emitted at the periastron $r_{\rm p}$. According to Keplerian mechanics, such a system becomes gravitationally bound when the emitted gravitational energy exceeds the system's kinetic energy. Under these circumstances, the maximum value for the periastron can be calculated as follows \citep{123456789p, 1989ApJ...343..725Q, 2002ApJ...566L..17M}:
\begin{equation}
r_{\rm mp}=\left[\frac{85\pi}{6\sqrt{2}}\frac{G^{7/2}m_{1}m_{2}(m_{1}+m_{2})^{3/2}}{c^{5}v_{\rm rel}^{2}}\right]^{2/7}.
\end{equation}
In the Newtonian limit, the impact parameter is related to the periastron as:
\begin{equation}
b^{2}(r_{\rm p})=\frac{2G(m_1+m_2)r_{\rm p}}{v_{\rm rel}^{2}}+r_{\rm p}^{2}.
\end{equation}
In regions where dark-matter spikes are gravitationally active, one can assume a strong gravitational focusing limit ($r_{\rm p}\ll b$). This allows us to neglect perturbations from surrounding compact objects on the formed binaries. Consequently, the cross-section for binary formation can be expressed as:
\begin{equation}
\xi(m_1,m_2, v_{\rm rel})=\pi b^{2}(r_{\rm mp})\simeq \frac{2\pi G(m_1+m_2)r_{\rm mp}}{v_{\rm rel}^{2}}.
\end{equation}

When we substitute the expression for $r_{\rm mp}$ into the cross-section equation, we obtain:
\begin{equation}
\xi \simeq 2\pi \left(\frac{85\pi}{6\sqrt{2}}\right)^{2/7}\frac{G^{2}(m_1+m_2)^{10/7}(m_1m_2)^{2/7}}{c^{10/7}v_{\rm rel}^{18/7}}.
\end{equation}

The merger rate of compact binaries in each dark-matter spike can be expressed as:
\begin{equation}\label{nsp}
N_{\rm sp} = 4\pi\int_{4r_{\rm s}}^{r_{\rm sp}} \mathcal{V}(\rho, m_1, m_2)\langle\xi v_{\rm rel}\rangle\, r^{2}dr,
\end{equation}
where for the PBH-PBH events:
\begin{equation}\label{vpp}
\mathcal{V}= \frac{1}{2}\frac{\left[f_{\rm PBH}\,\rho_{\rm sp}(r)\right]^{2}}{m_{1}m_{2}},
\end{equation}
and for the PBH-NS events:
\begin{equation}\label{vpn}
\mathcal{V}= \left(\frac{f_{\rm PBH}\,\rho_{\rm sp}(r)}{m_{1}}\right)\left(\frac{\rho_{\rm NS}(r)}{m_{2}}\right).
\end{equation}
In these expressions, $0<f_{\rm PBH}\leq 1$ denotes the fraction of PBHs contributing to dark matter\footnote{Eqs.\,(\ref{nsp})-(\ref{vpn}) reveal that PBH-PBH binary merger rates are proportional to $f_{\rm PBH}^{2}$, while PBH-NS event rates exhibit a linear relationship with $f_{\rm PBH}$.}, and the angle brackets indicate an average over the relative velocity distribution near the central SMBH. Additionally, $\rho_{\rm NS}(r)$ represents the NS profile, defined in a spherically symmetric form as:
\begin{equation}
\rho_{\rm NS} = \rho^{*}_{\rm NS}\exp\left(-\frac{r}{r^{*}_{\rm NS}}\right),
\end{equation}
where $ r^*_{\rm NS}$ and $\rho^*_{\rm NS}$ are characteristic radius and density of NSs, respectively, which must be determined. According to \cite{2022ApJ...931....2S}, the total range for the characteristic radius is considered to be $r^*_{\rm NS}\simeq (0.01\mbox{-}0.1)\,r_{0}$. To compute the merger rate of PBH-NS binaries in dark-matter spikes, we set the characteristic radius of NSs as $r^{*}_{\rm NS}\simeq0.1\,r_{0}$. 

To accurately represent the characteristic density of NSs, it is essential to normalize their distribution relative to their estimated abundance within a typical galaxy. This normalization process employs the time-independent variant of the initial Salpeter stellar mass function, which is approximated as $\chi(m_*) \propto m_*^{-2.35}$ \footnote{It should be note that other stellar mass functions like \cite{2002Sci...295...82K}, can be considered in future studies to further explore their impact on NS population estimates and merger rates.}. The fundamental premise is that stellar objects with masses ranging from $8$ to $20\, M_{\odot}$ culminate in supernova events, subsequently forming NSs. Based on this assumption, one can derive the number of NSs in a galaxy of stellar mass $M_{*}$ through appropriate integration of the mass function over the specified mass range 
\begin{equation}
n_{\rm NS} = M_*\int_{m^{\rm min}_*}^{m^{\rm max}_*} \chi(m_*)dm_*,
\end{equation}
where the stellar mass function $\chi(m_*)m_*$ is normalized to unity, $m^{\rm min}_{*}=8\,M_{\odot}$, and $m^{\rm max}_{*} =20\,M_{\odot}$. To characterize the galactic stellar mass $M_*$, it is necessary to establish the relation between stellar mass and halo mass, denoted as $M_*(M_{\rm halo})$. This relation can be derived from \cite{2013ApJ...770...57B}. The analysis assumes that the highest concentration of NSs is located at the galactic halo center.

In the vicinity of the central SMBH, the relative velocity is approximated using the circular velocity formula $v(r)=\sqrt{GM_{\rm SMBH}/r}$ at each radial distance within the dark-matter spike. This approximation is supported by the relatively small mass of the dark-matter spike compared to that of the central SMBH, which predominantly shapes the gravitational field in the region. The SMBH’s substantial gravitational influence dictates the dynamics within the spike area, where compact objects are concentrated. Consequently, the dark-matter spike’s mass contribution is negligible, allowing the gravitational field to be approximated as primarily that of the SMBH. In this context, compact objects within the spike can be assumed to have similar circular velocities determined by the SMBH's gravitational attraction \citep{2019PhRvD..99d3533N}. However, the relative velocity between compact objects can slightly exceed this estimate, depending on the angles of their velocities. Here, we have assigned specific mass values to the compact objects involved in the mergers, enabling a direct comparison with GW data from similar-mass compact object mergers. In PBH-PBH events, with a monochromatic mass distribution, $M_{\rm PBH}$ is set at $30$ solar masses, while in PBH-NS events, $M_{\rm PBH}$ and $M_{\rm NS}$ are fixed at $5$ and $1.4$ solar masses, respectively. As an initial assumption, we also deem that PBHs constitute the entirety of dark matter, i.e., $f_{\rm PBH}=1$. However, we will discuss more realistic assumptions for the abundance of PBHs later.

The cumulative merger rate of compact binaries is a key quantity for LVK detectors. To determine the overall merger rate of compact binaries per unit volume, it is necessary to multiply the SMBH mass function, $\phi(M_{\rm SMBH})$, with the merger rate of compact binaries per spike, $N_{\rm sp}(M_{\rm SMBH})$, and integrate over the mass of SMBHs:
\begin{equation}
\mathcal{R} = \int_{M_{\rm min}}^{M_{\rm max}} N_{\rm sp}(M_{\rm SMBH}) \phi(M_{\rm SMBH}) dM_{\rm SMBH}.
\end{equation}
As can be seen from Eqs.\,(\ref{massfunc1})-(\ref{massfunc3}), the SMBH mass functions exhibit a decreasing exponential term relative to SMBH mass, indicating that $M_{\rm max}$ can potentially have minimal impact on the final result. Conversely, the abundance of smaller central BHs in the Universe suggests that $M_{\rm min}$ may significantly influence the merger rate of compact binaries in dark-matter spikes.

With the computational tools now available to us, we will detail the specific findings of our analysis the following sections.
\subsection{PBH-PBH Merger Rate}

\begin{figure*}[ht!] 
\gridline{\fig{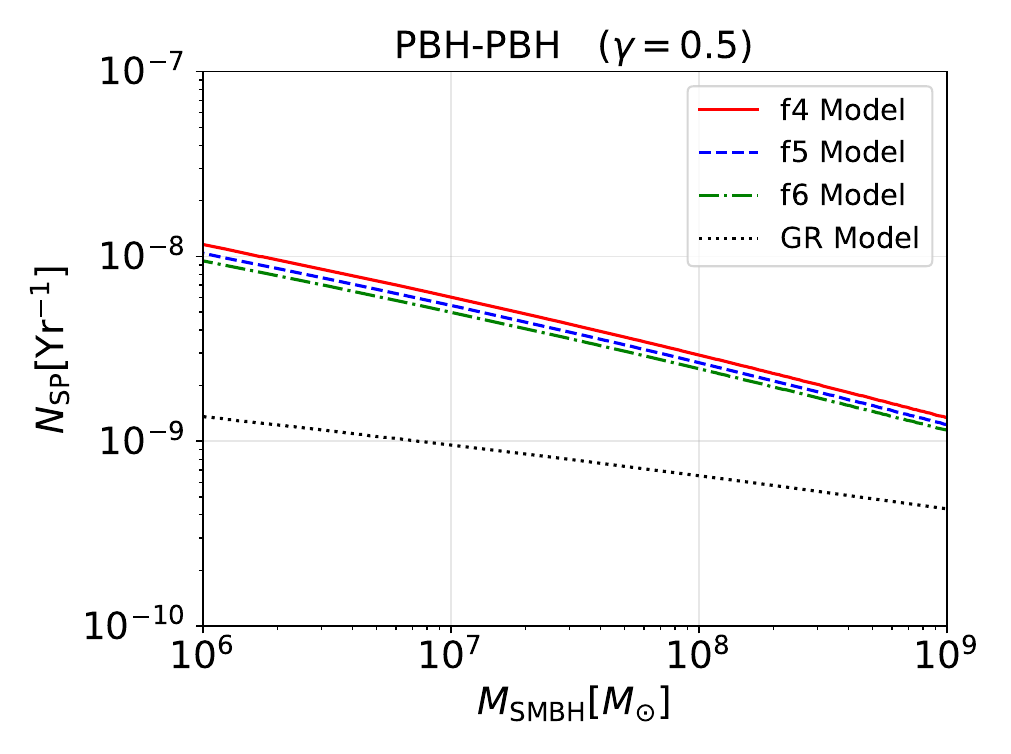}{0.33\textwidth}{(a)}
\fig{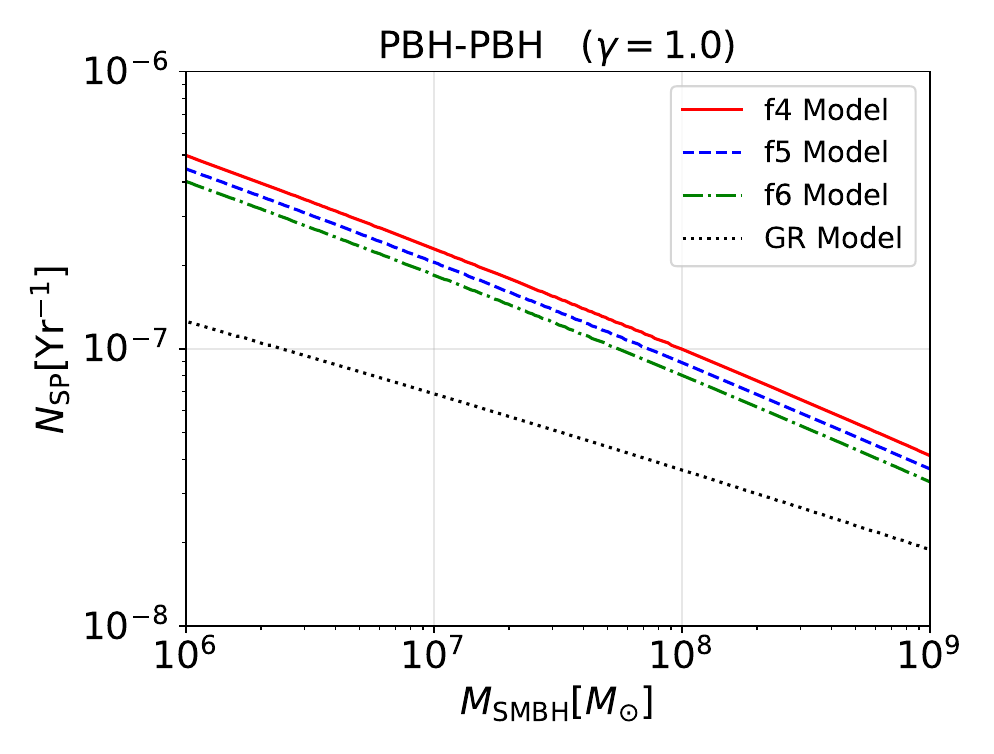}{0.33\textwidth}{(b)}
\fig{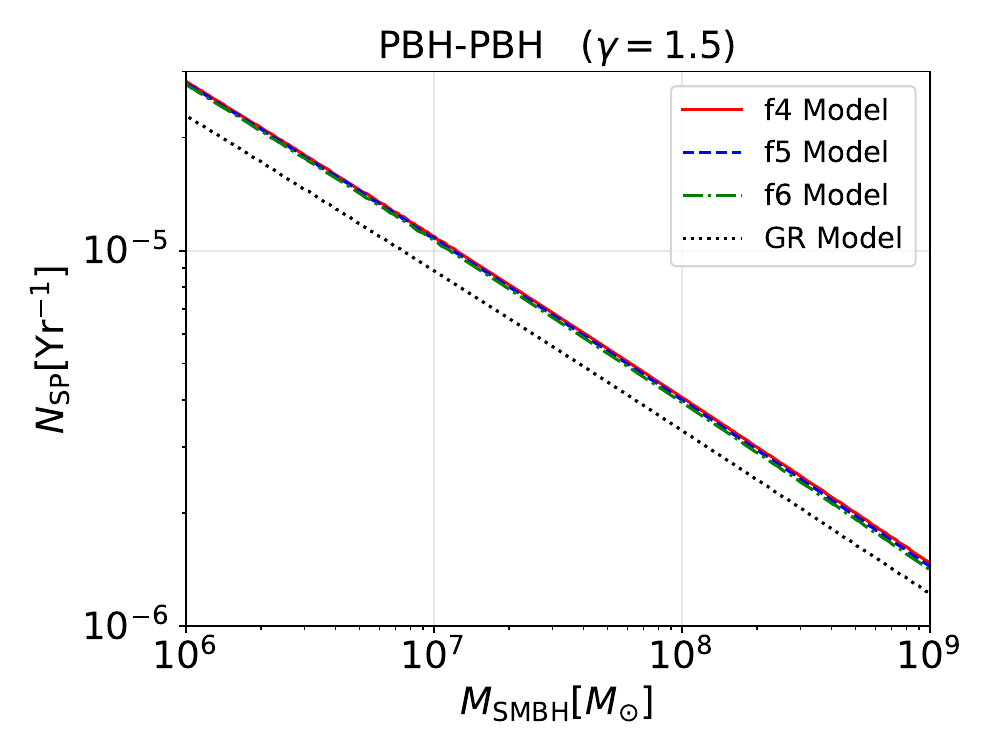}{0.33\textwidth}{(c)}
}
\gridline{\fig{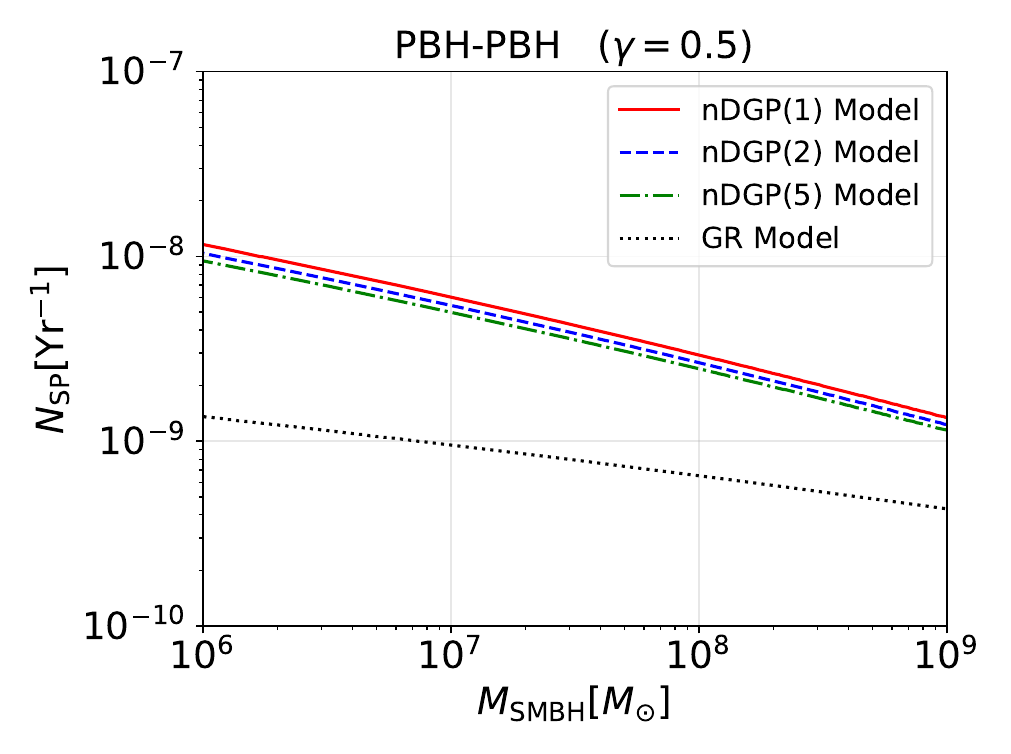}{0.33\textwidth}{(d)}
\fig{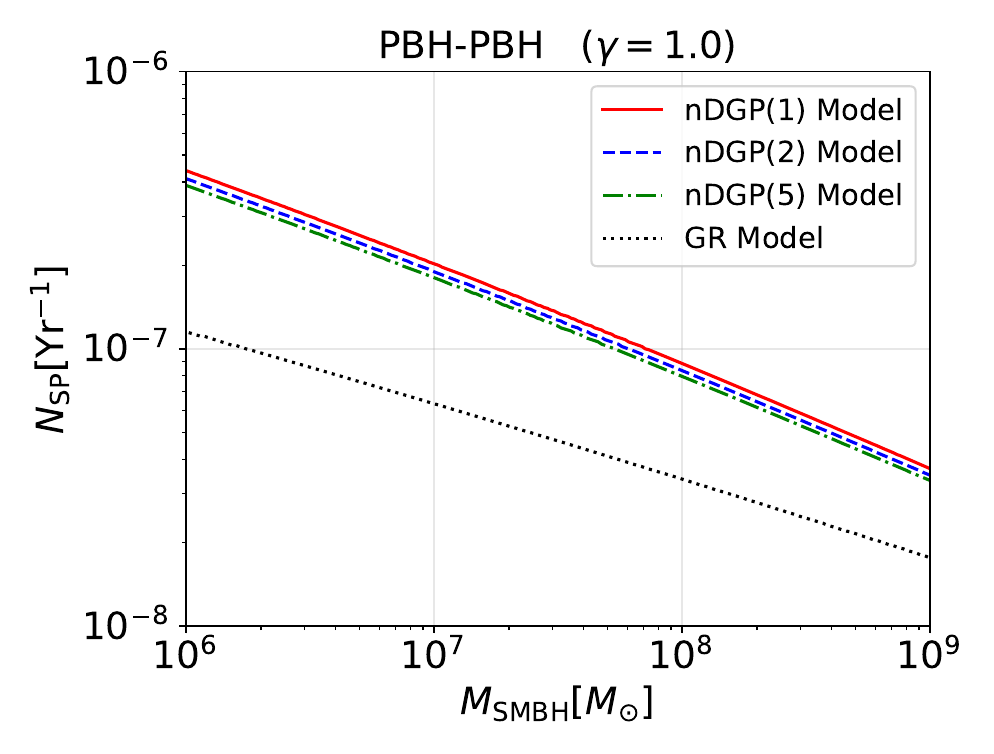}{0.33\textwidth}{(e)}
\fig{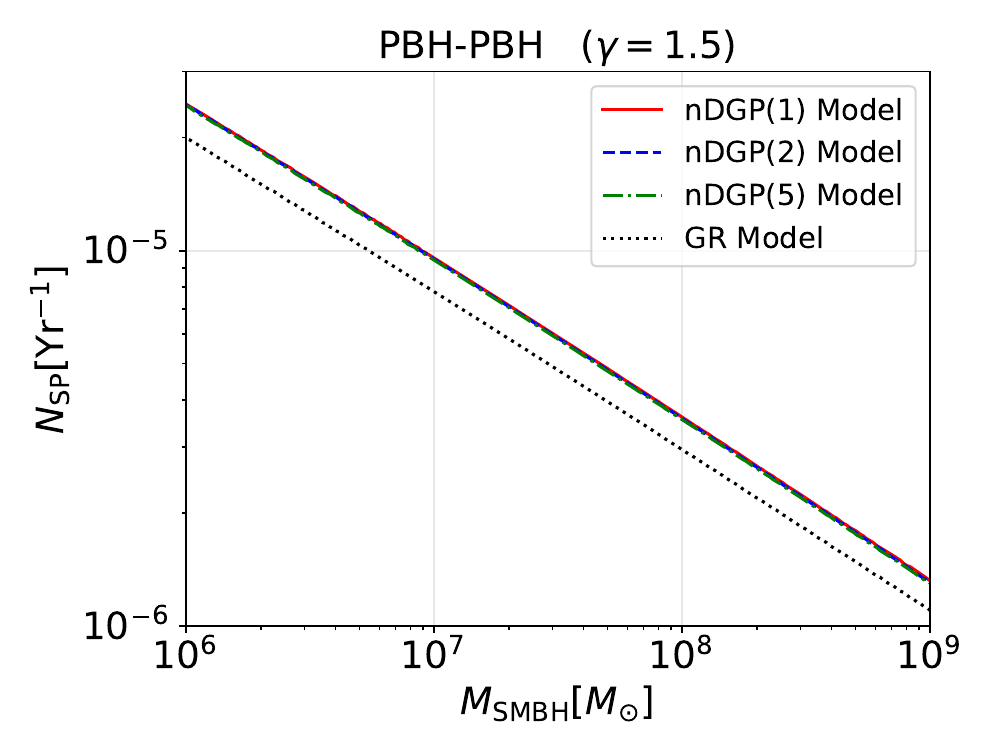}{0.33\textwidth}{(f)}
}
\caption{The merger rate of PBH-PBH binaries per dark-matter spike as a function of SMBH mass for different gravitational models and power-law indices $\gamma$. Top panels: Results for Hu-Sawicki $f(R)$ gravity models compared to GR. Bottom panels: Results for nDGP gravity models compared to GR. From left to right: Merger rates calculated for $\gamma = 0.5, 1.0$, and $1.5$, respectively.}
\label{fig2}
\end{figure*}

In Fig.\,\ref{fig2}, we have illustrated the merger rate of PBH-PBH binaries within a single dark-matter spike as a function of SMBH mass for various values of the power-law index $\gamma$. Our calculations incorporate different models of Hu-Sawicki $f(R)$ and nDGP MG and compare these results with those obtained from GR. All models consistently indicate that the merger rate of PBH-PBH binaries decreases as the mass of the SMBH increases. As the mass of central SMBH increases, the abundance of PBHs within the dark-matter spike decreases \citep{2019PhRvD..99d3533N}. However, this reduction is partially offset by an increase in the cross section. The MG models predict higher merger rates than GR, with more significant enhancements in models that deviate further from GR.

\begin{figure*}[ht!] 
\gridline{\fig{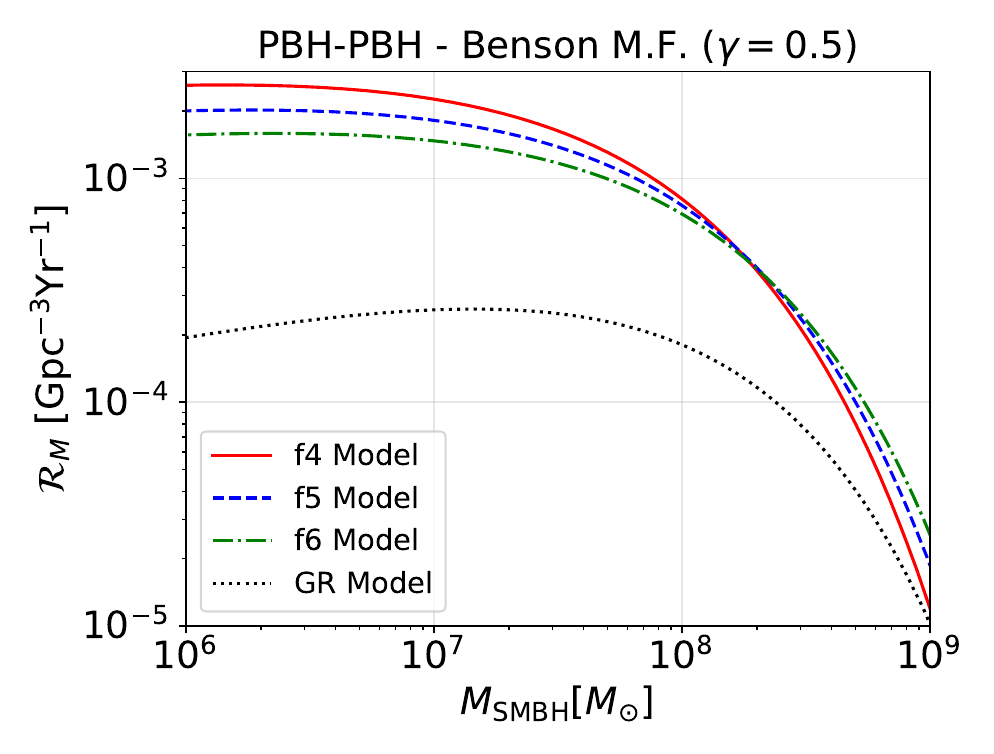}{0.33\textwidth}{(a)}
\fig{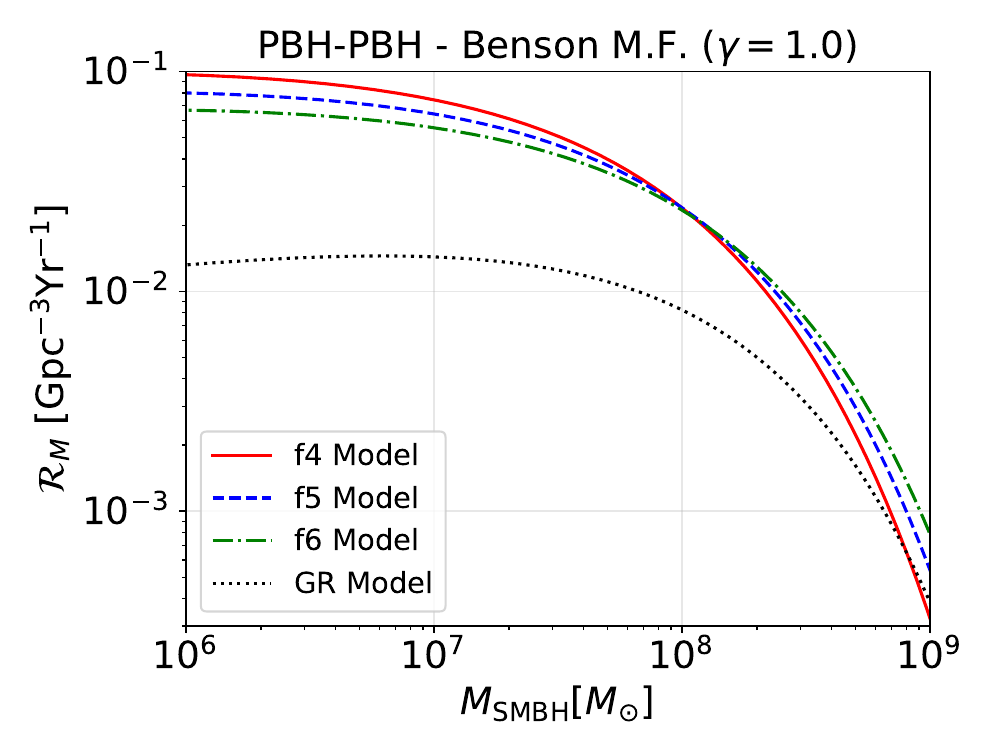}{0.33\textwidth}{(b)}
\fig{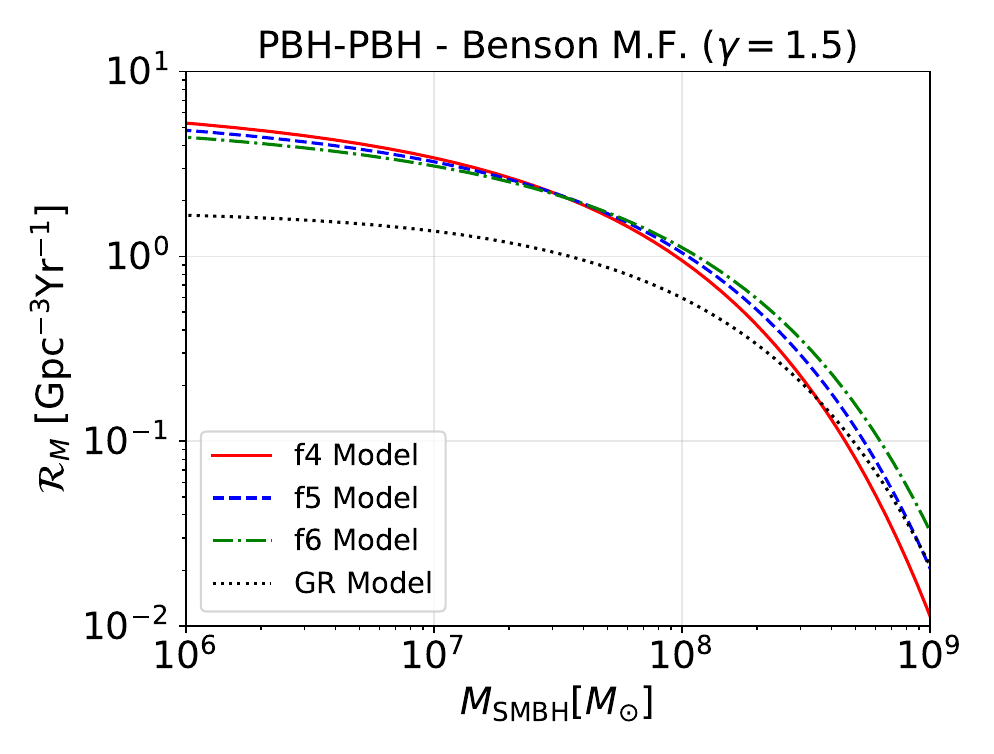}{0.33\textwidth}{(c)}
}
\gridline{\fig{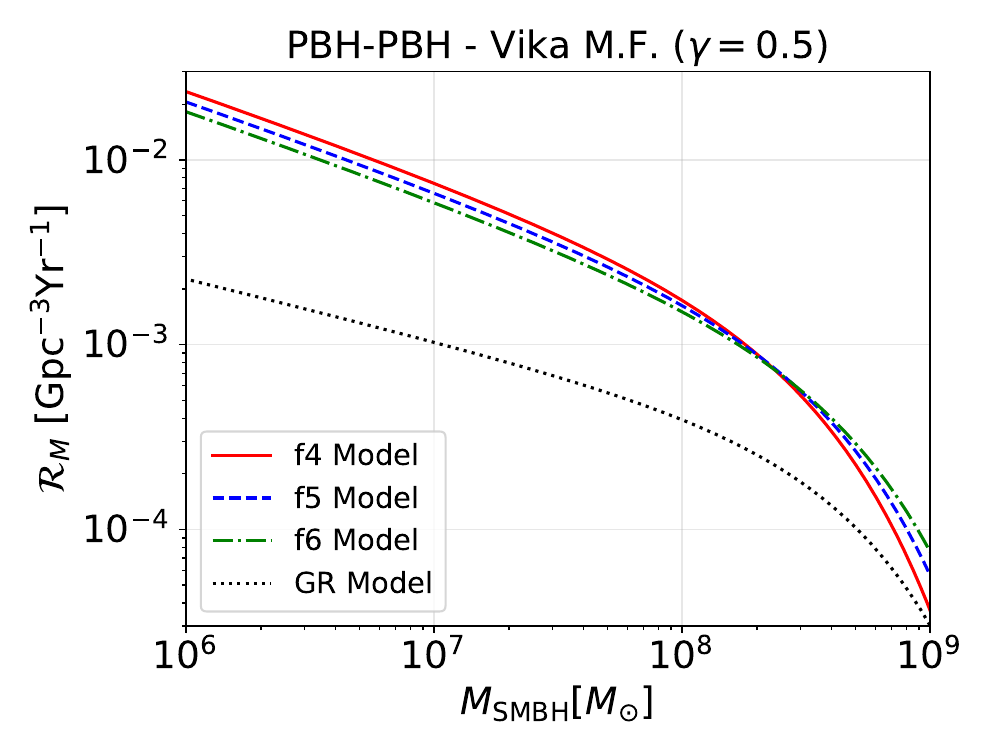}{0.33\textwidth}{(d)}
\fig{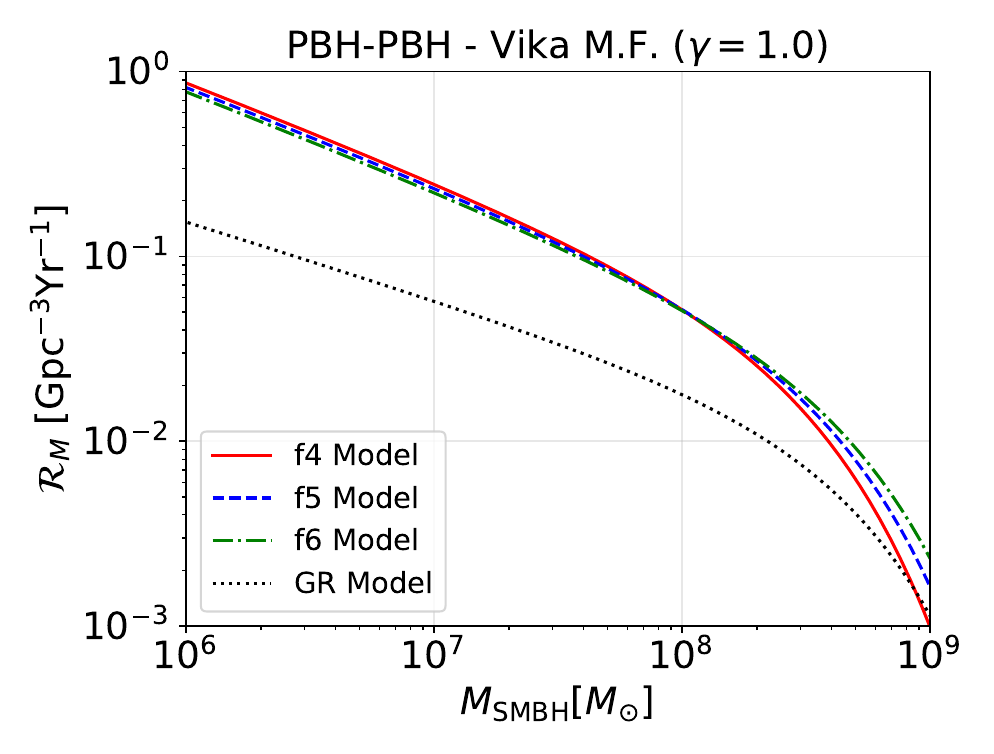}{0.33\textwidth}{(e)}
\fig{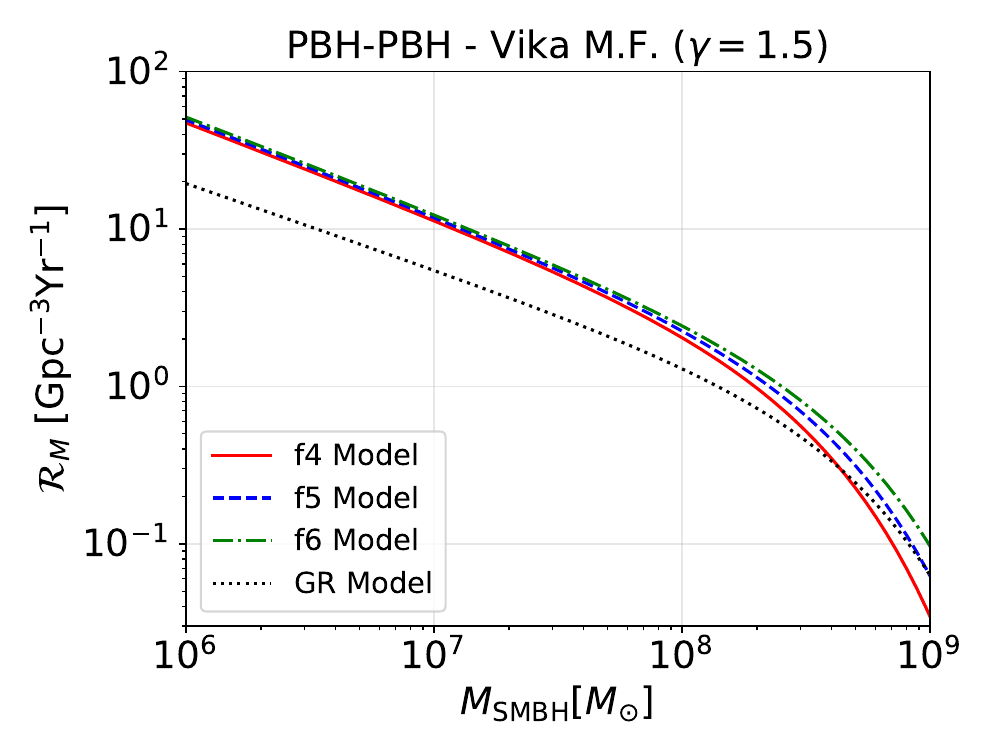}{0.33\textwidth}{(f)}
}
\gridline{\fig{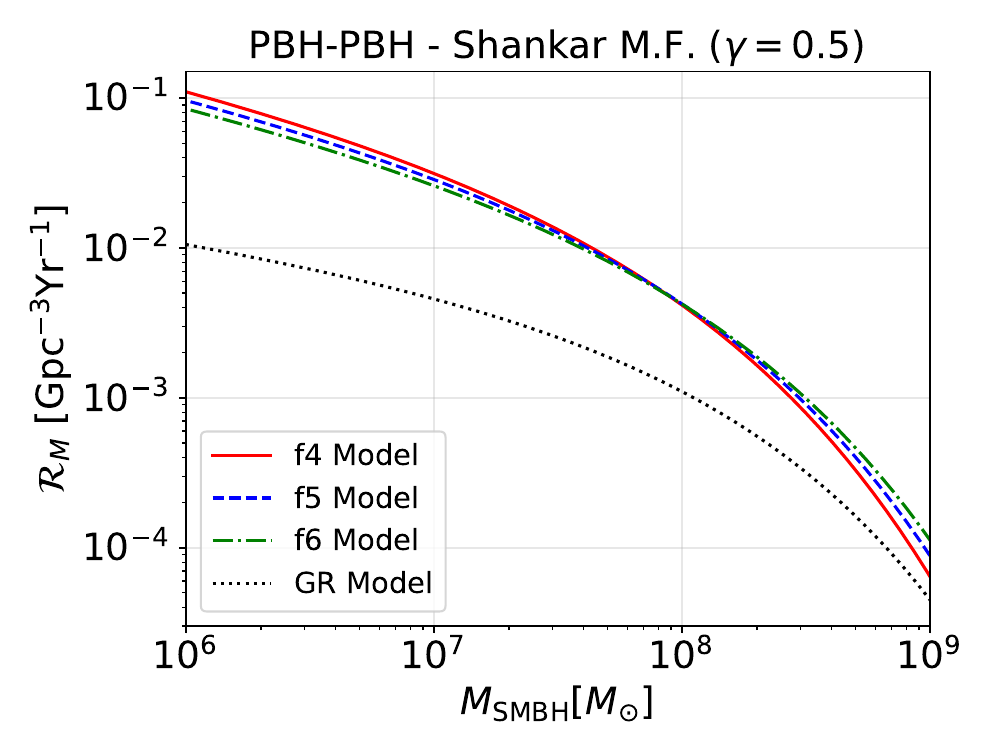}{0.33\textwidth}{(g)}
\fig{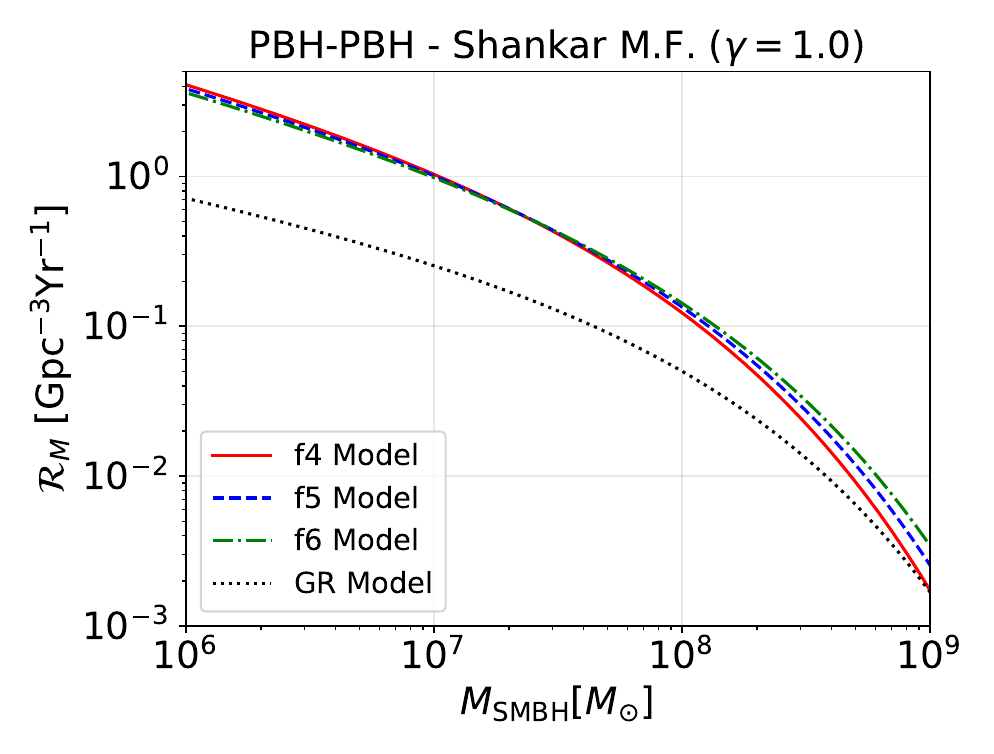}{0.33\textwidth}{(h)}
\fig{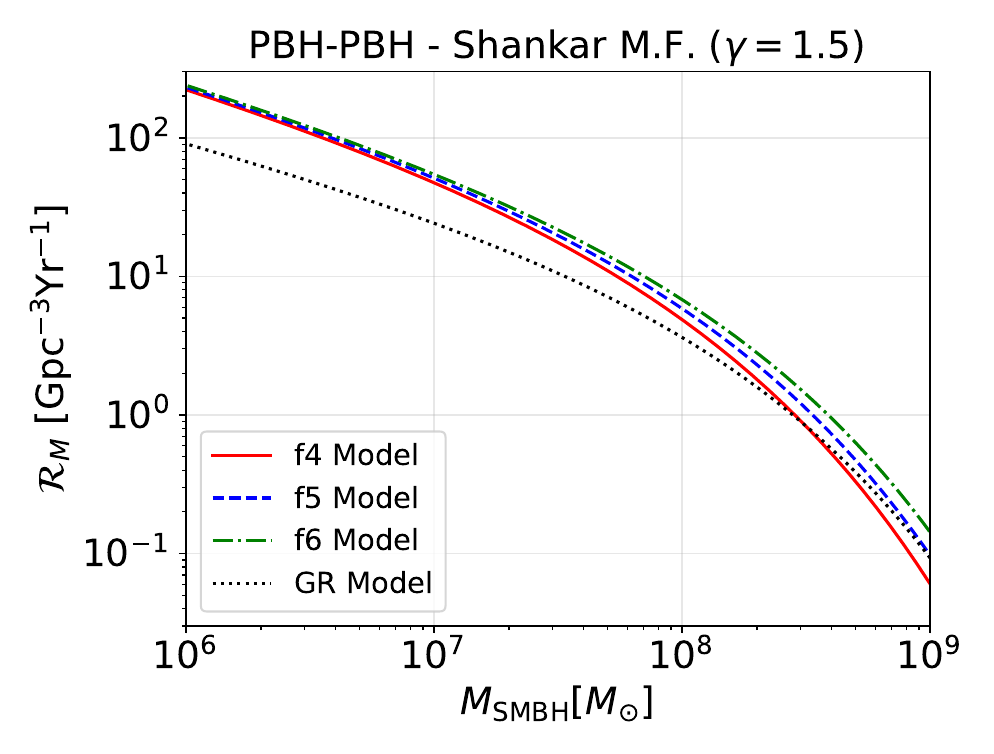}{0.33\textwidth}{(i)}
}
\caption{The merger rate of PBH-PBH binaries per unit volume as a function of SMBH mass for different Hu-Sawicki $f(R)$ gravity models compared to GR. Results are shown for three different SMBH mass functions and three values of the power-law index $\gamma$ characterizing the steepness of the dark-matter spike density profile. Top panels: Benson mass function; Middle panels: Vika mass function; Bottom panels: Shankar mass function. From left to right, each column represents $\gamma = 0.5, 1.0$, and $1.5$, respectively.}
\label{fig4}
\end{figure*}

\begin{figure*}[ht!] 
\gridline{\fig{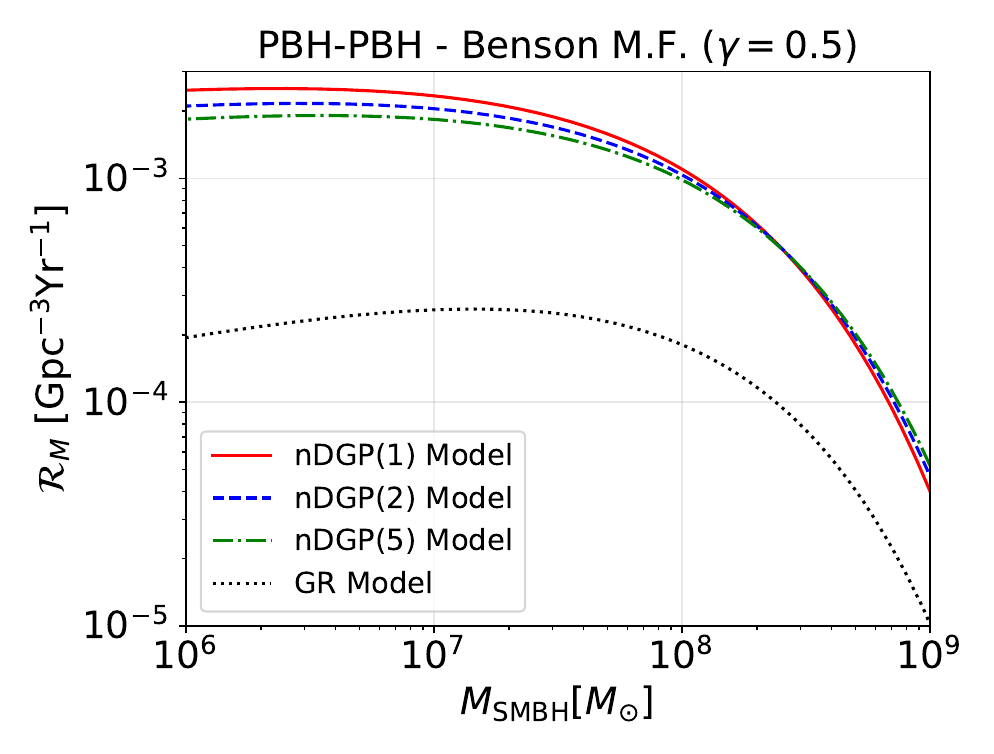}{0.33\textwidth}{(a)}
\fig{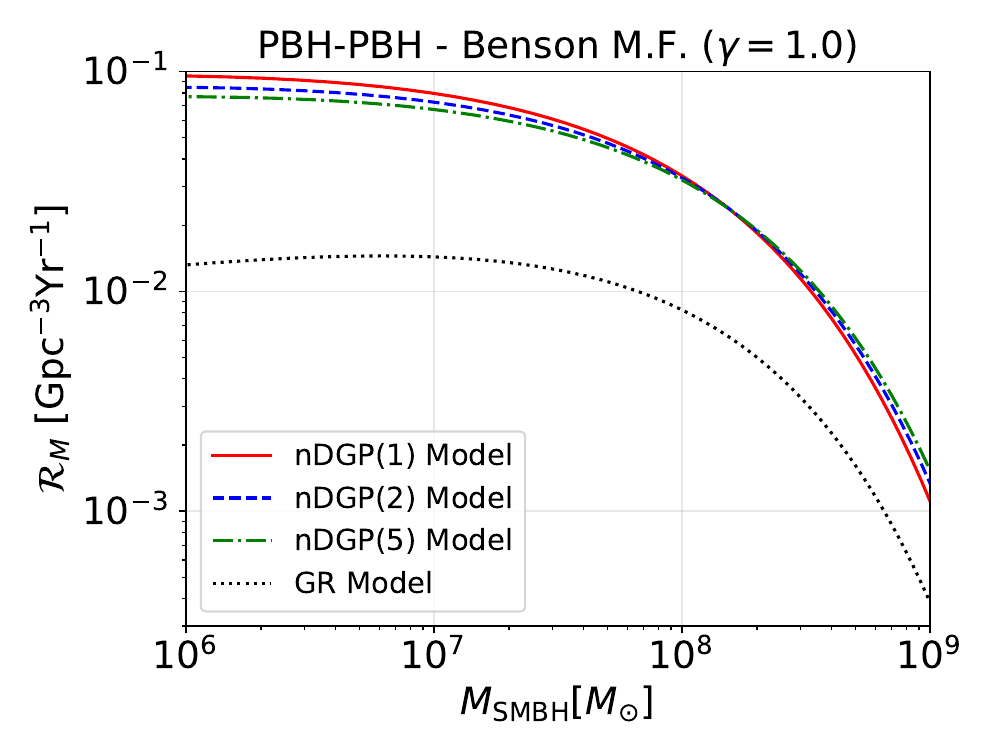}{0.33\textwidth}{(b)}
\fig{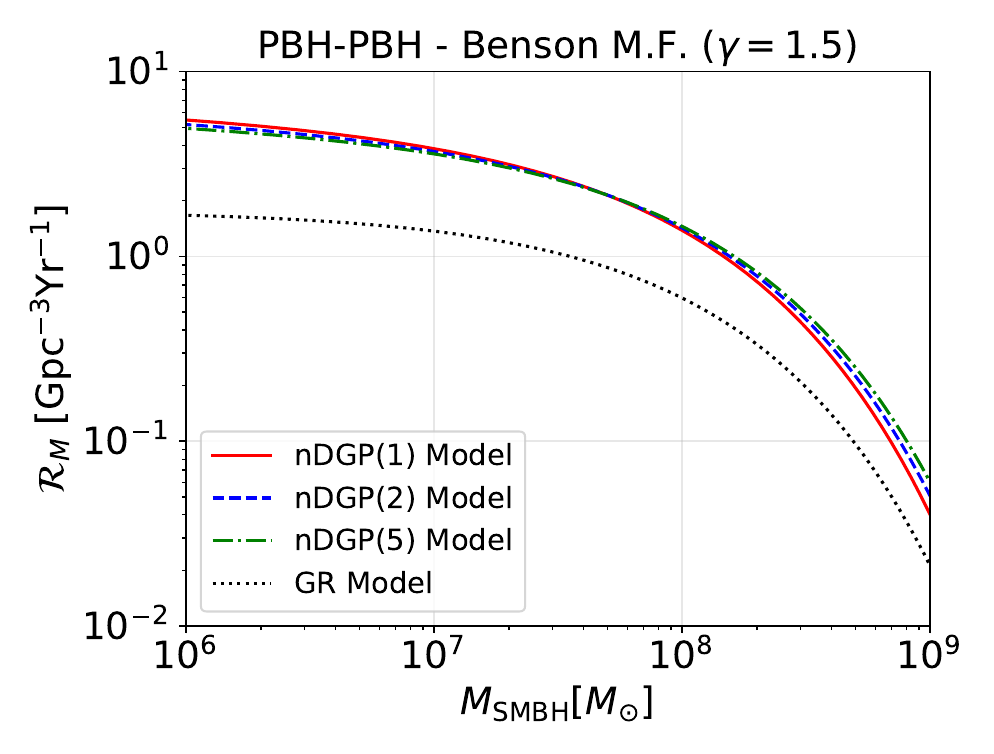}{0.33\textwidth}{(c)}
}
\gridline{\fig{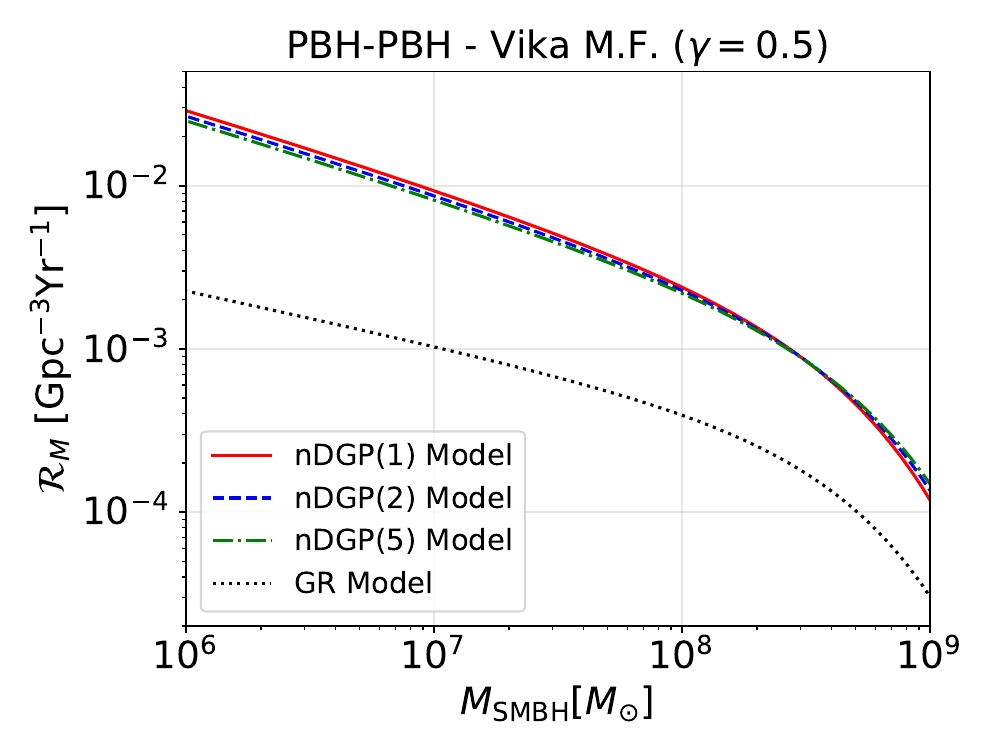}{0.33\textwidth}{(d)}
\fig{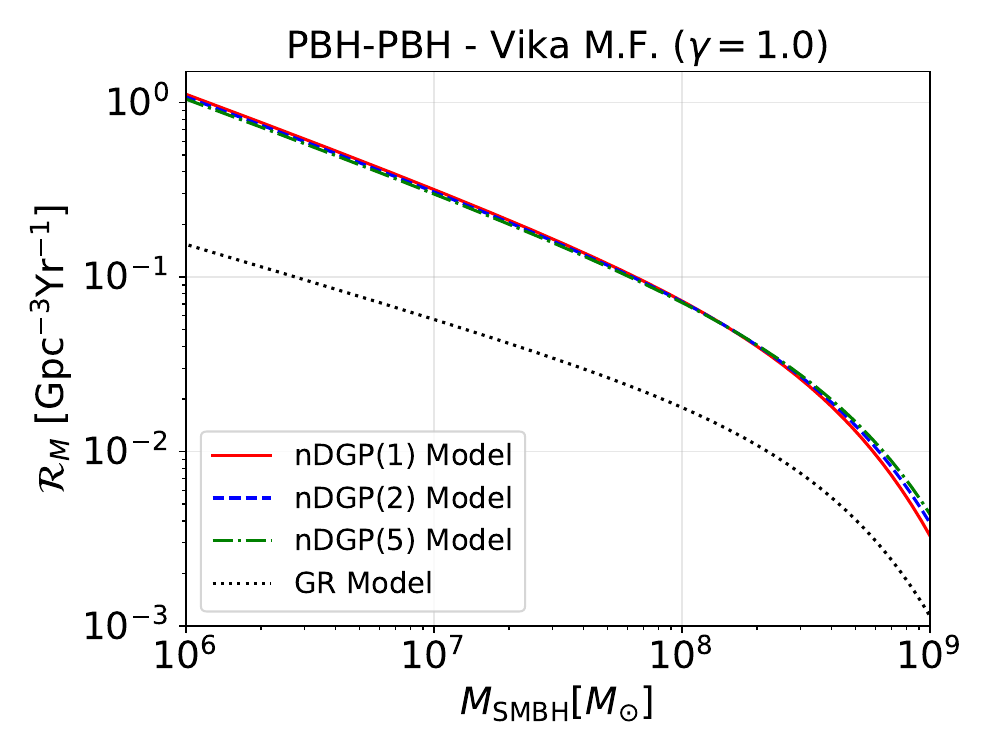}{0.33\textwidth}{(e)}
\fig{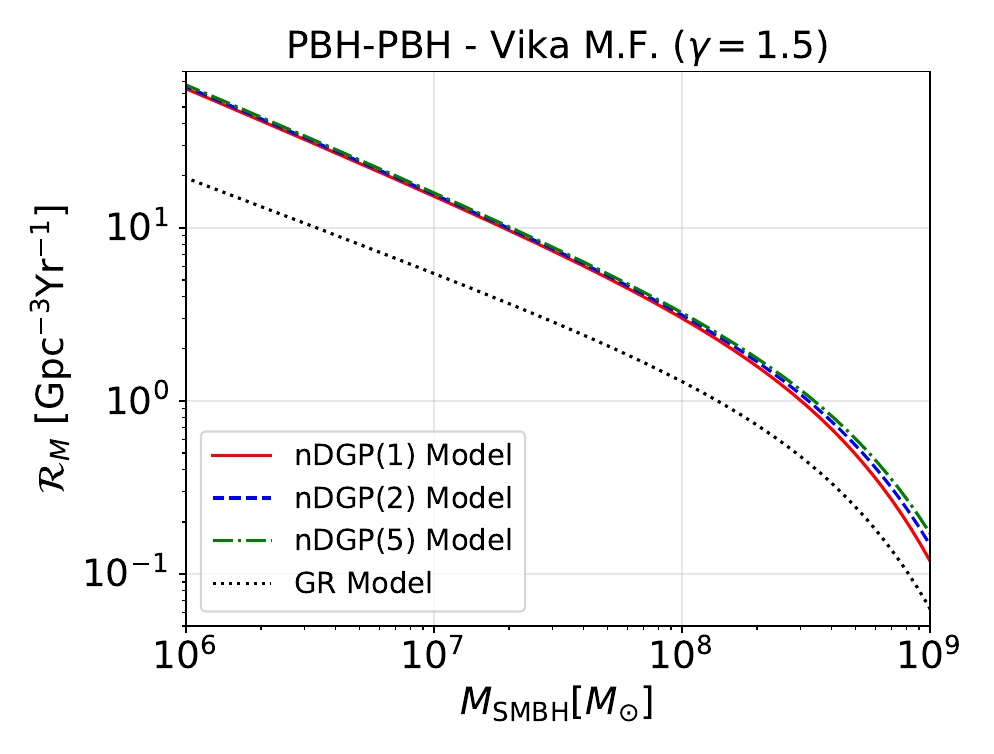}{0.33\textwidth}{(f)}
}
\gridline{\fig{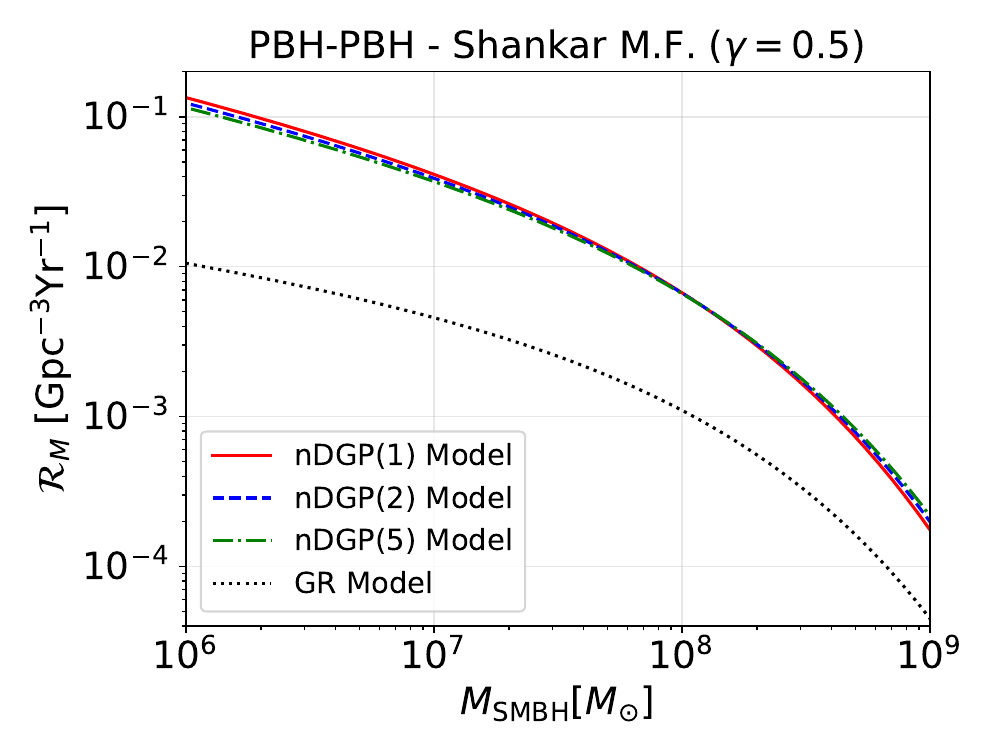}{0.33\textwidth}{(g)}
\fig{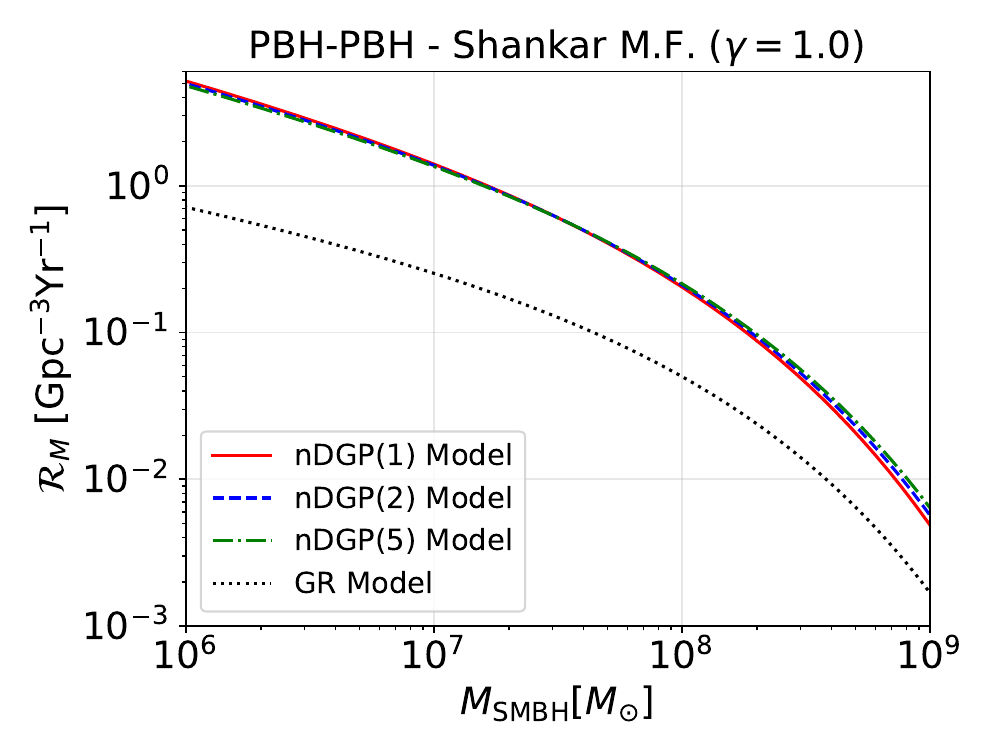}{0.33\textwidth}{(h)}
\fig{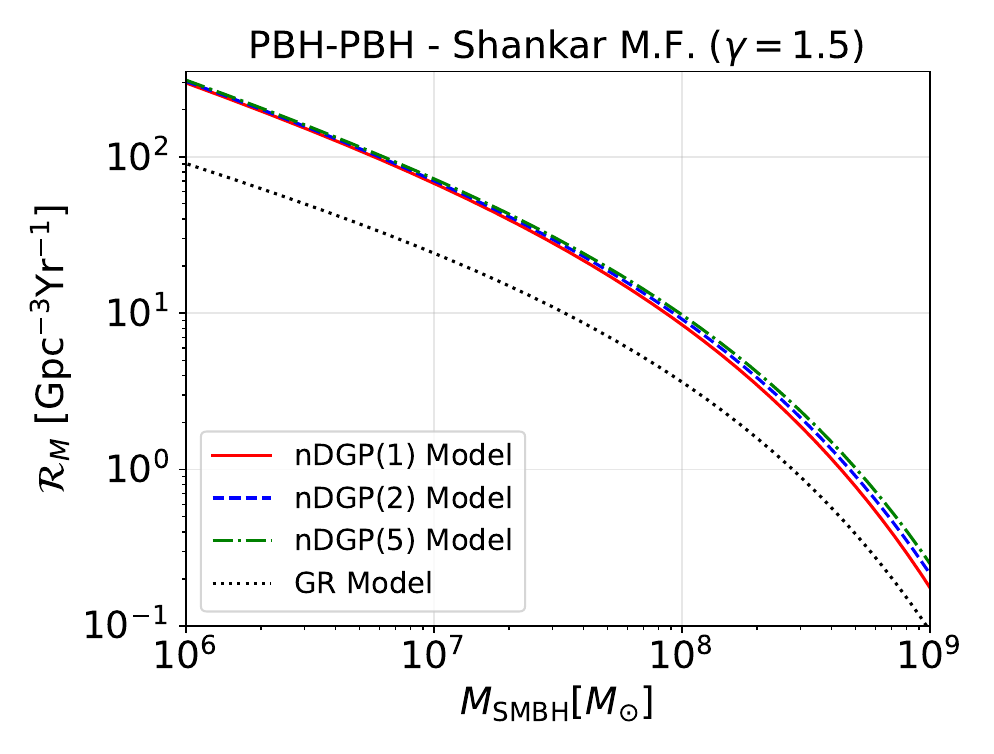}{0.33\textwidth}{(i)}
}
\caption{Similar to Fig.\,\ref{fig4} but for nDGP gravity models, i.e., nDGP(1), nDGP(2), nDGP(5).}
\label{fig5}
\end{figure*}

The upper panels reveal that the merger rate of PBH-PBH binaries within each dark-matter spike is higher in all Hu-Sawicki $f(R)$ models compared to GR. This can be attributed to the increased density distribution around central SMBHs, as predicted by the Hu-Sawicki $f(R)$ models. This enhancement is reflected in an elevated concentration parameter within dark matter halos, affecting the gravitational dynamics of dark-matter spikes. Additionally, it is noted that the influence of the field strength on the concentration parameter decreases progressively from f4 to f6. The bottom panels show similar results, indicating that the PBH-PBH binary merger rate in each dark-matter spike in nDGP models is significantly higher than in GR. The enhancement of the merger rate is most pronounced in the nDGP(1) model, followed by the nDGP(2) model, and least in the nDGP(5) model.

The power-law index $\gamma$, characterizing the steepness of the spike density profile, substantially impacts merger rates across all gravitational models. As $\gamma$ increases from $0.5$ to $1.5$, the merger rate of PBH-PBH binaries rises by several orders of magnitude. This significant enhancement underscores the critical role of the spike density profile in determining the merger rate of compact binaries. Furthermore, the relative advantage of MG models over GR becomes more pronounced at higher $\gamma$ values, indicating that more voluminous spikes can contain more PBH-PBH merger events. However, the discrepancy between the results derived from Hu-Sawicki $f(R)$ and nDGP models and those obtained from GR becomes more pronounced at lower $\gamma$ values, aligning with theoretical expectations. This phenomenon can be attributed to the fact that both Hu-Sawicki $f(R)$ and nDGP models predict higher central densities in dark matter halos compared to the GR model, leading to an elevated probability of PBH-PBH binary mergers within these regions. The comparison between $f(R)$ and nDGP models reveals that both types of MG theories lead to similar qualitative effects on the merger rate of PBH-PBH binaries, although the quantitative enhancements differ.

In Fig.\,\ref{fig4}, we have depicted the merger rate of PBH-PBH binaries per unit volume as a function of SMBH mass. This analysis spans various Hu-Sawicki $f(R)$ gravity models and GR. The study elucidates the intricate interactions between MG theories and SMBH mass functions, highlighting the different effects on the PBH-PBH binary merger rate. Across all models, the merger rate shows a declining trend with increasing SMBH mass.

The top panels display results using the Benson mass function. For $\gamma = 0.5$, f4, f5, and f6 present the highest merger rates across all masses, significantly exceeding GR predictions, particularly at lower SMBH masses. As $\gamma$ increases to $1.0$ and $1.5$, the decreasing trend in merger rates with increasing SMBH mass intensifies, with all MG models showing higher merger rates compared to GR, while f4 remains the most distinct. Notably, beyond \(M_{\text{SMBH}} \approx 10^8 M_{\odot}\), the merger rates begin to decrease more sharply.

The middle panels, utilizing the Vika mass function, reveal a similar pattern but with generally higher merger rates than the Benson mass function. The f4 model consistently yields the highest rates, followed by f5 and f6, with GR predicting the lowest rates for $\gamma = 0.5$ and $1.0$. For $\gamma = 1.5$, and smaller SMBHs, f4, f5, and f6 produce nearly identical results. Nevertheless, for larger SMBHs, the reverse trend remains. This underscores the significant influence of the mass function choice on predicted merger rates, especially under MG theories.

The bottom panels illustrate results based on the Shankar mass function, which predicts the highest merger rates compared to the Vika and Benson mass functions. The hierarchical order among the models is consistent, with f4 showing the highest rates and GR the lowest. However, the overall trend indicates a more gradual decline in merger rates with SMBH mass than the Vika function suggests. This implies that the Shankar function, potentially reflecting a different SMBH growth history or environment, moderates the influence of MG on merger rates. For $\gamma = 1.5$, the f4, f5, and f6 models yield approximately the same merger rates at lower SMBH masses before diverging at higher masses. The decreasing trend in merger rates with increasing SMBH mass continues, with deviations from GR being more pronounced at lower masses but diminishing as SMBH mass increases.

In Fig.\,\ref{fig5}, we have exhibited a comprehensive analysis of the merger rate of PBH-PBH binaries per unit volume as a function of SMBH mass, comparing various nDGP gravity models to GR across three SMBH mass functions and three values of the power-law index $\gamma$. A significant distinction between the nDGP models and the previously discussed Hu-Sawicki $f(R)$ models is that the nDGP models exhibit substantially higher merger rates compared to GR at higher SMBH mass ranges. This indicates that the nDGP gravity modifications exert a more enduring influence on merger rates across a broader spectrum of SMBH masses. The PBH-PBH binary merger rate with respect to SMBH mass follows a pattern similar to that in Fig.\,\ref{fig4}. Initially, there is a declining trend as SMBH mass increases; however, the merger rates begin to decrease more sharply after $M_{\rm SMBH}=10^{8} M_{\odot}$.

In the top panels, utilizing the Benson mass function, all nDGP models predict higher merger rates than GR, with nDGP(1), nDGP(2), and nDGP(5) showing the most significant enhancements, respectively. The differences between models become less pronounced as $\gamma$ increases, and the reversal trend at $M_{\rm SMBH}=10^{8} M_{\odot}$ is apparent. A higher $\gamma$ value indicates a steeper profile, meaning the dark matter density increases more sharply near the SMBH. The plots demonstrate that as $\gamma$ increases, the merger rate also rises significantly. This trend occurs because steeper density profiles generate stronger gravitational potentials, enhancing the likelihood of compact objects, such as PBHs and NSs, encountering each other and merging. Thus, regions with high $\gamma$ values facilitate more frequent mergers due to the increased density and gravitational interactions within the dark-matter spikes.

The middle panels, employing the Vika mass function, exhibit similar trends but with higher overall merger rates. The distinction between nDGP models and GR is more pronounced, particularly for lower SMBH masses. The reversal point is consistent, but the subsequent increase in merger rates is steeper compared to the Benson function results. The nDGP(1) model consistently predicts the highest rates, followed by nDGP(2) and nDGP(5), while GR forecasts the lowest rates for $\gamma=0.5$ and $1.0$. When $\gamma = 1.5$, the nDGP models produce nearly identical outcomes for smaller SMBHs. In contrast, for larger SMBHs, the initial trend persists, with the nDGP(1) model showing the highest rates.

The bottom panels, using the Shankar mass function, display the highest merger rates among all three mass functions. The separation between nDGP models and GR is most evident here. Furthermore, when considering the Shankar mass function, the slope of the merger rate concerning SMBH mass is significantly steeper compared to the results derived from the Vika and Benson mass functions.

\subsection{PBH-NS Merger Rate}
\begin{figure*}[ht!] 
\gridline{\fig{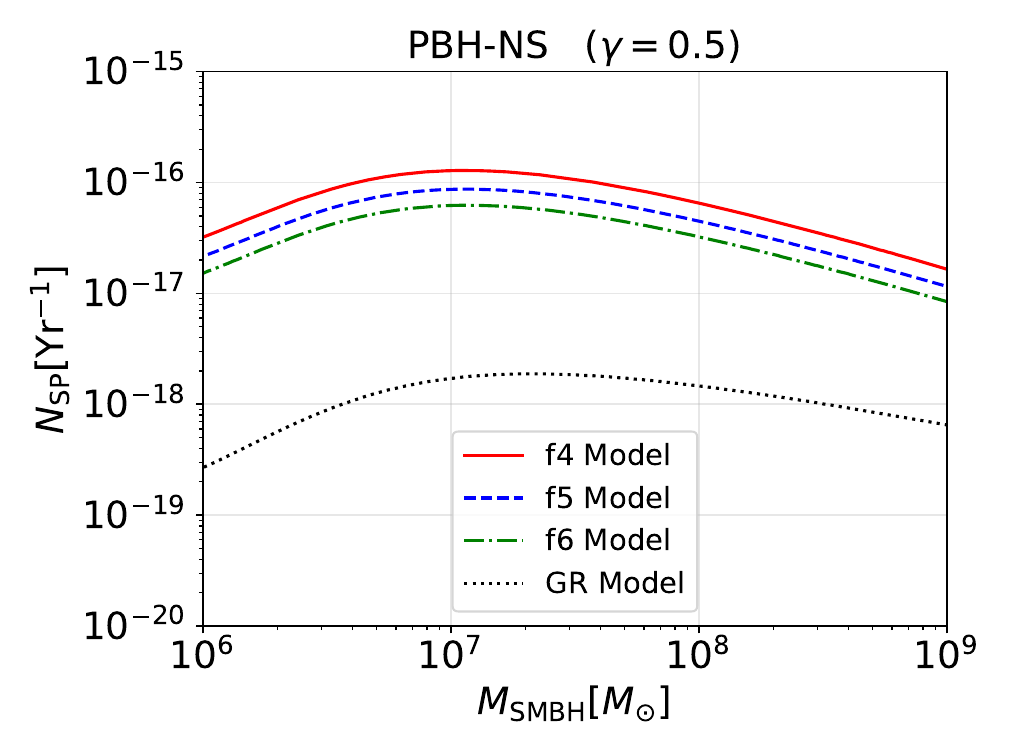}{0.33\textwidth}{(a)}
\fig{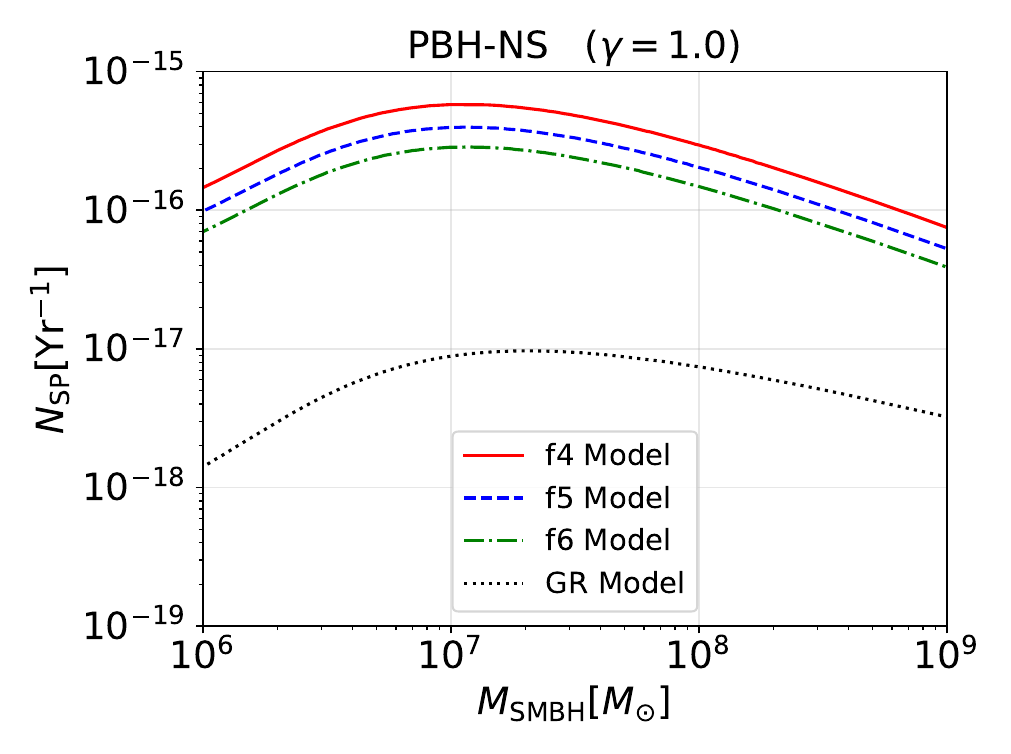}{0.33\textwidth}{(b)}
\fig{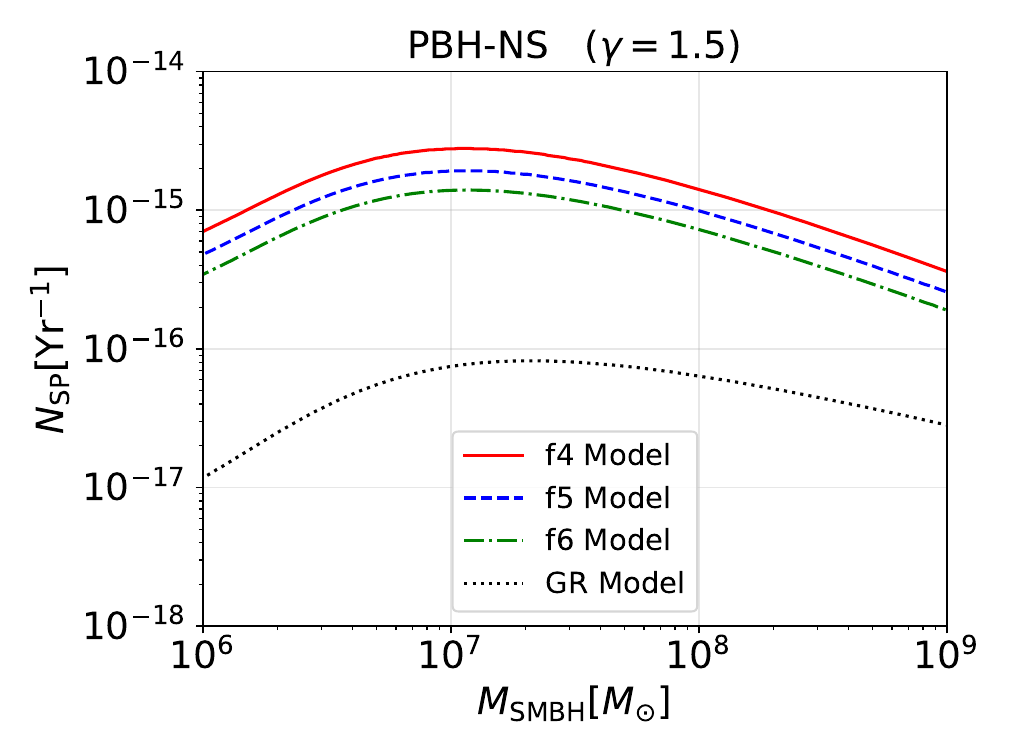}{0.33\textwidth}{(c)}
}
\gridline{\fig{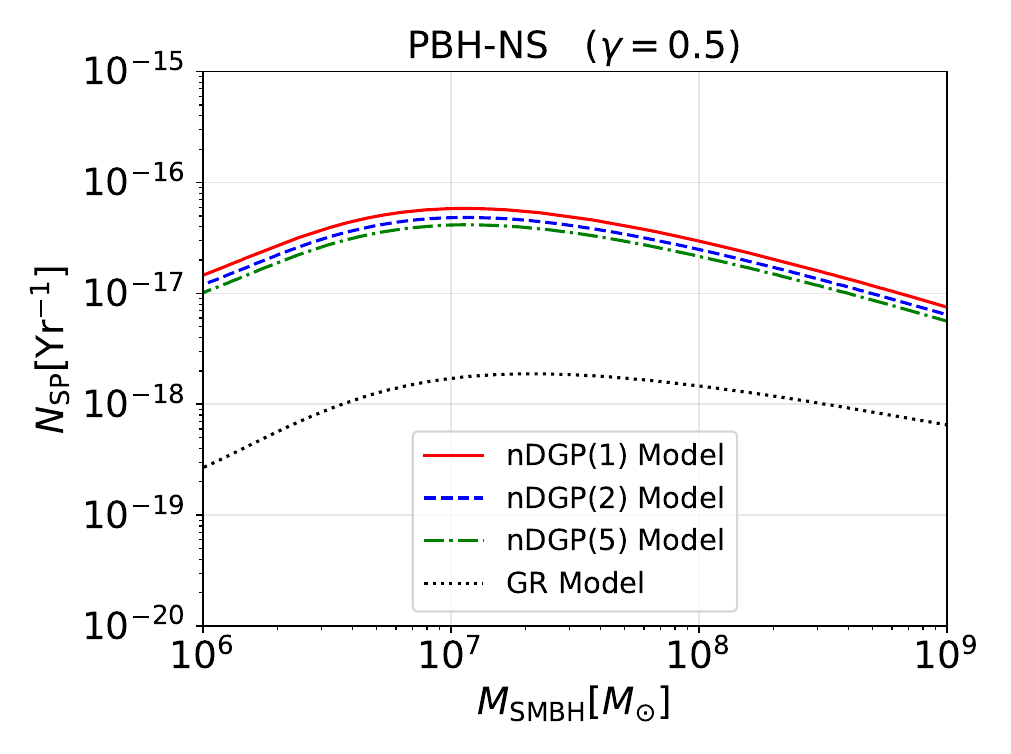}{0.33\textwidth}{(d)}
\fig{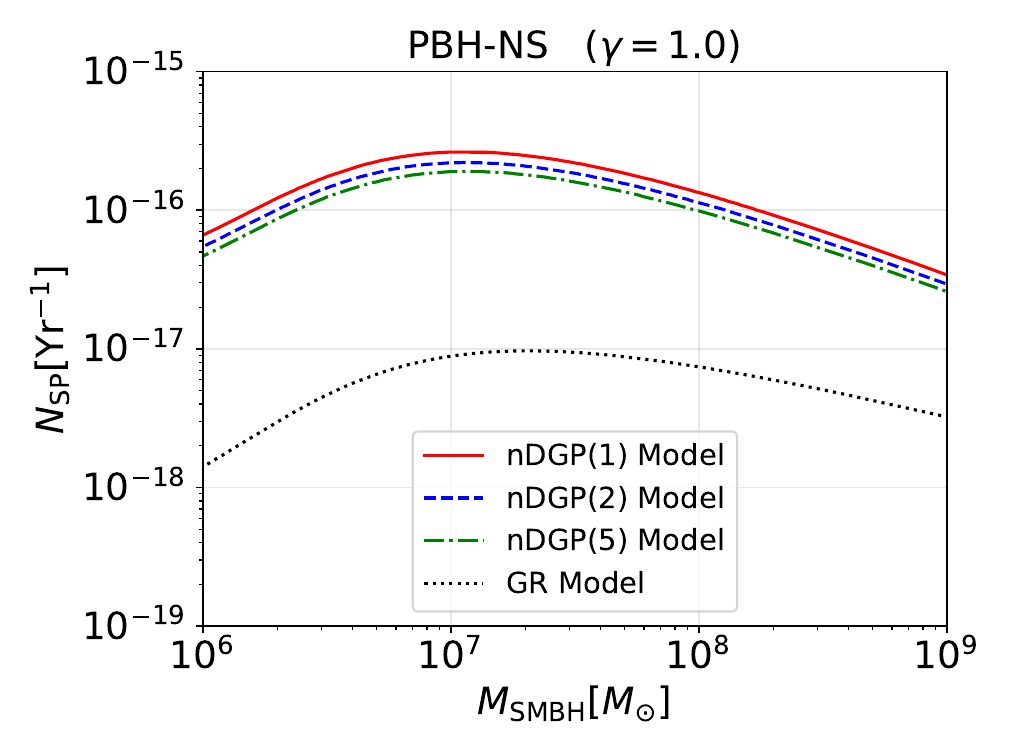}{0.33\textwidth}{(e)}
\fig{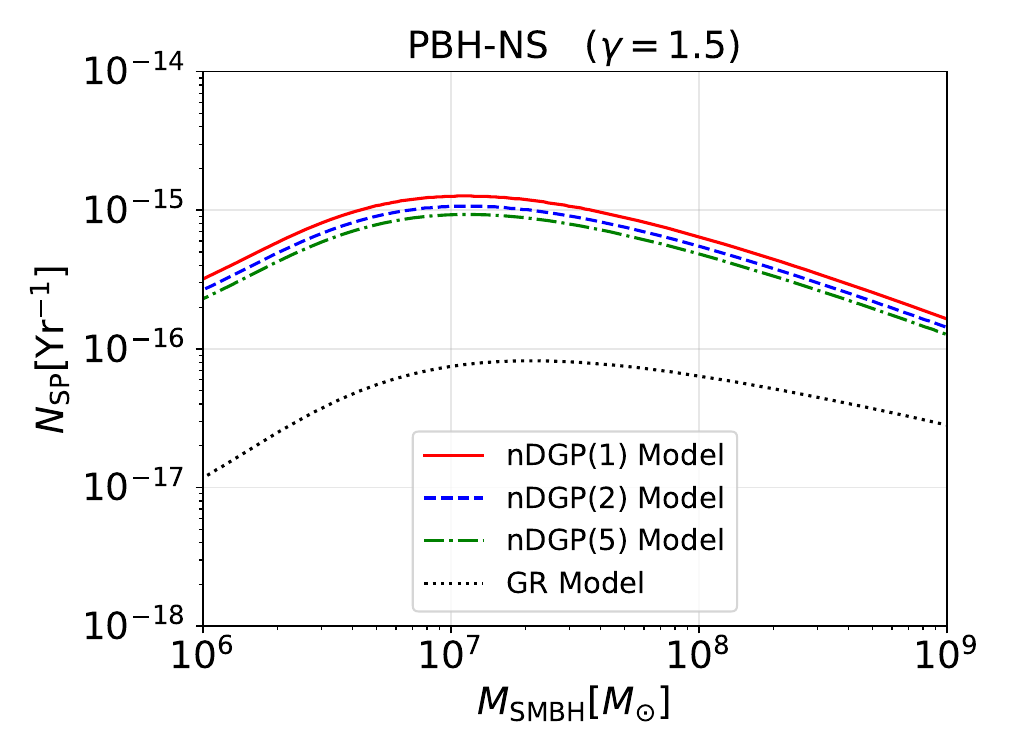}{0.33\textwidth}{(f)}
}
\caption{Similar to Fig.\,\ref{fig2} but for the PBH-NS binaries.}
\label{fig3}
\end{figure*}

In Fig.\,\ref{fig3}, we have presented the merger rate of PBH-NS binaries per dark-matter spike as a function of SMBH mass, comparing various Hu-Sawicki $f(R)$ and nDGP models with GR. A significant characteristic that sets these results apart from the PBH-PBH scenario is the concave down shape of the functions, which exhibit a maximum near the SMBH clumps at $M_{\rm SMBH} \approx 10^7 M_{\odot}$. This characteristic can be attributed to the distribution of NSs within nuclear star clusters \citep{2018MNRAS.473..927A}, where object density peaks near the gravitational influence radius of central SMBHs \citep{2014A&A...570A...2F, 2015MNRAS.447..948C}. Additionally, the hilltop value for the PBH-NS binary merger rate corresponds well with the mass of the Milky Way's dark-matter spike, associated with its central SMBH of approximately $10^7 M_{\odot}$ \citep{2019A&A...625L..10G}. Therefore, it can be inferred that the highest event rate for PBH-NS binaries occurs in dark-matter spikes within galactic halos. Moreover, the increased density of the dark matter spike at lower SMBH masses may lead to a higher merger rate of PBH-NS binaries. However, at higher SMBH masses, the dark matter density in the spike surrounding the SMBH decreases, which can naturally reduce the merger rate of PBH-NS binaries.

As $\gamma$ increases from $0.5$ to $1.5$, the merger rates rise dramatically across all models, spanning several orders of magnitude. This indicates the sensitivity of merger rates to the steepness of the spike density profile. Physically, a higher $\gamma$ implies a higher concentration in dark matter distribution near the central SMBH, leading to more frequent interactions and higher merger rates. 

Both the Hu-Sawicki $f(R)$ and nDGP models consistently predict higher merger rates than GR across all SMBH masses and $\gamma$ values. This enhancement is particularly noticeable at lower $\gamma$ values, suggesting that MG effects are more significant in regions with less steep dark matter profiles. Among the Hu-Sawicki $f(R)$ models, f4 shows the largest enhancement over GR, followed by f5 and f6. Similarly, for the nDGP models, nDGP(1) exhibits the greatest increase, followed by nDGP(2) and nDGP(5).

The superior performance of these generalized gravitational models over GR in predicting higher merger rates can be attributed to their prediction of higher central densities in dark matter halos. The enhancement of central density, or equivalently of concentration parameter, in MG models compared to GR stems from several key factors that alter gravitational interactions and structure formation. In MG theories, the effective gravitational force can be significantly stronger, allowing for greater mass accumulation in halos, with concentrations potentially increased by up to a factor of $4/3$. Additionally, MG models often feature non-linear dynamics that enable more efficient mass aggregation as structures evolve. The concentration is sensitive to cosmological parameters, leading to variations based on halo mass and redshift, while local environmental conditions further influence density distributions, with stronger MG variants showing higher densities in dense regions and lower in voids. Together, these elements create a more complex structure formation scenario in MG models compared to GR \citep[see, e.g.,][]{2019MNRAS.487.1410M, 2021MNRAS.508.4140M}. Such increased density facilitates more PBH-NS interactions and subsequent mergers. The varying degrees of enhancement among the models reflect their different deviations from GR, with models deviating more strongly (e.g., f4 and nDGP(1)) showing larger increases in merger rates.

\begin{figure*}[ht!] 
\gridline{\fig{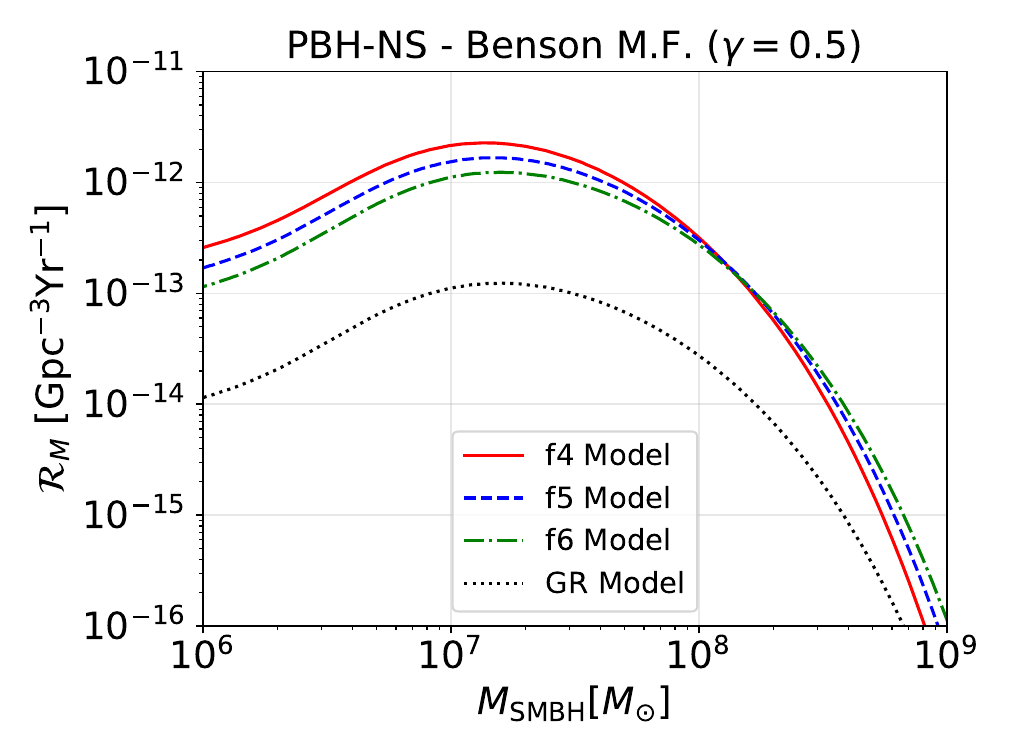}{0.33\textwidth}{(a)}
\fig{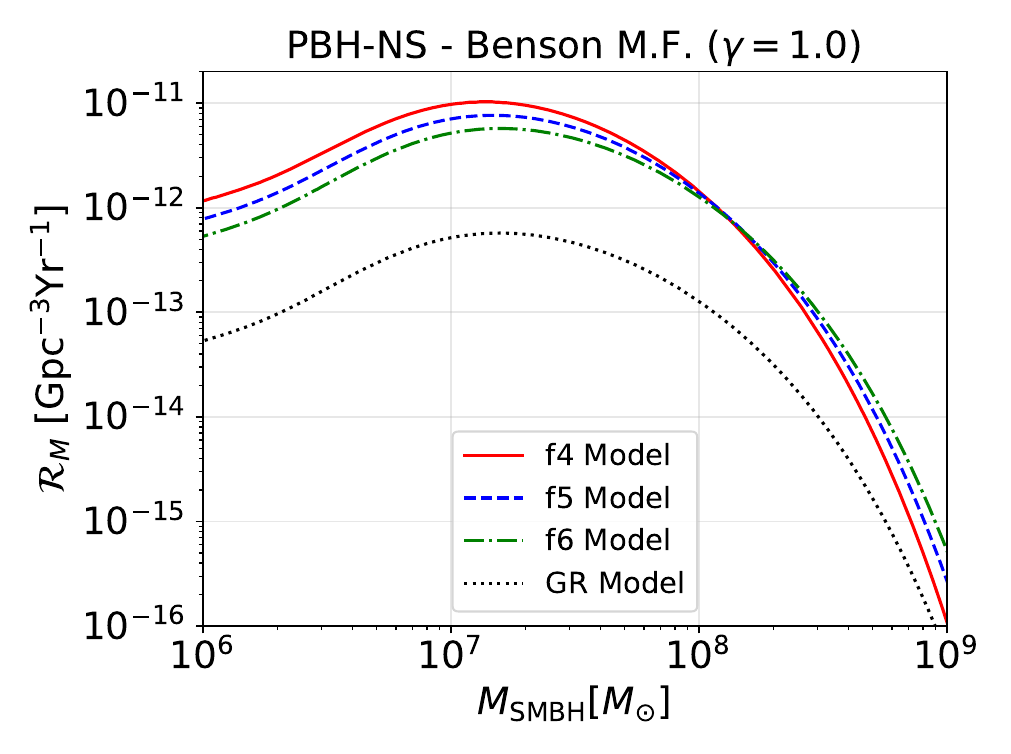}{0.33\textwidth}{(b)}
\fig{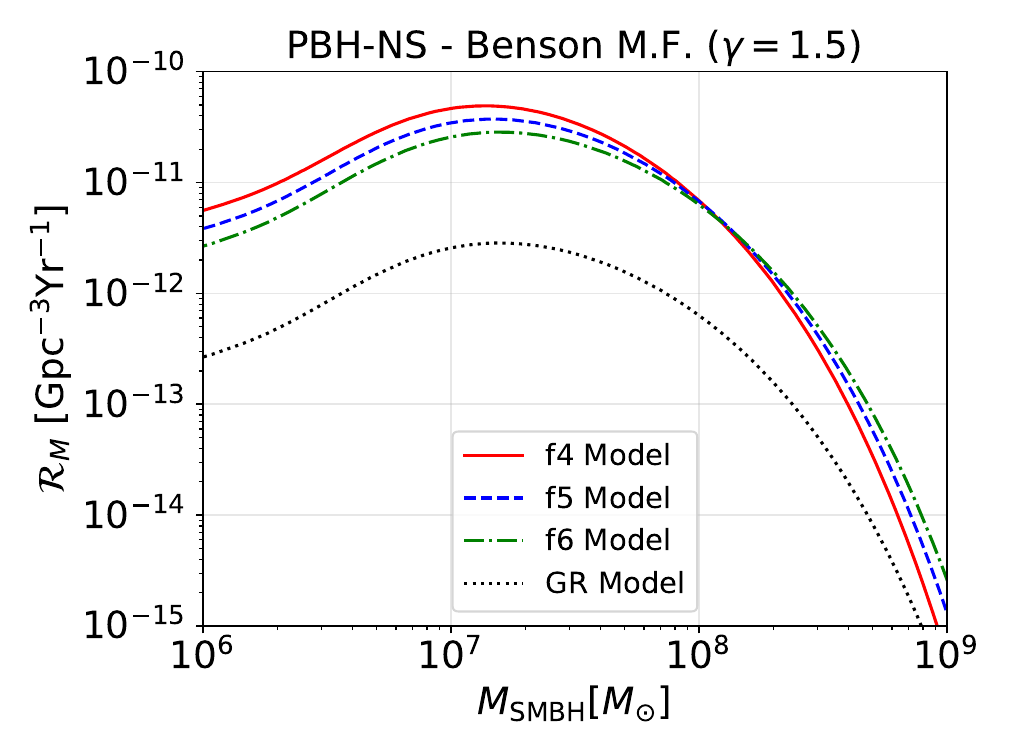}{0.33\textwidth}{(c)}
}
\gridline{\fig{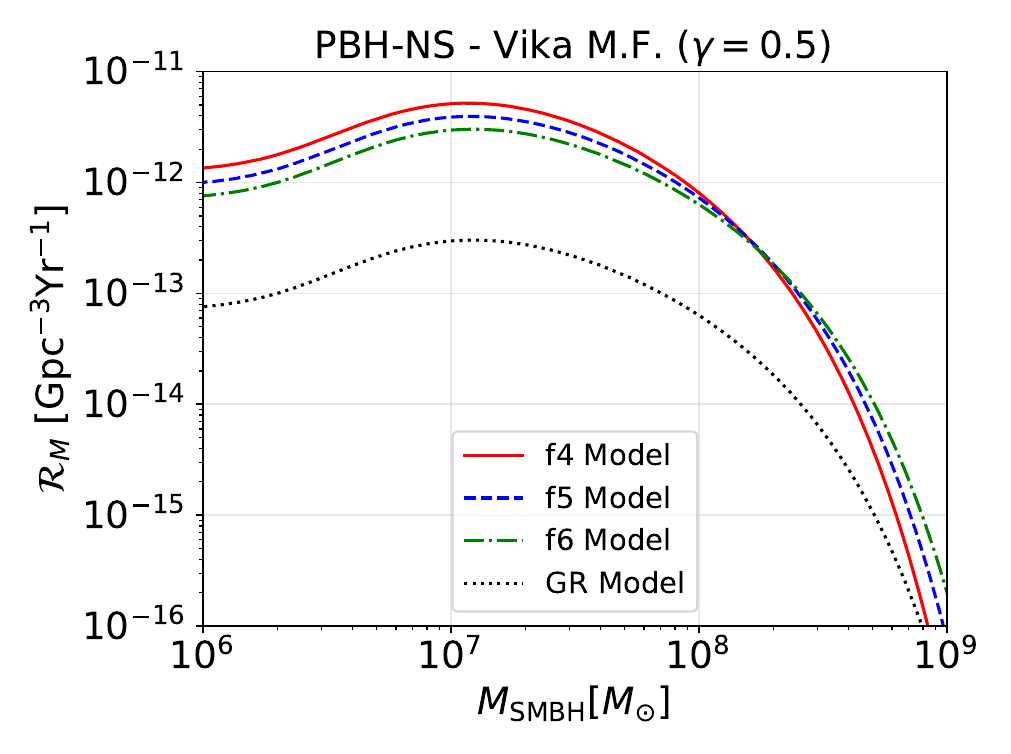}{0.33\textwidth}{(d)}
\fig{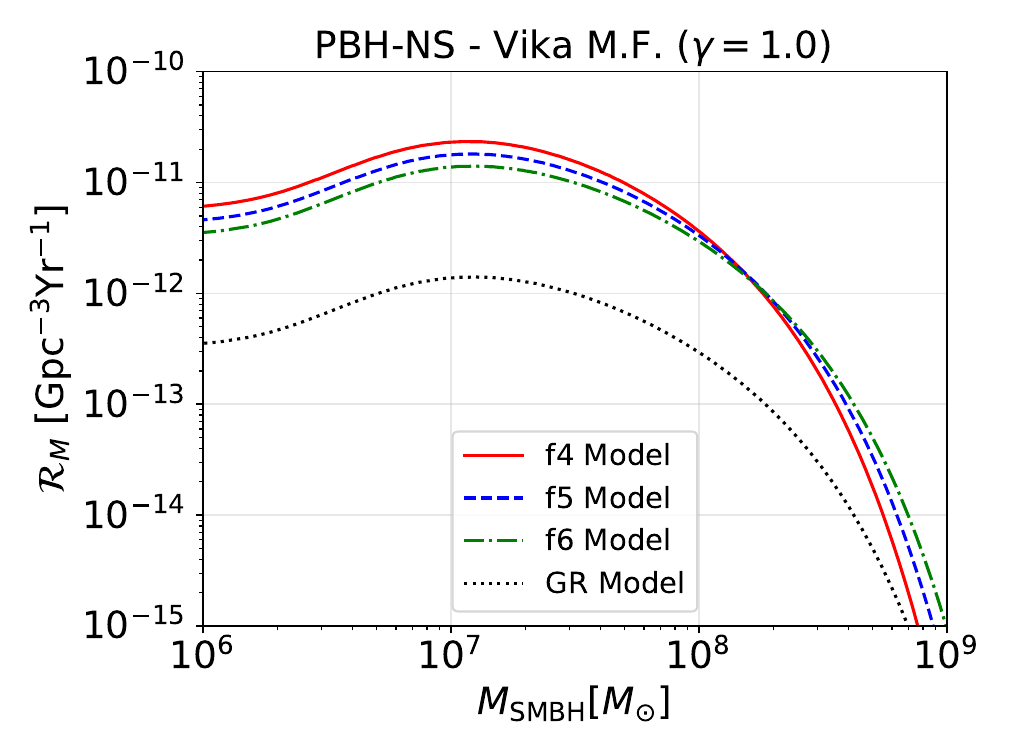}{0.33\textwidth}{(e)}
\fig{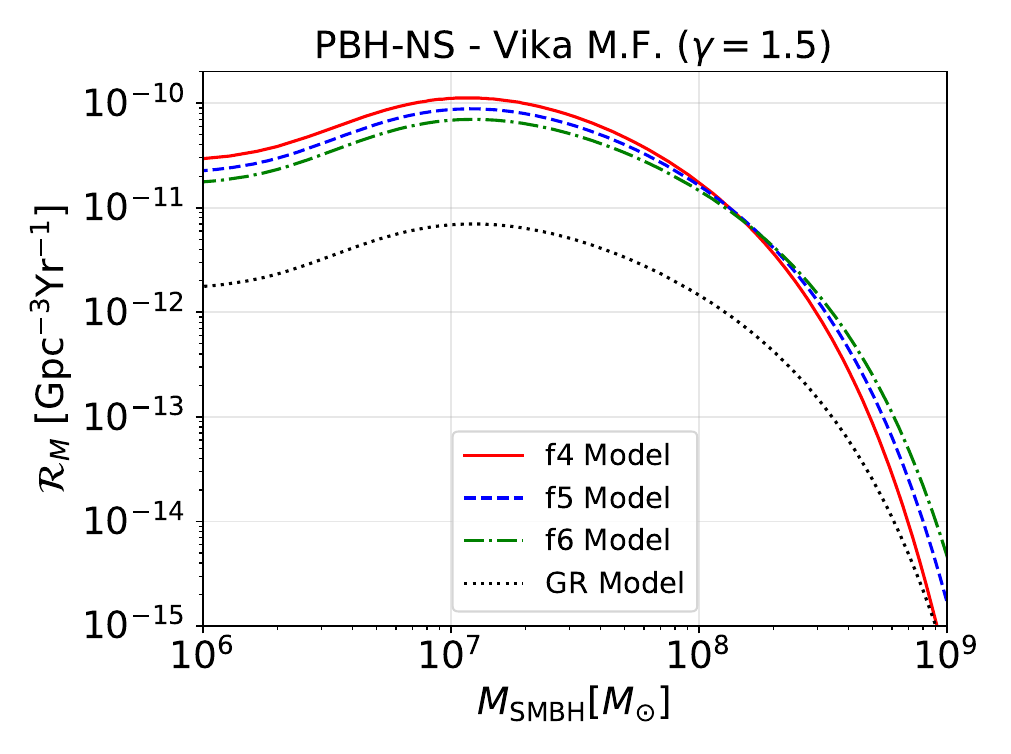}{0.33\textwidth}{(f)}
}
\gridline{\fig{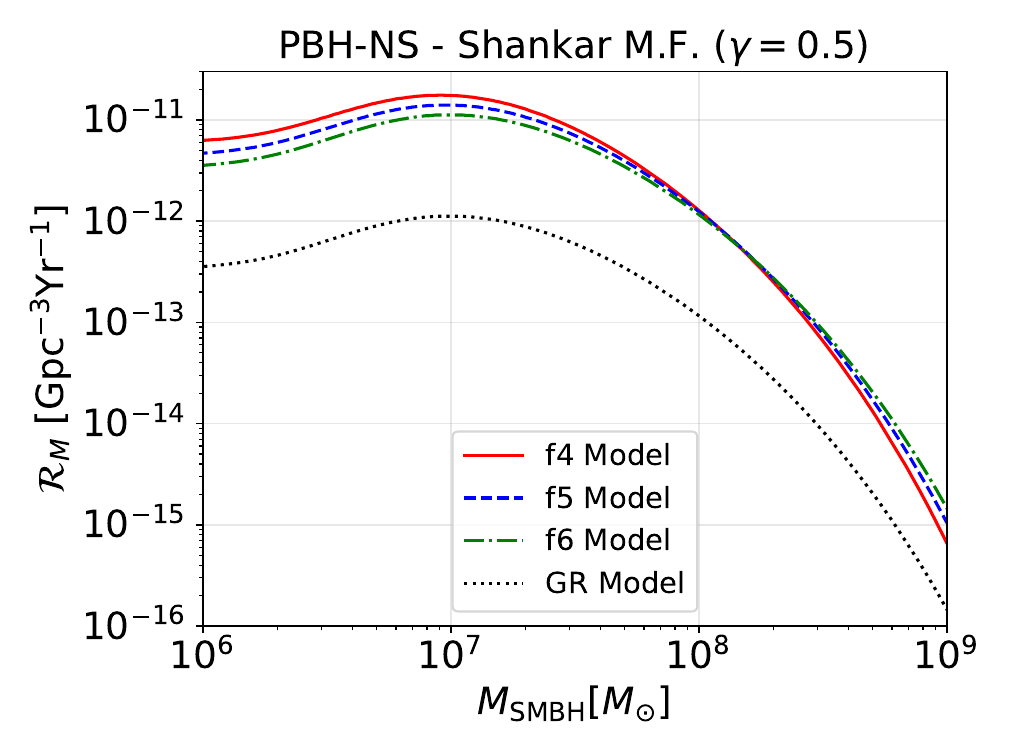}{0.33\textwidth}{(g)}
\fig{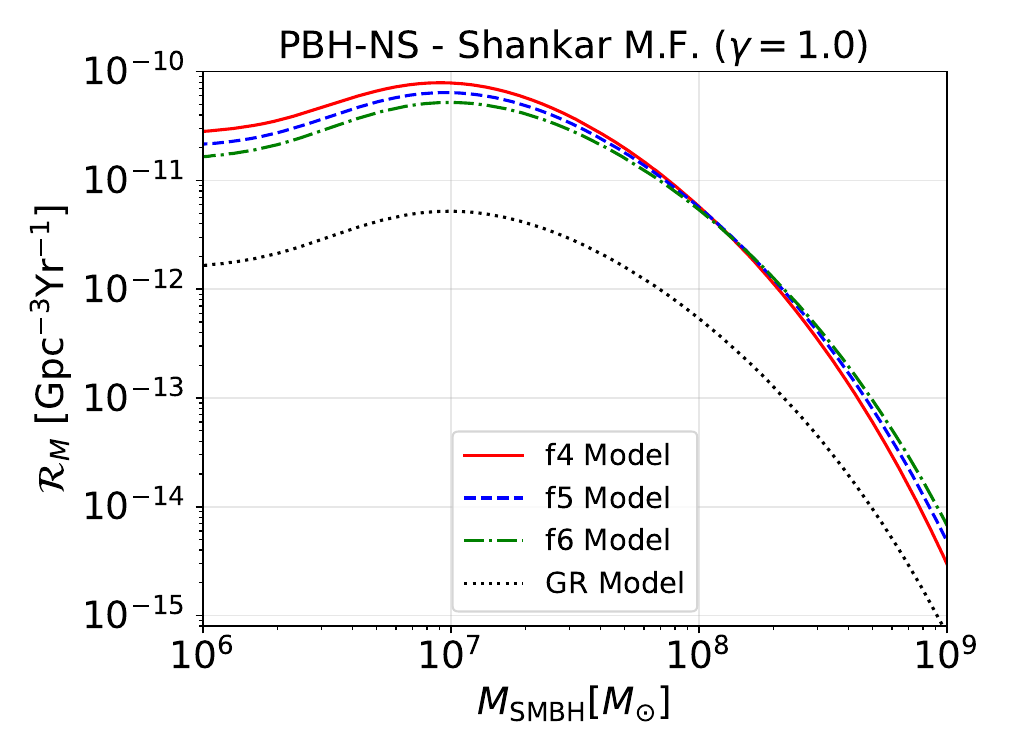}{0.33\textwidth}{(h)}
\fig{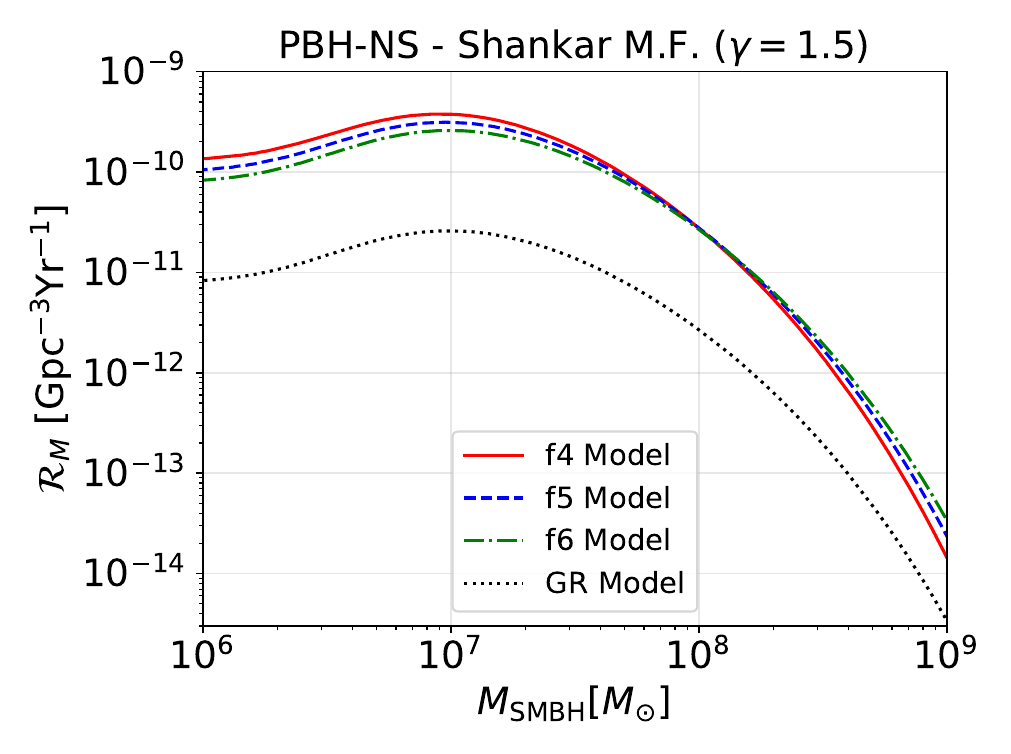}{0.33\textwidth}{(i)}
}
\caption{The merger rate of PBH-NS binaries per unit volume as a function of SMBH mass for different Hu-Sawicki $f(R)$ gravity models compared to GR. Results are presented for three different SMBH mass functions and three values of the power-law index $\gamma$ characterizing the steepness of the dark-matter spike density profile. Top panels: Benson mass function; Middle panels: Vika mass function; Bottom panels: Shankar mass function. From left to right, each column represents $\gamma = 0.5, 1.0$, and $1.5$, respectively.}
\label{fig6}
\end{figure*}

\begin{figure*}[ht!] 
\gridline{\fig{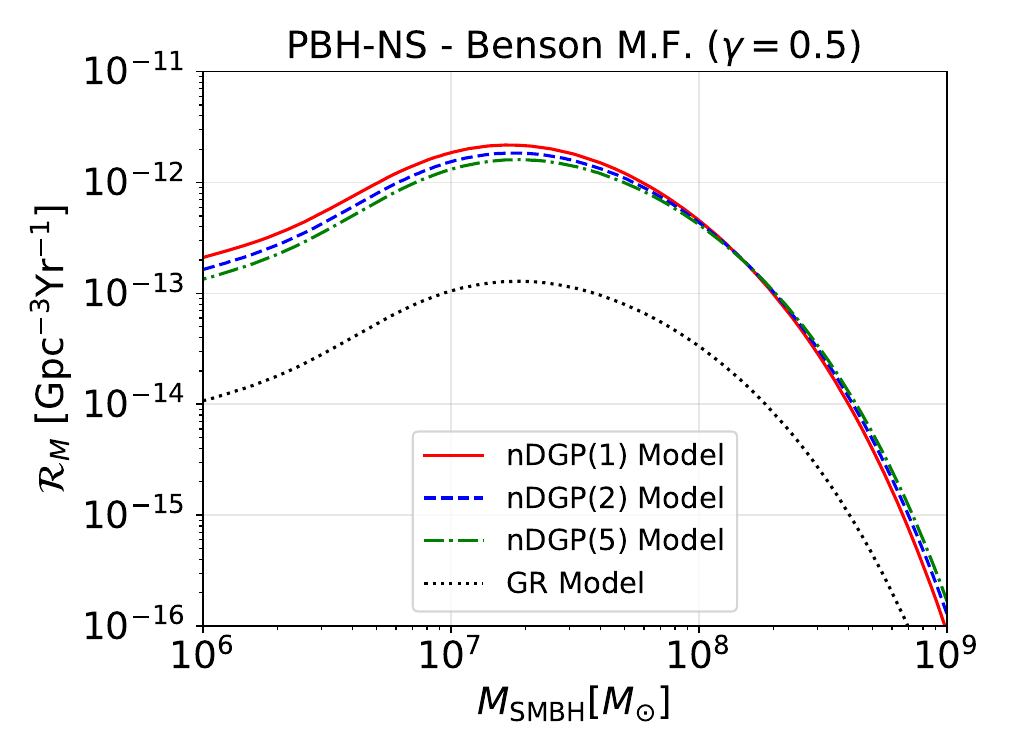}{0.33\textwidth}{(a)}
\fig{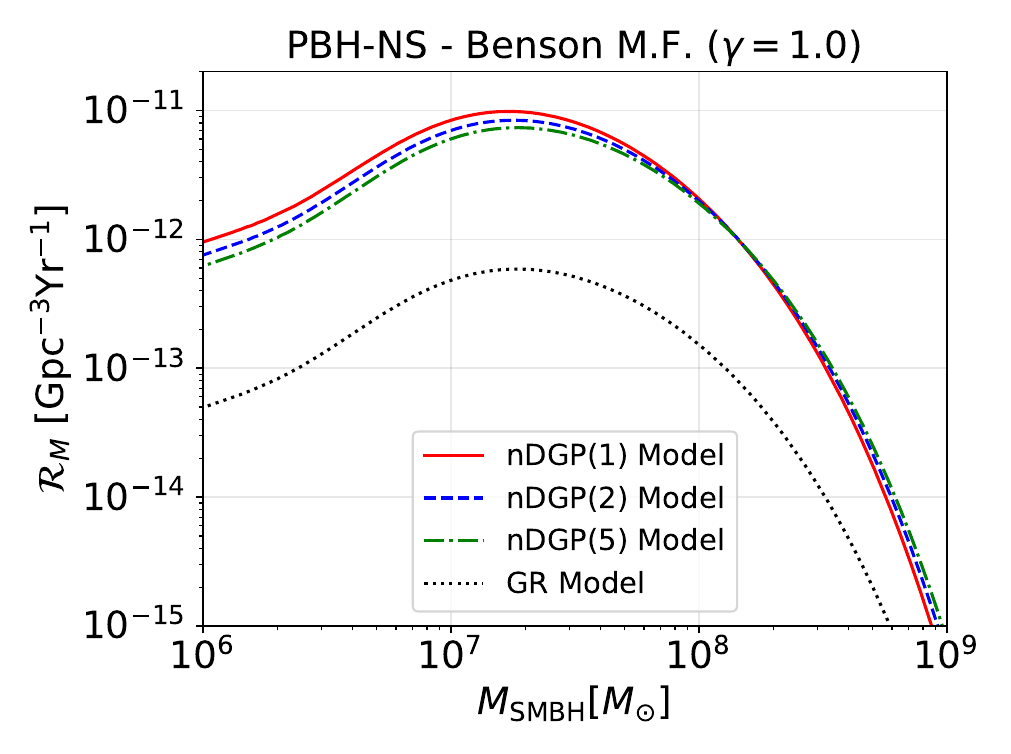}{0.33\textwidth}{(b)}
\fig{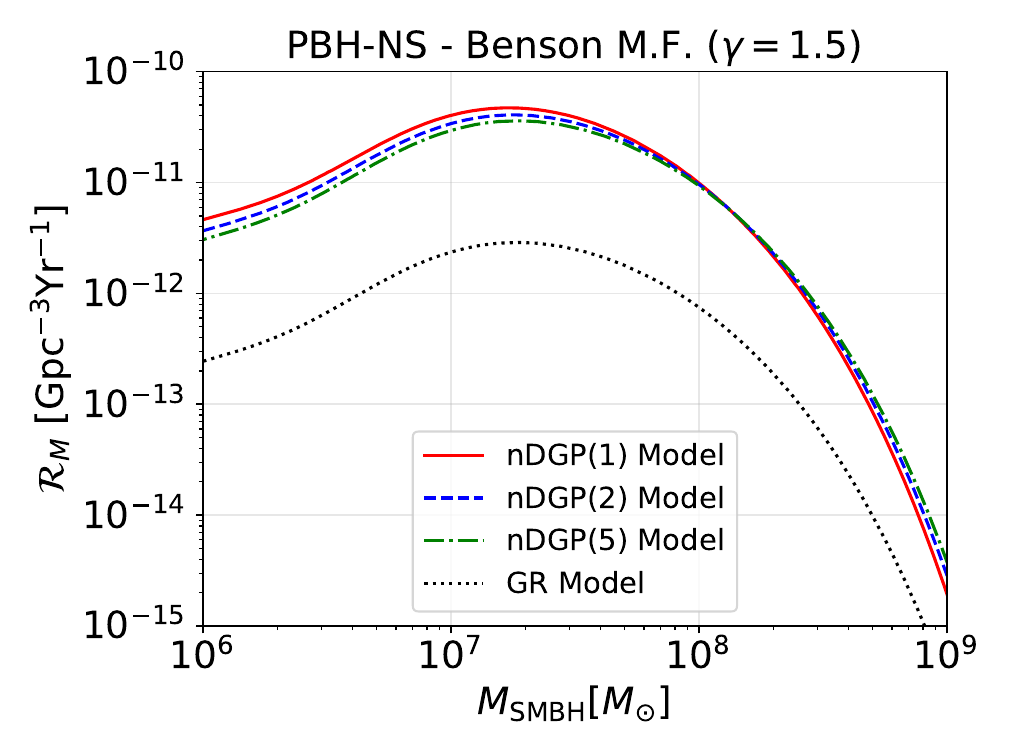}{0.33\textwidth}{(c)}
}
\gridline{\fig{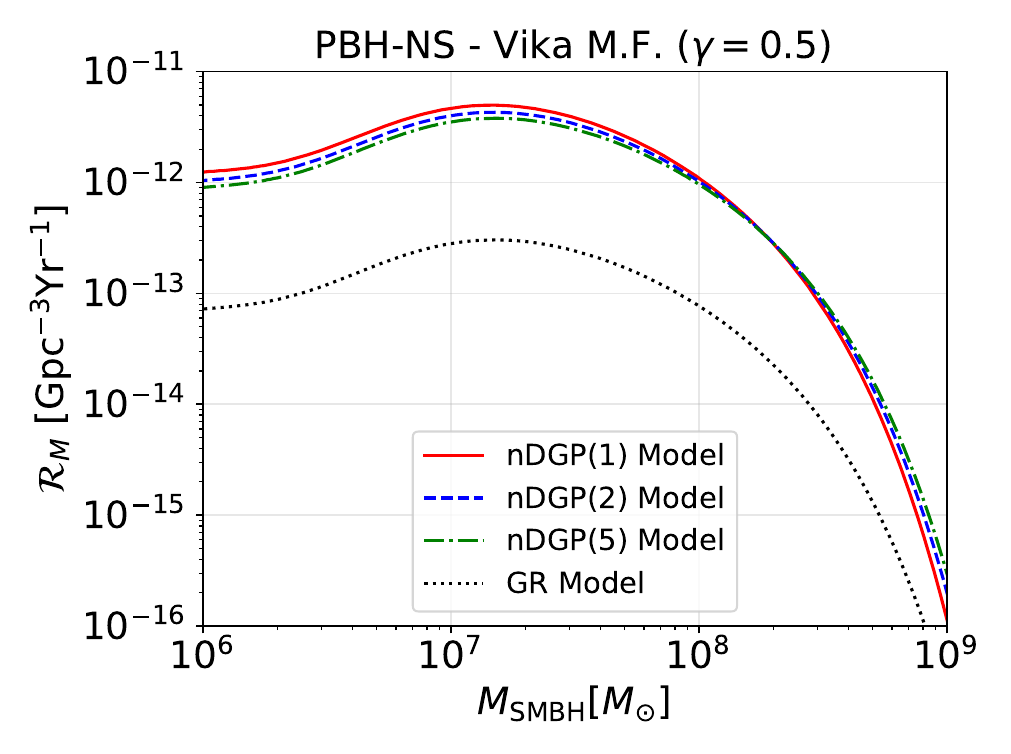}{0.33\textwidth}{(d)}
\fig{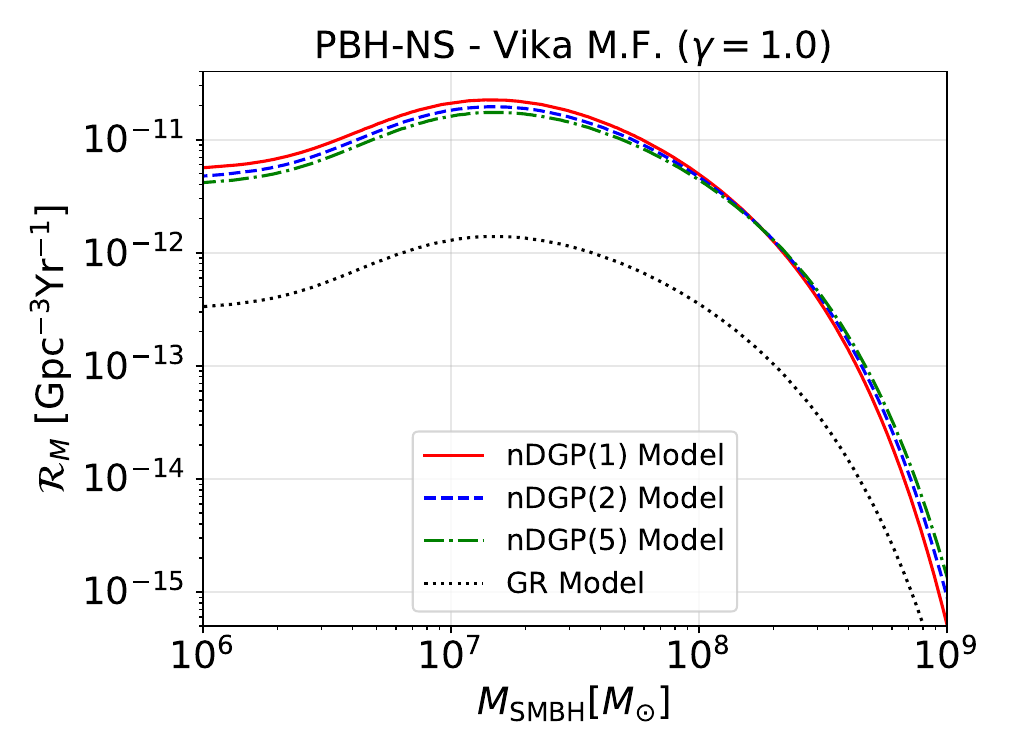}{0.33\textwidth}{(e)}
\fig{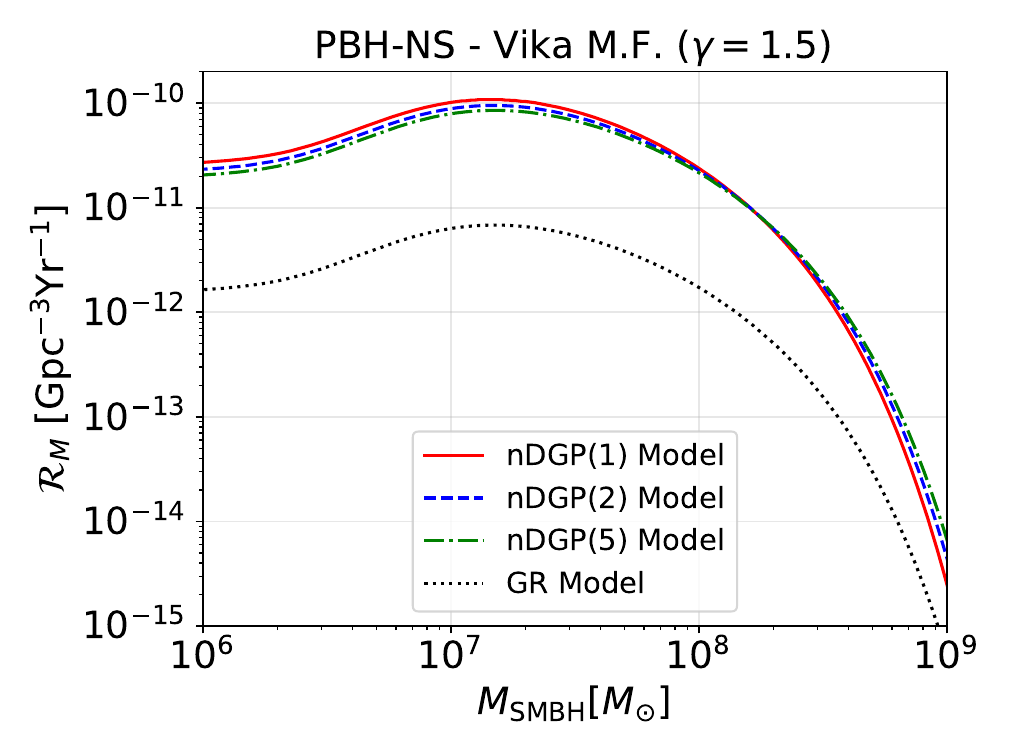}{0.33\textwidth}{(f)}
}
\gridline{\fig{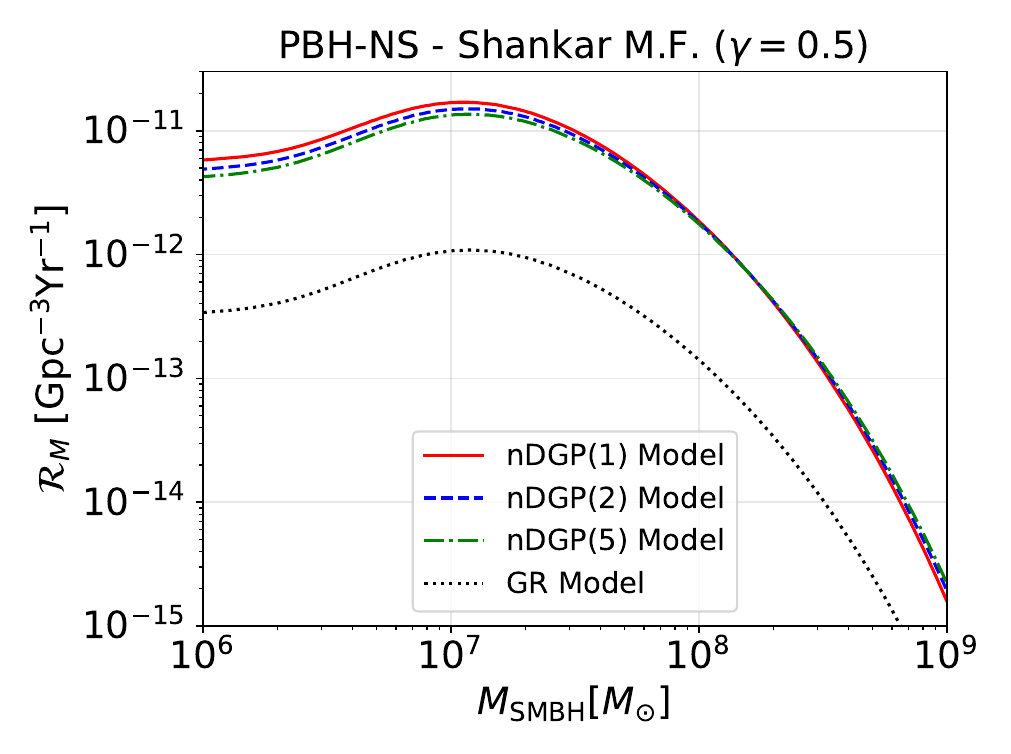}{0.33\textwidth}{(g)}
\fig{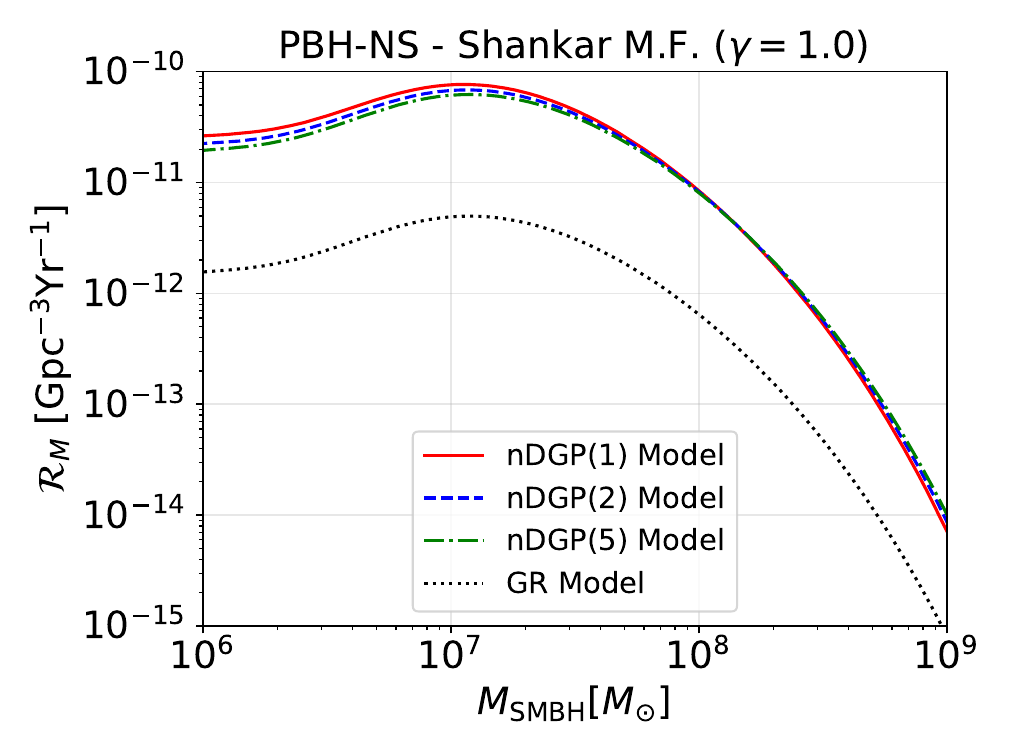}{0.33\textwidth}{(h)}
\fig{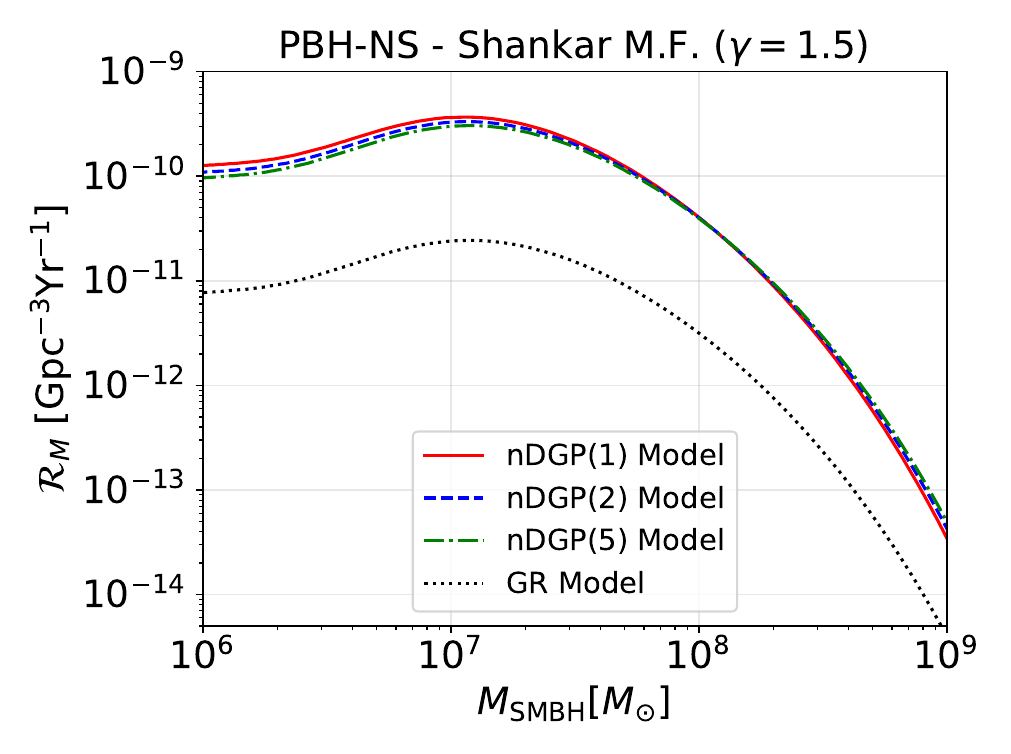}{0.33\textwidth}{(i)}
}
\caption{Similar to Fig.\,\ref{fig6} but for nDGP gravity models, i.e., nDGP(1), nDGP(2), nDGP(5).}
\label{fig7}
\end{figure*}

In Fig.\,\ref{fig6}, we have indicated the merger rate of PBH-NS binaries per unit volume as a function of SMBH mass for various Hu-Sawicki $f(R)$ gravity models, compared to GR. The results include three SMBH mass functions: Benson, Vika, and Shankar, and three values of the power-law index $\gamma$, which characterizes the steepness of the dark-matter spike density profile. This comprehensive analysis aims to discern the impact of different gravity models and SMBH mass functions on the predicted merger rate of PBH-NS binaries.

In the top panels, using the Benson mass function, all Hu-Sawicki $f(R)$ models predict higher merger rates than GR across the SMBH mass range. Among these models, f4 consistently shows the highest rates, followed by f5 and f6. An increase in $\gamma$ from $0.5$ to $1.5$ results in a significant rise in merger rates, indicating that the steepness of the dark-matter spike greatly influences these rates. This trend underscores the sensitivity of merger rates to the characteristics of the dark matter density profile.

The middle panels, employing the Vika mass function, exhibit similar trends but generally higher merger rates compared to the Benson function, implying that the choice of SMBH mass function significantly affects the predicted rates. Various models of $f(R)$ gravity diverge more from GR, especially at lower SMBH masses. This separation suggests that modifications in gravity theories have a pronounced effect when combined with certain mass functions, impacting the overall merger rate predictions.

The bottom panels, using the Shankar mass function, demonstrate analogous trends but show the highest merger rates among all three functions, highlighting the critical role of SMBH mass function selection. The distinction between $f(R)$ models and GR is most pronounced here. This pattern indicates that the variations in predicted merger rates are highly dependent on the SMBH mass function, with certain functions amplifying the effects of MG models more than others.

Across all panels, a consistent feature emerges: a peak in the merger rate around $M_{\rm SMBH} \approx 10^{7} M_{\odot}$, suggesting an optimal SMBH mass for PBH-NS mergers due to a balance between the gravitational influence of the SMBH and the dark-matter spike density. After this peak, the rates generally decline with increasing SMBH mass.

In Fig.\,\ref{fig7}, we have displayed the merger rate of PBH-NS binaries per unit volume as a function of SMBH mass for various nDGP gravity models, compared to GR.

The top panels utilize the Benson mass function. As the power-law index $\gamma$ increases from 0.5 to 1.5, the merger rates rise significantly across all nDGP models and GR. This demonstrates the strong dependence of merger rates on the steepness of the dark-matter spike density profile. A higher $\gamma$ value indicates a steeper profile, resulting in a more pronounced gravitational potential and enhanced likelihood of compact object interactions and mergers within the dark-matter spike. A consistent feature across these panels is the presence of a peak in the merger rate around $M_{\rm SMBH} \approx 10^{7} M_{\odot}$. This suggests an optimal SMBH mass range for PBH-NS mergers, where the balance between the gravitational influence of the SMBH and the dark-matter spike density is most favorable.

Comparing the nDGP models, the nDGP(1) model consistently predicts the highest merger rates, followed by nDGP(2) and nDGP(5), while GR forecasts the lowest rates. This hierarchy is maintained across the different values of $\gamma$, indicating that the nDGP(1) model, which deviates most strongly from GR, enhances the merger rates the most. This can be attributed to the nDGP(1) model's prediction of higher central densities in dark matter halos, facilitating more frequent PBH-NS interactions and mergers.

The middle panels, employing the Vika mass function, exhibit similar trends to the top panels but with generally higher merger rates. The distinction between the nDGP models and GR is more pronounced, especially at lower SMBH masses. The peak around $M_{\rm SMBH} \approx 10^{7} M_{\odot}$ is consistent with the top panels, but the subsequent decrease in merger rates is steeper for the Vika function compared to the Benson function.

The bottom panels, using the Shankar mass function, display the highest merger rates among the three mass functions considered. The separation between the nDGP models and GR is most evident in this case, indicating that the choice of SMBH mass function significantly impacts the predicted merger rates. The slope of the merger rate decrease with increasing SMBH mass is also the steepest for the Shankar function compared to the Benson and Vika functions.

Across all panels, the nDGP(1) model consistently predicts the highest merger rates, followed by nDGP(2) and nDGP(5), while GR forecasts the lowest rates. This hierarchy reflects the varying degrees of deviation from GR in these MG models, with the nDGP(1) case exhibiting the most significant enhancement in merger rates compared to the standard GR predictions. The deviations from GR are most pronounced at lower SMBH masses and become negligible at higher masses, suggesting that the influence of MG theories on PBH-NS merger rates diminishes as SMBH mass increases.

Note that mass segregation, particularly the concentration of more massive objects like compact remnants in the central regions of stellar systems, plays a crucial role in dense environments, where massive objects tend to drift toward the center \citep[see, e.g.,][]{2009ApJ...697.1861A, 2011MNRAS.413.2345H}. We have incorporated this well-established phenomenon into our analysis by assuming a spherically symmetric distribution of neutron stars. We have adopted a simplified exponential density profile, characterized by a radius $r^{*}_{\rm NS}$ and density $\rho^{*}_{\rm NS}$, to account for neutron star abundance within a typical galactic halo. We have constrained the free parameters through careful consideration of observational data, selecting the initial mass function and the stellar mass-halo mass relation to capture the effects of mass segregation. We have validated this approach through results showing that the merger rate of PBH-NS binaries in regions with dark matter spikes is significantly lower than in dark matter halo environments, as noted by \cite{2022ApJ...941...36F, 2024ApJ...966..235F}.

\begin{figure*}[ht!] 
\gridline{\fig{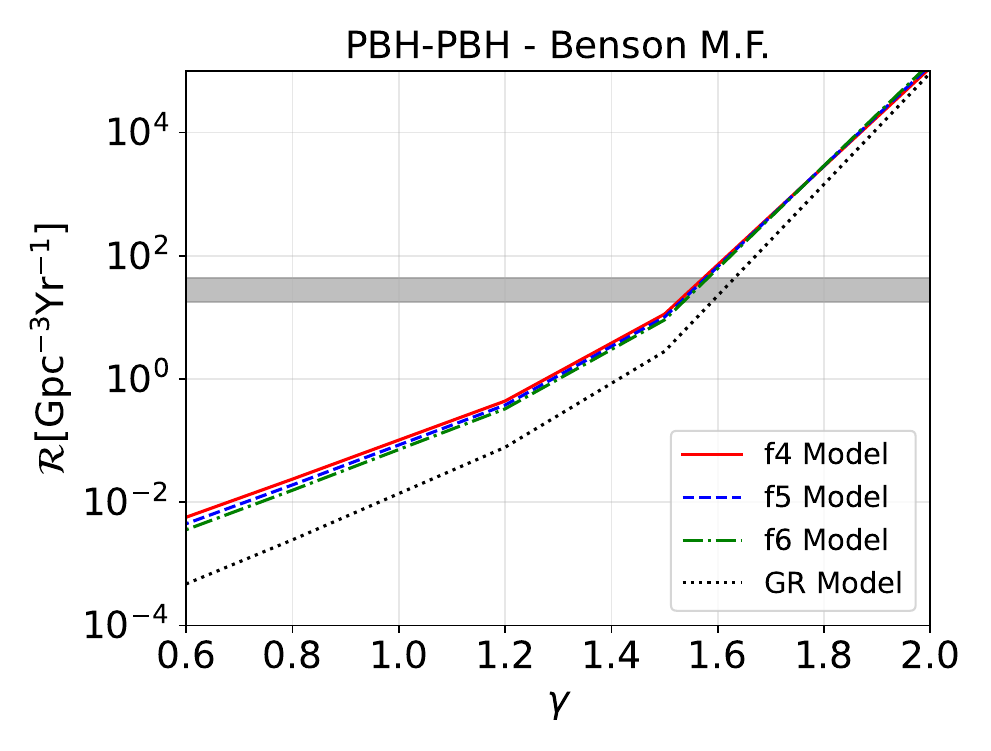}{0.33\textwidth}{(a)}
\fig{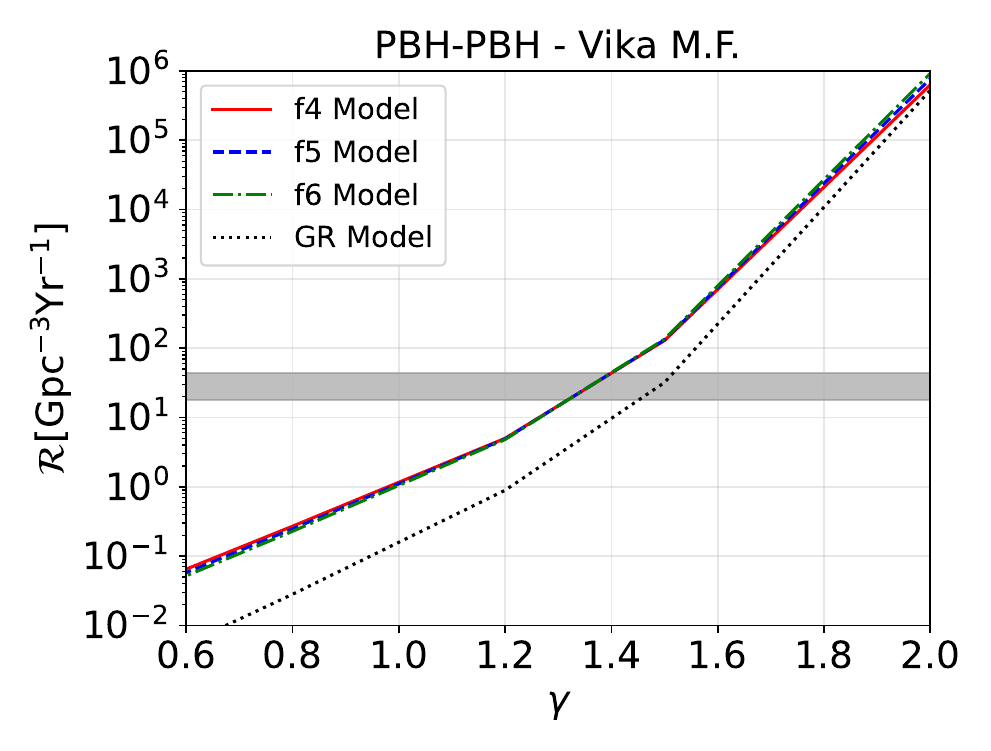}{0.33\textwidth}{(b)}
\fig{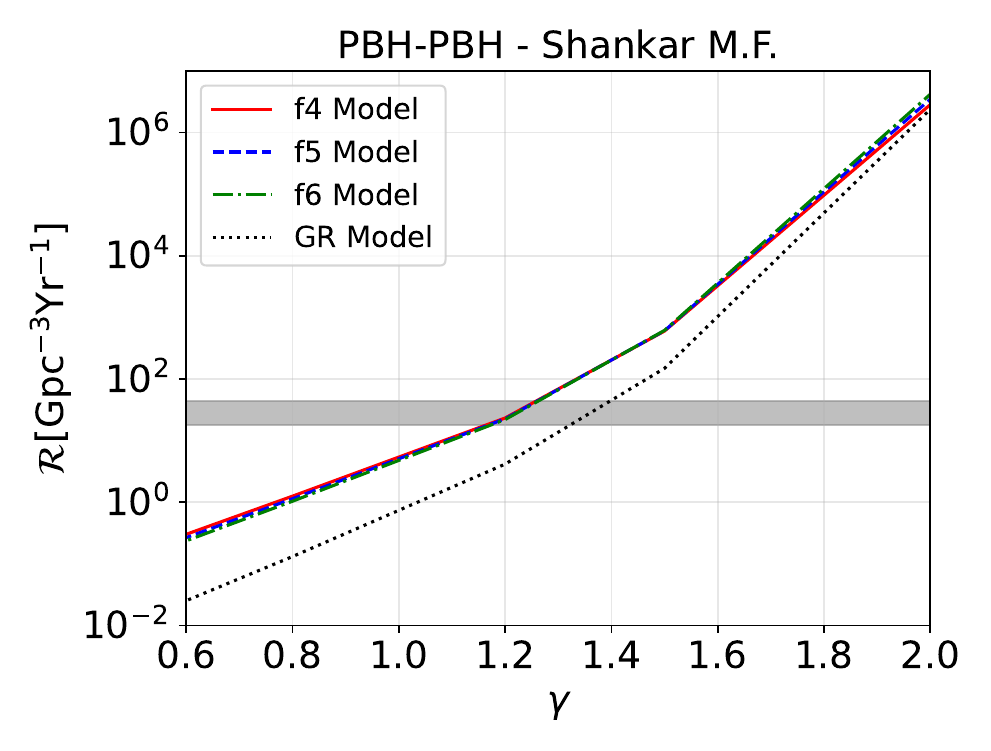}{0.33\textwidth}{(c)}
}
\gridline{\fig{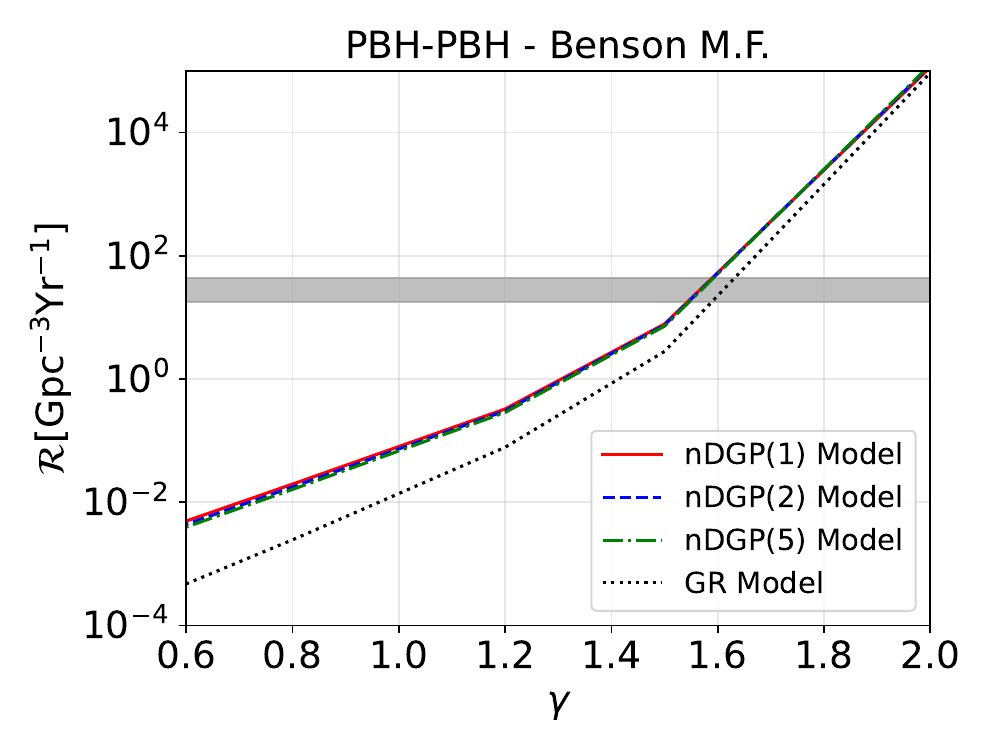}{0.33\textwidth}{(d)}
\fig{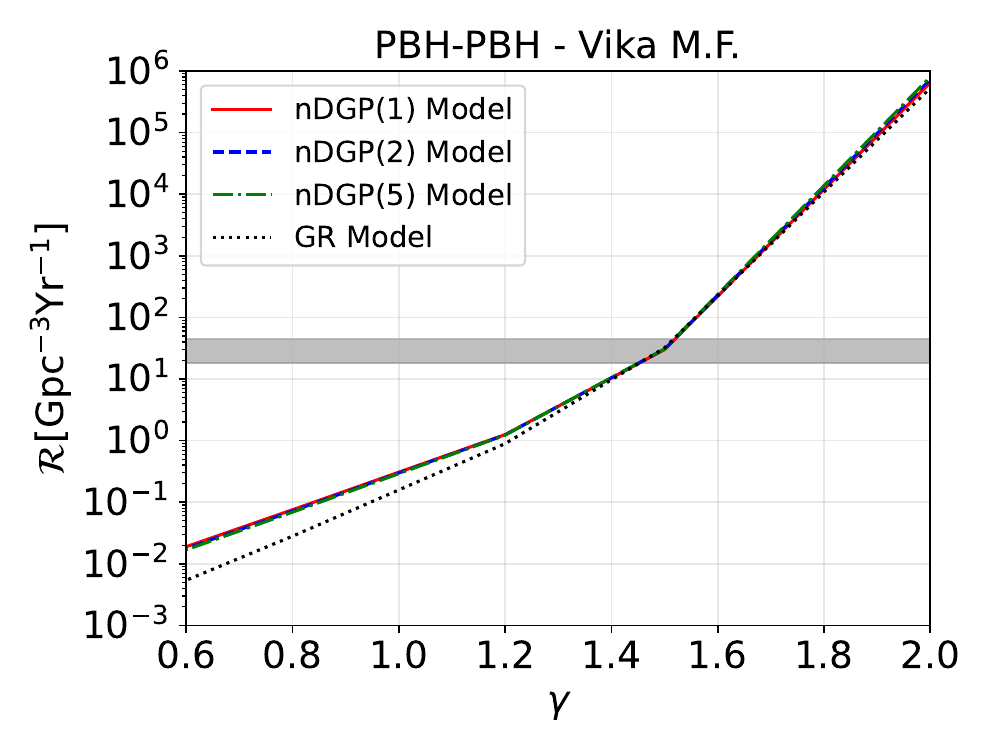}{0.33\textwidth}{(e)}
\fig{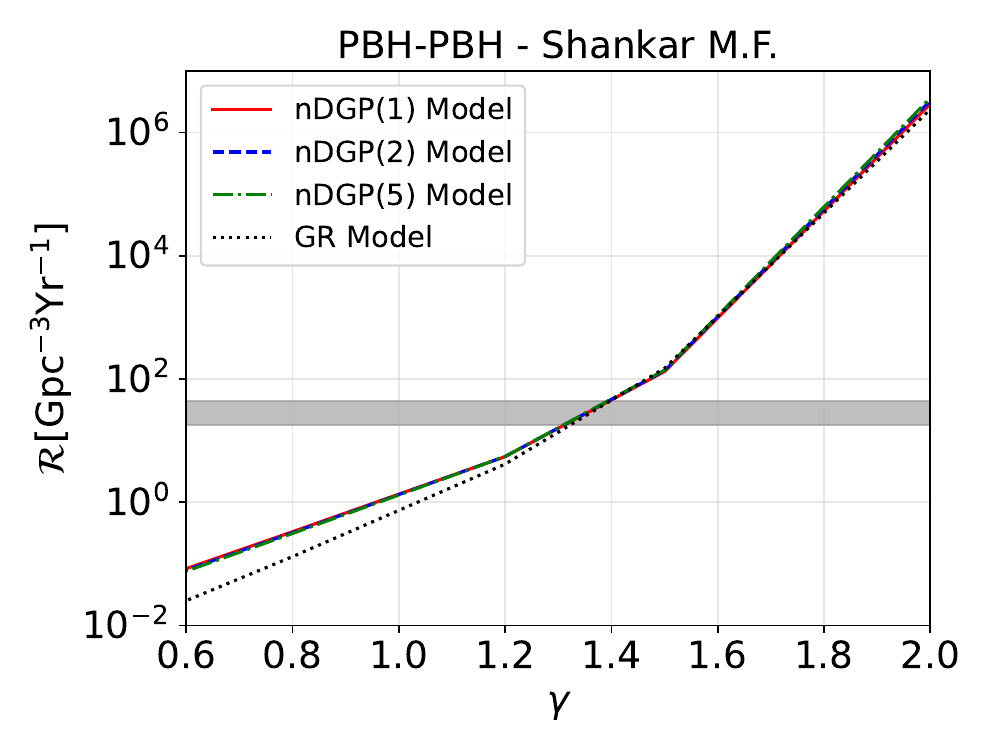}{0.33\textwidth}{(f)}
}
\caption{The cumulative merger rate of PBH-PBH binaries as a function of the power-law index $\gamma$ for different gravitational models. Top panels: Results for Hu-Sawicki $f(R)$ gravity models compared to GR. Bottom panels: Results for nDGP gravity models compared to GR. From left to right: Merger rates calculated using the Benson, Vika, and Shankar SMBH mass functions, respectively. The shaded gray bands represent the BH-BH mergers estimated by the LVK detectors during the latest observing run, i.e., $(17.9\mbox{-}44)\, {\rm Gpc^{-3} yr^{-1}}$ \citep{2021arXiv211103606T}.}
\label{fig8}
\end{figure*}

\subsection{Constraint on Power-Law Index}

In Fig.\,\ref{fig8}, we have demonstrated the cumulative merger rate of PBH-PBH binaries as a function of the power-law index $\gamma$, for different Hu-Sawicki $f(R)$ and nDGP gravity models compared to GR. The shaded gray bands represent the BH-BH mergers estimated by the LVK detectors during the latest observing run, i.e., $(17.9-44)\, {\rm Gpc^{-3} yr^{-1}}$ \citep{2021arXiv211103606T}. 

The top panels show this analysis for different Hu-Sawicki $f(R)$ gravity models in comparison with GR. As $\gamma$ increases, the divergence between the Hu-Sawicki $f(R)$ gravity models and GR becomes less pronounced in terms of the merger rates. 

In the left top panel, using the Benson mass function, the merger rates for Hu-Sawicki $f(R)$ models, f4, f5, and f6, respectively, are significantly higher than the GR prediction, especially at lower values of $\gamma$ around $(0.6-1.5)$. In this range, Hu-Sawicki $f(R)$ gravity models predict merger rates that are several orders of magnitude higher than GR. However, as $\gamma$ increases further, the differences between the models start to diminish. For $\gamma$ values above $1.5$, the Hu-Sawicki $f(R)$ models converge closer to the GR predictions, indicating that the impact of the MG theories becomes less dominant in regions with extremely steep dark matter density profiles. This trend indicates that the enhanced central densities forecasted by $f(R)$ models significantly contribute to the rise in merger rates. As the volume of the dark-matter spike decreases (with lower values of $\gamma$), the gravitational dynamics become increasingly governed by the high-density medium. Conversely, when the spike profiles encompass a larger volume (higher values of $\gamma$), the relative influence of the underlying Hu-Sawicki $f(R)$ models diminishes. This inference can be drawn by comparing the predictions of Hu-Sawicki $f(R)$ models and GR with the sensitivity range of the LVK detectors, based on the merger rate of BH-BH binaries. Specifically, the predictions of Hu-Sawicki $f(R)$ models in intervals with lower values than GR align with the sensitivity range of the LVK detectors.

The middle top panel uses the Vika mass function and shows a similar trend, but with overall higher merger rates compared to the Benson function. The separation between the $f(R)$ models and GR is more pronounced, particularly at lower $\gamma$ values. Again, the f4, f5, and f6 models predict the higher merger rates than GR, respectively. It is important to note that when incorporating the Vika mass function, the predictions of Hu-Sawicki $f(R)$ models and GR for the merger rate of PBH-PBH binaries, at smaller $\gamma$ values compared to the Benson mass function, fall within the compatibility range of the LVK detectors.

In the right top panel, calculations were performed using the Shankar mass function, yielding the highest merger rate of PBH-PBH binaries among the three mass functions considered. The comparative process of these results is similar to the previous two cases, owing to the compatibility of the prediction models with the sensitivity range of theLVK detectors. Consequently, when using the Shankar mass function, the predictions of Hu-Sawicki $f(R)$ models and GR for the merger rate of PBH-PBH binaries for lower $\gamma$ values (compared to the previous two mass functions) fall within the sensitivity window of the LVK detectors.

The bottom panels of Fig.\,\ref{fig8} present the cumulative merger rate of PBH-PBH binaries as a function of $\gamma$, but this time for different nDGP gravity models compared to GR. Similar to the $f(R)$ models, the divergence between the nDGP models and GR strengthens as $\gamma$ decreases. However, compared to the Hu-Sawicki $f(R)$ models, the nDGP models exhibit a lower ability to enhance the merger rate of PBH-PBH binaries relative to GR, as illustrated in the bottom panels.

The left bottom panel, using the Benson mass function, shows that nDGP models predict higher merger rates than GR, with nDGP(1) exhibiting the most significant enhancement, followed by nDGP(2) and nDGP(5). As $\gamma$ increases, the merger rate of PBH-PBH binnaries increase dramatically, consistent with the trend observed in the top panels. When compared to the sensitivity range of LVK detectors, the merger rate predictions for PBH-PBH binaries derived from nDGP models, across $\gamma$ values lower than those in GR, fall within the acceptable range.

\begin{figure*}[ht!] 
\gridline{\fig{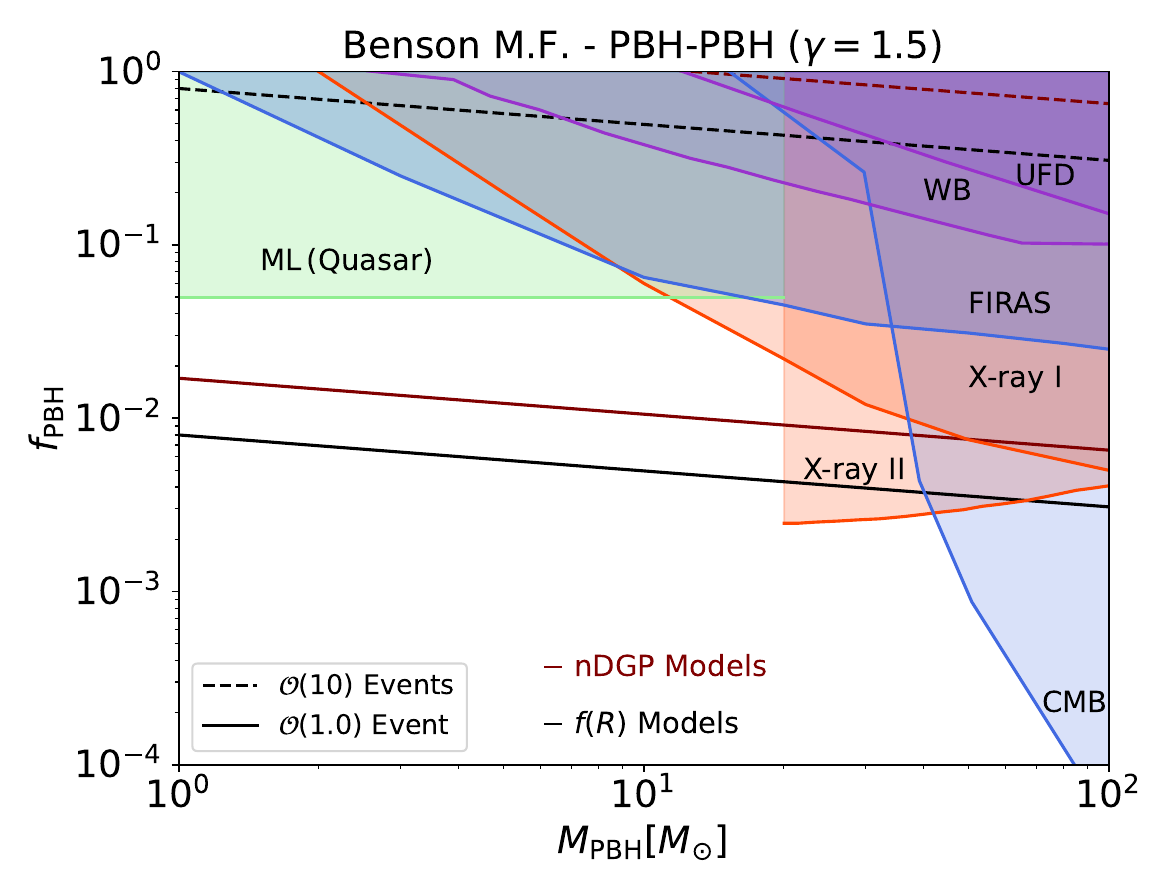}{0.33\textwidth}{(a)}
\fig{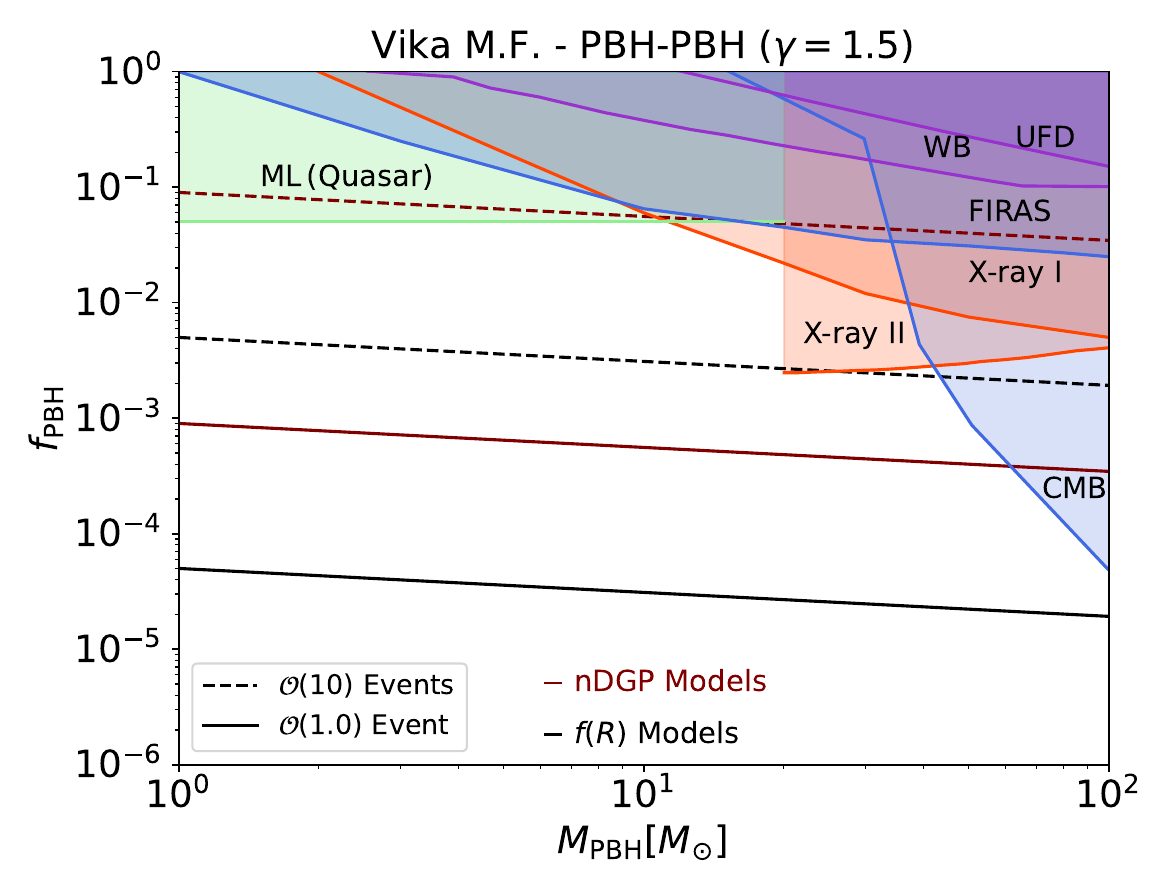}{0.33\textwidth}{(b)}
\fig{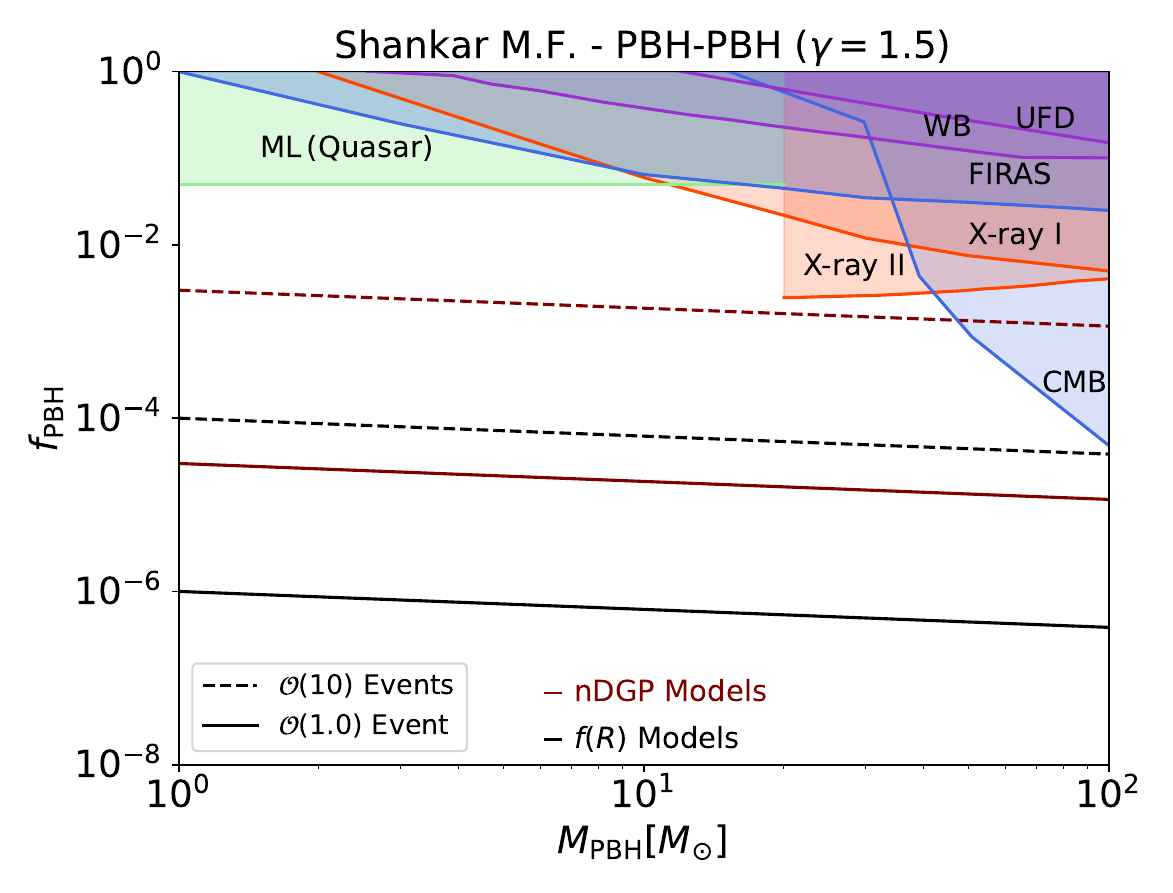}{0.33\textwidth}{(c)}
}
\caption{The expected bounds on the fraction of PBHs as a function of their masses are shown, assuming the detection of approximately $\mathcal{O}(1)$ event (solid lines) and approximately $\mathcal{O}(10)$ events (dashed lines) of PBH-PBH binaries within a comoving volume of $1\,{\rm Gpc^{3}}$. These calculations consider the Hu-Sawicki $f(R)$ and nDGP models and incorporate the Benson, Vika, and Shankar mass functions. The power-law index is set to be $\gamma=1.5$. Observational constraints on PBH dark matter include: gravitational microlensing constraints from quasars (ML Quasar) \citep{2009ApJ...706.1451M}; constraints from the disruption of wide binaries (WB) \citep{2014ApJ...790..159M}; constraints from the disruption of ultra-faint dwarfs (UFD) \citep{2016ApJ...824L..31B}; X-ray constraints related to accreting PBHs (X-rayI) \citep{2017JCAP...10..034I}; corresponding X-ray constraints in the Milky Way (X-rayII) \citep{2019JCAP...06..026M}; and constraints from modifications to the cosmic microwave background (CMB) spectrum due to accreting PBHs \citep{2020PhRvR...2b3204S}, which include particle dark matter accretion and FIRAS data \citep{2008ApJ...680..829R}.}
\label{fig9}
\end{figure*}

The middle bottom panel, which employs the Vika mass function, exhibits comparable trends but indicates an even higher overall merger rate. The deviation of nDGP models from GR is less pronounced than with the Benson mass function. The predicted merger rates of PBH-PBH binaries for both gravitational models fall within a roughly similar range of $\gamma$ within the permitted regions of the LVK detector observations.

In the right bottom panel, the Shankar mass function yields the highest merger rate of PBH-PBH binaries compared to the other two mass functions. The differences between the nDGP models and GR are negligible, particularly at higher values of $\gamma$. Additionally, as with the scenario where the Vika mass function is used, the predictions of the merger rate of PBH-PBH binare from both gravitational models are nearly identical across the same range of $\gamma$ within the sensitivity range of LVK detectors.

Across all panels, the MG models, both Hu-Sawicki $f(R)$ and nDGP, consistently predict higher merger rates than GR, especially at lower $\gamma$ values. This enhancement can be attributed to the higher central densities in dark matter halos predicted by these models, which facilitate more frequent PBH-PBH interactions and mergers. As $\gamma$ increases, the divergence between the MG theories and GR becomes negligible, indicating that the enhanced central densities play a more crucial role in driving up the merger rates when the dark-matter spike has a more gradual slope (lower $\gamma$'s). However, for very steep spike profiles (higher $\gamma$'s), the gravitational dynamics become increasingly dominated by the high-density environment, reducing the relative influence of the underlying MG.

Notwithstanding the comparative augmentation in merger rates of PBH-NS binaries when considering Hu-Sawicki $f(R)$ and nDGP models relative to those derived from GR, the calculated rates, which fall within the range of $\mathcal{O}(10^{-7}-10^{-9})$, remain substantially disparate from the projected detection window of LVK observatories for BH-NS binary mergers, estimated at $(7.8\mbox{-}140)\,{\rm Gpc^{-3}yr^{-1}}$ \citep{2021arXiv211103606T}. 

\subsection{Constraint on PBH Fraction}

In Fig.\,\ref{fig9}, we have plotted the expected bounds on the fraction of PBHs as a function of their masses, assuming the detection of approximately one and ten PBH-PBH binary events within a co-moving volume of $1\,{\rm Gpc^{3}}$. In these calculations, we have considered the Hu-Sawicki $f(R)$ and nDGP models and incorporated the Benson, Vika, and Shankar mass functions. We have also set the power-law index to $\gamma = 1.5$, which is close to the LVK sensitivity band in almost all models of Fig.\,\ref{fig8}. For a convenient comparison, we have also provided observational constraints on PBH dark matter, including gravitational microlensing constraints from quasars (ML Quasar), constraints from the disruption of wide binaries (WB) and ultra-faint dwarfs (UFD), X-ray constraints related to accreting PBHs (X-rayI and X-rayII), and constraints from modifications to the cosmic microwave background (CMB) spectrum. At first glance, Hu-Sawicki $f(R)$ models predict PBH abundances lower than nDGP models.

The left figure, using the Benson mass function, shows that both the Hu-Sawicki $f(R)$ and nDGP models predict bounds on the fraction of PBHs that are tighter when $\mathcal{O}(1)$ event can be detected by LVK detectors. The black and maroon lines representing the predictions of these models lie below many observational constraints, indicating that these predictions suggest lower abundances of PBHs. Specifically, the predicted lines lead to more stringent constraints than ML Quasar, WB, UFD, and X-ray I for $M_{\rm PBH}<20 M_{\odot}$, which is $f_{\rm PBH}\leq 10^{-2}$. However, these models cannot overcome observational constraints on the fraction of PBHs when $\mathcal{O}(10)$ events of PBH-PBH binaries are expected to be detected by LVK detectors.

In the middle figure, with the Vika mass function, the predicted bounds on the fraction of PBHs are more stringent than those obtained from the Benson mass function while maintaining a similar trend, highlighting the prominent effect of the Vika mass function in limiting the allowed parameter space for PBHs. Theoretical predictions show that the constraints derived from the $f(R)$ and nDGP models can be more robust than most of the observational constraints provided if one can expect to detect $\mathcal{O}(1)$ event by LVK detectors, i.e., $f_{\rm PBH}\leq 10^{-4} \sim 10^{-3}$. On the other hand, the predictions of $f(R)$ models regarding the abundance of PBHs, with an expectation of detecting around $\mathcal{O}(10)$ merger events, also appear promising, i.e., $f_{\rm PBH}\leq 10^{-2}$. However, similar conditions cannot be established for the predictions of nDGP models.

In the right figure, utilizing the Shankar mass function, the most rigorous projected limitations are shown among the three mass functions, encompassing both the $f(R)$ and nDGP models. These models anticipate PBH abundances of approximately $f_{\rm PBH}\leq 10^{-6} \sim 10^{-4}$ and $f_{\rm PBH}\leq 10^{-5} \sim 10^{-3}$, respectively, alongside expectations of detecting $\mathcal{O}(1)$ and $\mathcal{O}(10)$ PBH-PBH events annually by LVK detectors. These predictions significantly exceed observational constraints, highlighting the robustness of the models.

At the end of this section, we recall that our study has developed a theoretical model centered on the formation of BHs of primordial origin. Consequently, we have excluded formation pathways for BHs of similar masses that result from the remnants of baryonic matter, such as the collapse of massive stars \citep[see, e.g., ][]{2012ApJ...749...91F}. It is crucial to highlight that the constraints on the abundance fraction of PBHs presented in this analysis represent the upper limits allowed by current observational data. Because some BHs within this mass range could originate from non-primordial channels and may contribute to merger events.
\section{Conclusions}\label{sec:v}
In this study, we have investigated the merger rate of compact binaries within dark-matter spikes surrounding SMBHs in the context of two widely studied MG models: Hu-Sawicki $f(R)$ gravity and the nDGP gravity. By employing three SMBH mass functions, Benson, Vika, and Shankar, we have examined how these MG models and mass functions influence the predicted merger rates of binaries involving PBHs and/or NSs.

Our results show that both Hu-Sawicki $f(R)$ and nDGP models predict consistently higher merger rates for PBH-PBH and PBH-NS binaries compared to GR. This enhancement is particularly significant at lower SMBH masses and for lower values of the power-law index. This suggests that MG models may play a crucial role in environments with high dark-matter densities, such as those surrounding SMBHs.

When comparing different SMBH mass functions, we have found that the choice of mass function significantly affects the predicted merger rates. Specifically, the Shankar and Vika mass functions provide the lowest abundance of PBHs across both MG models. This highlights the importance of accurately modeling the SMBH mass distribution when studying the merger rates of compact binaries in dark-matter spikes. We have shown that the structure of dark-matter spikes, influenced by the growth of the SMBH and the dark-matter halo model, is crucial in determining the merger rates of compact binaries. The enhanced density of dark-matter particles within these spikes implies a significant number density of PBHs, which, under MG models, leads to higher merger rates than those predicted by GR.

Our results also demonstrate that the Hu-Sawicki $f(R)$ models generally predict a lower abundance of PBHs than the nDGP models, especially when utilizing the Vika and Shankar mass functions. This suggests that the Hu-Sawicki $f(R)$ models may be more effective in explaining the observed merger rates and constraining the properties of dark-matter spikes. We have found that the orders of enhancement in the Hu-Sawicki $f(R)$ models, i.e., f4, f5, f6, respectively, relative to GR show the importance of the field strength $f_{R0}$ in our analysis. Similarly, the nDGP models, i.e., nDGP(1), nDGP(2), nDGP(5), respectively, indicate the importance of the crossover distance $r_{\rm c}$, with higher orders of nDGP models showing more significant deviations from GR.

Our findings confirm that the predictions for the merger rate of PBH-PBH binaries under MG models, especially at lower power-law index values than those predicted by GR, fall within the sensitivity range of the LVK detectors. This indicates that current and future GW observations could potentially distinguish between GR and MG scenarios by analyzing the merger rates of compact binaries in these environments. Our analysis reveals that the observed discrepancies in merger rates between GR and MG models underscore the necessity of incorporating MG effects when studying compact binary mergers in dark-matter spikes. This approach provides a more comprehensive understanding of the dynamics of these systems and their implications for GW astronomy. Furthermore, our comparison with observational data on the abundance of PBHs and those from the LVK detectors suggests that MG models can offer valuable insights into the formation and evolution of compact binaries. This underscores the potential of GW observations to test and constrain alternative theories of gravity.

This analysis highlights the critical role of MG models in accurately predicting the merger rates of compact binaries within dark-matter spikes. By carefully selecting SMBH mass functions and considering the effects of MG, one can better understand the dynamics of these systems and their implications for cosmology and GW astronomy. Our findings pave the way for future research to further explore the interplay between dark matter, SMBHs, and MG in shaping the merger rates of compact binaries.

It is intresting to note that the enhanced density spike around SMBHs can significantly affect the dynamics of nearby PBHs, potentially creating a gravitational well that facilitates their capture and merger with SMBHs. This interaction leads to extreme mass ratio inspirals (EMRIs), where a smaller BHs spirals into a larger one, producing unique GW signatures. The dense environment increases the likelihood of close encounters, enhancing the merger rate of PBHs with SMBHs. MG models, such as Hu-Sawicki $f(R)$ and nDGP gravities, introduce an effective fifth force that can enhance gravitational interactions, further increasing merger rates compared to general relativity. As these systems evolve, energy loss through gravitational radiation causes smaller BHs to lose orbital energy and spiral inward. The anticipated rise in EMRI rates is particularly exciting for space-based observatories like LISA \citep{2017arXiv170200786A}, which could provide critical insights into fundamental physics and the conditions of the early Universe, including the nature of dark matter. These MG theories may also alter our understanding of cosmic evolution by influencing the formation mechanisms of both SMBHs and PBHs, making them a compelling focus for future studies in astrophysics and cosmology.

In this work, while we have stated that LVK observations could differentiate between GR and MG scenarios based on merger rate of compact binaries in dark matter spikes, it is crucial to recognize the multifaceted nature of these event rates, which arise from various channels such as isolated binary evolution, dynamical mergers in star clusters, and contributions from PBHs. Each of these channels can be further subdivided into numerous subchannels, complicating the interpretation of merger rate data \citep{2021NatAs...5..749G}. The uncertainty surrounding the proportion of mergers attributable to PBHs adds another layer of complexity, as it remains unclear which specific events contribute to the overall merger rate. Consequently, distinguishing between different MG models may be constrained by these astrophysical uncertainties. Our findings suggest that these uncertainties could overshadow the potential to differentiate between various MG models based solely on merger rates. Therefore, while LVK observations provide valuable data, the inherent complexities and uncertainties in astrophysical processes necessitate a cautious approach when interpreting the implications for MG theories.
\section*{ACKNOWLEDGMENTS} 
SF acknowledges the Deputy for Research and Technology, Shahid Beheshti University.
\section*{List of Abbreviations}
\begin{table}[ht!]
	\begin{center}
		\label{table 1}
		\begin{tabular}{c|c}
			\hline
			{\bf Abbreviation} & {\bf Full Term} \\
			\hline
		BBH & Binary Black Hole\\
		BNS & Binary Neutron Star\\
		BH & Black Hole\\
		BH-NS & Black Hole-Neutron Star\\
		GR & General Relativity\\
		GW & Gravitational Waves\\
		LVK & LIGO-Virgo-KAGRA\\
		MG & Modified Gravity\\
		nDGP & Normal Branch of Dvali-Gabadadze-Porrati\\
		NS & Neutron Star\\
		NFW & Navarro-Frenk-White\\
		PBH & Primordial Black Hole\\
		sDGP & Self-accelerating Dvali-Gabadadze-Porrati\\
		SMBH & Supermassive Black Hole\\
			\hline
			\hline
		\end{tabular}
	\end{center}
\end{table}

\noindent 

\bibliography{sample631}{}
\bibliographystyle{aasjournal}



\end{document}